\documentclass[preprint,12pt]{elsarticle}
\usepackage{amssymb}
\usepackage{amsmath}
\usepackage{slashed}
\usepackage{undertilde}
\usepackage{multirow}
\usepackage{color}
\usepackage{framed}
\usepackage{graphicx}
\usepackage{multicol}
\usepackage{colortbl}

  \newcounter{bla}

 \definecolor{lightgrey}{gray}{0.9}
\def\btabu#1\etabu{\begin{tabular}{p{125mm}}#1\end{tabular}}
\def\btab#1\etab{\begin{tabular}{p{50mm}p{70mm}}#1\end{tabular}}
\def\btabnn#1\etabnn{\begin{tabular}{p{45mm}p{75mm}}#1\end{tabular}}
\def\btabx#1\etabx{\begin{tabular}{p{65mm}p{55mm}}#1\end{tabular}}
\def\btaby#1\etaby{\begin{tabular}{p{40mm}p{80mm}}#1\end{tabular}}
\def\btabyy#1\etabyy{\begin{tabular}{p{20mm}p{100mm}}#1\end{tabular}}
\def\btabzz#1\etabzz{\begin{tabular}{p{35mm}p{85mm}}#1\end{tabular}}
\def\btabyyy#1\etabyyy{\begin{tabular}{p{10mm}p{110mm}}#1\end{tabular}}
\def\btabyyyy#1\etabyyyy{\begin{tabular}{p{2mm}p{118mm}}#1\end{tabular}}
\def\btabwide#1\etabwide{\begin{tabular}{p{82mm}p{38mm}}#1\end{tabular}}
\def\bcen{\begin{center}}
\def\ecen{\end{center}}
 
\def\bgfb#1\egfb{\bcen\fcolorbox{black}{lightgrey}{\parbox{130mm}{\btabu#1\etabu}}\ecen}
\def\bgfbn#1\egfbn{\bcen\fcolorbox{black}{lightgrey}{\parbox{130mm}{\btab#1\etab}}\ecen}
\def\bgfbnn#1\egfbnn{\bcen\fcolorbox{black}{lightgrey}{\parbox{130mm}{\btabnn#1\etabnn}}\ecen}

\def\bgfbx#1\egfbx{\bcen\fcolorbox{black}{lightgrey}{\parbox{130mm}{\btabx#1\etabx}}\ecen}
\def\bgfbyy#1\egfbyy{\bcen\fcolorbox{black}{lightgrey}{\parbox{130mm}{\btabyy#1\etabyy}}\ecen}
\def\bgfbyyyy#1\egfbyyyy{\bcen\fcolorbox{black}{lightgrey}{\parbox{130mm}{\btabyyyy#1\etabyyyy}}\ecen}
\def\bgfbzz#1\egfbzz{\bcen\fcolorbox{black}{lightgrey}{\parbox{130mm}{\btabzz#1\etabzz}}\ecen}
\def\bgfbyyy#1\egfbyyy{\bcen\fcolorbox{black}{lightgrey}{\parbox{130mm}{\btabyyy#1\etabyyy}}\ecen}

\def\bgfbalign#1\egfbalign{\bcen\fcolorbox{black}{lightgrey}{\parbox{130mm}{\btaby#1\etaby}}\ecen}
\def\bgfbwide#1\egfbwide{\bcen\fcolorbox{black}{lightgrey}{\parbox{130mm}{\btabwide#1\etabwide}}\ecen}

\newcommand{\etc}{{\it etc}}
\newcommand{\ie}{{\it i.e.}}
\newcommand{\eg}{{\it e.g.}}

\newcommand{\feynrules}{{\sc FeynRules}}

\newcommand{\ROOT}{{\sc Root}}
\newcommand{\python}{{\sc Python}}
\newcommand{\cpp}{{\sc C++}}

\newcommand{\be}{\begin{equation*}}
\newcommand{\bed}{\begin{equation}}

\newcommand{\ee}{\end{equation*}}
\newcommand{\eed}{\end{equation}}
\def\bsp#1\esp{\begin{split}#1\end{split}}
\def\bpm{\begin{pmatrix}}
\def\epm{\end{pmatrix}}

\definecolor{purple}{rgb}{0.62,0.12,0.94}
\definecolor{grey}{rgb}{0.3,0.3,0.3}
\definecolor{orange}{rgb}{1,0.5,0}

\journal{Computer Physics Communications}

\begin{document}

\newcommand{\madanalysis}{{\sc MadAnalysis}}
\newcommand{\madgraph}{{\sc MadGraph}}
\newcommand{\alpgen}{{\sc Alpgen}}
\newcommand{\comix}{{\sc Comix}}
\newcommand{\calchep}{{\sc CalcHep}}
\newcommand{\comphep}{{\sc CompHep}}
\newcommand{\helac}{{\sc Helac}}
\newcommand{\mgme}{{\sc MadGraph/MadEvent}}
\newcommand{\sherpa}{{\sc Sherpa}}
\newcommand{\whizard}{{\sc Whizard}}
\newcommand{\gosam}{{\sc GoSam}}
\newcommand{\herwig}{{\sc Herwig}}
\newcommand{\pythia}{{\sc Pythia}}
\newcommand{\stdhep}{{\sc StdHep}}
\newcommand{\lanhep}{{\sc LanHep}}
\newcommand{\lhe}{{\sc LHE}}
\newcommand{\lhco}{{\sc LHCO}}
\newcommand{\hepmc}{{\sc HepMC}}
\newcommand{\delphes}{{\sc Delphes}}
\newcommand{\pgs}{{\sc PGS}}
\newcommand{\rooot}{{\sc Root}}
\newcommand{\sampleanalyzer}{{\sc SampleAnalyzer}}
\newcommand{\fortran}{{\sc Fortran}}
\newcommand{\perl}{{\sc Perl}}
\newcommand{\xml}{{\sc XML}}
\newcommand{\fastjet}{{\sc FastJet}}
\newcommand{\exroot}{{\sc ExRootAnalysis}}
\newcommand{\gpp}{{\sc g++}}
\newcommand{\html}{{\sc html}}
\newcommand{\latex}{{\sc latex}}
\newcommand{\pdflatex}{{\sc pdflatex}}
\newcommand{\lhapdf}{{\sc LHAPDF}}
\newcommand{\pdflib}{{\sc PDFLib}}
\newcommand{\doxy}{{\sc Doxygen}}
\newcommand{\maweb}{{\texttt{http://madanalysis.irmp.ucl.ac.be}}}

\begin{frontmatter}
  \title{\madanalysis\ 5, a user-friendly framework for collider phenomenology}

  \author[a]{Eric Conte}
  \address[a]{Groupe  de Recherche de Physique des Hautes Energies (GRPHE),
    Universit\'e de Haute-Alsace, IUT Colmar, 34 rue du Grillenbreit BP 50568, 
    68008 Colmar Cedex, France\\ E-mail: eric.conte@iphc.cnrs.fr}
  \author[b]{Benjamin Fuks}
  \address[b]{Institut Pluridisciplinaire Hubert Curien/D\'epartement Recherches
     Subatomiques, Universit\'e de Strasbourg/CNRS-IN2P3, 23 Rue du Loess, F-67037
     Strasbourg, France\\E-mail: benjamin.fuks@iphc.cnrs.fr, 
     guillaume.serret@iphc.cnrs.fr}
  \author[b]{Guillaume Serret}

  \begin{abstract}
  \begin{flushright}
    \vspace{-10cm}  IPHC-PHENO-12-06 \vspace{10cm}
  \end{flushright}
  We present \madanalysis\ 5, a new framework for phenomenological
  investigations at particle colliders. Based on a \cpp\ kernel, this program
  allows us to efficiently perform, in a straightforward and user-friendly fashion,
  sophisticated physics analyses of event files such as those generated by a large
  class of Monte Carlo event generators. \madanalysis\ 5 comes with two modes of
  running. The first one, easier to handle, uses the strengths of a powerful
  \python\ interface in order to implement physics analyses by means of a set of
  intuitive commands. The second one requires one to implement the analyses in the
  \cpp\ programming language, directly within the core of the analysis
  framework. This opens unlimited possibilities concerning the level of
  complexity which can be reached, being only limited by the
  programming skills and the originality of the user.
  \end{abstract}
  

\end{frontmatter}
\newpage

\noindent {\bf PROGRAM SUMMARY}                                               \\
  \begin{small}
  {\em Manuscript Title:} \madanalysis\ 5, a user-friendly framework for collider
    phenomenology. \\
  {\em Authors:} Eric Conte, Benjamin Fuks, Guillaume Serret.                 \\
  {\em Program Title: \madanalysis\ 5}                                        \\
  {\em Licensing provisions:} Permission to use, copy, modify and distribute
this program is granted under the terms of the GNU General Public License.    \\
  {\em Programming language:} \python, \cpp.                                  \\
  {\em Computer:} All platforms on which \python\ version 2.7, \ROOT\ version
5.27 and the \gpp\ compiler are available. Compatibility with newer versions of
these programs is also ensured. However, the \python\ version must be below
version 3.0. \\ 
  {\em Operating system:} {\sc Unix}, {\sc Linux} and {\sc Mac OS} 
    operating systems on which
    the above-mentioned versions of \python\ and \ROOT, as well as \gpp, are
    available. 
  {\em Classification:} 11.1 General, High Energy Physics and Computing.      \\
  {\em Nature of problem:} Implementing sophisticated phenomenological analyses
    in high-energy physics through a flexible, efficient and straightforward
    fashion,
    starting from event files as those produced by Monte Carlo event generators. The
    event files can have been matched or not to parton-showering and can have
    been processed or not by a (fast) simulation of a detector. According to the
    sophistication level of the event files (parton-level, hadron-level,
    reconstructed-level), one must note that several input formats are possible.\\
  {\em Solution method:} We implement an interface allowing to produce
    predefined as well as user-defined histograms for a large class of
    kinematical distributions after applying a set of event selection cuts
    specified by the user. This therefore allows to devise robust and novel
    search strategies for collider experiments, such as those currently running
    at the Large Hadron Collider at CERN, in a very efficient way. \\
  {\em Restrictions:} Unsupported event file format. \\
  {\em Unusual features:} The code is fully based on object representations for
    events, particles, reconstructed objects and cuts, which facilitates the
    implementation of an analysis. \\
  {\em Running time:} It depends on the purposes of the user and on the number of
    events to process. It varies from a few seconds to the order of the minute
    for several millions of events.\\
\end{small}
\newpage
\tableofcontents
\newpage
\section{Introduction} \label{sec:intro}

Among the key topics of the present experimental program of high-energy physics
lies the quest for new physics and the identification of the fundamental building
blocks of matter, together with their interactions. In
these prospects, the Large Hadron Collider (LHC) is currently exploring the TeV
scale and multi-purpose experiments, such as ATLAS or CMS, are currently pushing
the limits on beyond the Standard Model physics to a further and further
frontier. Discoveries from these experiments, together with their
interpretation, are in general challenging and strongly rely on our ability to
accurately simulate both the possible candidate signals and the backgrounds.
This task is however rendered quite complicated due to the complexity of the
typical final states to be produced at the LHC, which contain large numbers of
light and heavy flavor jets, charged leptons and missing transverse energy.
Consequently, the overwhelming sources of Standard Model background require the
development of robust, and possibly novel, search strategies. In this context,
tools allowing us to compute predictions for large classes of models are central. 

This has triggered, during the last twenty years, a lot of efforts dedicated to
the development of multi-purpose matrix-element based event generators such as
\alpgen\ \cite{Mangano:2002ea}, \comix\ \cite{Gleisberg:2008fv},
\comphep/\calchep\ \cite{Pukhov:1999gg,Boos:2004kh,Pukhov:2004ca}, \helac\
\cite{Cafarella:2007pc}, \mgme\
\cite{Stelzer:1994ta,Maltoni:2002qb,Alwall:2007st,Alwall:2008pm,Alwall:2011uj},
\sherpa\ \cite{Gleisberg:2003xi,Gleisberg:2008ta} and \whizard\
\cite{Moretti:2001zz,Kilian:2007gr}. As a result, the problem of the generation
of parton-level events, at the leading-order accuracy, for many renormalizable
or non-renormalizable new physics theories has been solved. More recently,
progress has also been achieved in the automation of next-to-leading-order
computations. On
the one hand, the generation of the real emission contributions with the
appropriate 
subtraction terms has been achieved in an automatic way \cite{Gleisberg:2007md,%
Seymour:2008mu,Hasegawa:2008ae,Frederix:2008hu,Czakon:2009ss,Frederix:2009yq}.
On the other hand, several algorithms addressing the numerical calculation of loop
amplitudes have been proposed \cite{Ossola:2007ax,Zanderighi:2008na,%
Hirschi:2011pa,Cullen:2011ac,Cascioli:2011va} and successfully applied to the computation of
Standard Model processes of physical interest
\cite{Ellis:2009zw,Berger:2009zg,vanHameren:2009dr,Berger:2010vm,Berger:2010zx}.

Even if each of the above-mentioned tools is based on a different philosophy,
uses a specific programming language and requires a well-defined
input format for the physics models under consideration, programs such
as \feynrules\ \cite{Christensen:2008py,Christensen:2009jx,Christensen:2010wz,%
Duhr:2011se,Fuks:2012im} and \lanhep\ \cite{Semenov:1998eb,Semenov:2008jy}
have alleviated the time-consuming and error-prone task
of implementing new physics theories in the Monte Carlo tools. Furthermore, the
introduction of the Universal \feynrules\ Output (UFO) format \cite{Degrande:2011ua}
and the development of an automated tool computing helicity amplitudes
\cite{aloha}
have also streamlined the communication between the construction of a new
physics theory and its implementation in the matrix-element generators through 
a standardized fashion.

Parton-level physics analyses based on event samples produced by 
matrix-element generators are far from describing the reality of what is
observed in any existing
detector. This kind of phenomenological work can however be useful in the
prospects of
investigating new types of signature and devising original search
strategies, preliminary to more complete phenomenological investigations. In
order to facilitate the information transfer from the matrix-element generators,
a generic format for storing (parton-level) events
and their properties has been proposed ten years ago, the
so-called Les Houches Event (\lhe) file format \cite{Boos:2001cv,Alwall:2006yp}.
Following the \lhe\ standards, events are stored in a single file, following an
\xml-like structure, together with additional information related to the way in
which the events have been generated. 

Monte Carlo generator-based physics analyses at the parton-level 
then consist in first parsing \lhe\ event files related to both the signal and
the different
sources of background, then implementing various selection cuts on the objects
contained in the events, \ie, quarks, leptons, neutrinos, new stable particles,
\etc, and finally in creating histograms representing several kinematical
quantities. By means of signal over background ratios, optimization of the
selection cuts can be achieved in order to maximize the chances to unveil the 
signal of interest. Of
course, the only sensible conclusions which could be stated at this stage would
be to motivate (or not) a more realistic analysis, including at least parton 
showering and hadronization.

An accurate simulation of the collision to be observed at hadron colliders
indeed requires a proper modeling of the strong interaction, including parton
showering, fragmentation and hadronization. This is efficiently provided by
packages such as \pythia\ \cite{Sjostrand:2006za,Sjostrand:2007gs} and \herwig\
\cite{Corcella:2000bw,Bahr:2008pv} and several algorithms matching parton
showering to hard scattering matrix elements have been recently developed
\cite{Catani:2001cc,Krauss:2002up,Mrenna:2003if,Mangano:2006rw}.

Even if not directly necessary for the analysis of the event samples, a large
part of the information present in a parton-level LHE event file allows for a
further matching
procedure. Consequently, the generation of hadron-level events for a
multitude of Standard Model and beyond the Standard Model processes can be (and
has been) done in a systematic fashion. This starts from
parton-level events stored in \lhe\ format-compliant files, as generated by
matrix-element based event generators. These files are then further
processed by a parton showering and hadronization code which allows us to match the
strengths of both the description of the physics embedded into the matrix
elements and the one modeled by the parton showering. Consequently, physics
analyses at the hadron-level are far more sophisticated than their counterparts
at the parton-level.

When analyzing hadron-level events, one must note that the key difference
with respect to the parton-level case lies in the objects which are contained
in the event files. In contrast to partonic events, hadronic events consist in
general in a huge collection of hadrons given together with their
four-momentum. The particle content of the final state is thus much richer and
the storing of the information at the time of parsing the event file is hence
rendered rather cumbersome.

In order to streamline this procedure, dedicated event class libraries have
been developed and several common formats for outputting hadron-level event
samples exist. Event files compliant with such formats
can in general be straightforwardly read by fast detector simulation programs,
necessary for a more advanced level of sophistication of the phenomenological
analysis. As examples, one finds the \stdhep\ \cite{stdhep} or \hepmc\
\cite{Dobbs:2001ck} structure for event files, both widely used in the
high-energy physics community. According to these conventions, an event
contains, in addition to the final state
particles and their properties, the whole event history as a set of
mother-particle to daughter-particle relations. Let us also note that the task
of parsing hadron-level event files can be highly simplified after clustering
the hadrons into jets, using jet algorithms such as, \eg, those included in the
\fastjet\ program
\cite{Cacciari:2005hq}. This simplified picture allows for the usage of a
simpler event format, such as the \lhe\ format introduced above,
which subsequently renders an analysis easier to implement.

State-of-the-art phenomenological analyses require in general fast (and
realistic)
detector simulation in order to correctly estimate both the signals under
consideration and the backgrounds. Starting from hadronized event samples,
several frameworks, such as \pgs\ 4 \cite{pgs} and \delphes\ \cite{Ovyn:2009tx}
are dedicated to this task. They produce event samples containing reconstructed 
objects such as photons, jets,
electrons, muons or missing energy, together with their properties. 
Whereas the description of a specific signature as provided by a fast detector
simulation tool is still far from what could have been obtained using a full
detector
simulation, including among others the transport of the particles through the
detector material, fast simulation of collider experiments is often sufficient
to study the feasibility of a specific analysis on realistic grounds. The results
then motivate (or not) the associated analysis in the context of a
full detector simulation, the latter being embedded in generally complex
experimental software such as those used by large collaborations.

As stated above, after their processing by a fast detector simulation, the
original sets of hadrons are replaced by high-level reconstructed objects
stored together with properties
such as their (smeared) four-momentum or if they have been tagged as a $b$-jet
or not. The structure of the events and their properties can be very efficiently
saved into a file following the \lhco\ conventions \cite{lhco}. Let us note that
this format has been designed in particular for that purpose.

As outlined above, the study of the hadron collider phenomenology of a
given signature can be performed at several levels of sophistication. Sometimes,
very preliminary works at the parton-level are necessary in order to motivate
further investigations. Hadron-level analyses give a clearer hint about the
expected sensitivity of colliders to the signal under consideration. This
assumes a perfect and ideal detector. In contrast, state-of-the-art
phenomenology includes a fast and semi-realistic simulation of the detectors.
This allows us to derive conclusions closer to the expectations 
estimated with the help of a full simulation of the detector as performed by 
the large collaboration software. The major drawback of the
latter is that such tools consist in very heavy, complex, time-consuming and
non-public 
algorithms. This therefore motivates pioneering works under the parton-level,
hadronic-level or reconstructed-level assumptions\footnote{In this
paper, we denote by \textit{reconstructed-level} events which have already been
processed by a fast detector simulation, in contrast to \textit{hadron-level}
events which have only been showered, fragmented and hadronized and
\textit{parton-level} events where the matching to a parton showering algorithm
is fully absent.}. It is however important to keep in mind that the final word
is always included in the data. 

Performing a phenomenological analysis on the basis of the results provided by
Monte Carlo generators always starts with the reading of
several event samples. Since partonic or hadronic events (the latter including
or not a fast detector simulation) are in general stored under different formats,
this step demands to use appropriate routines capable of understanding the file 
structure. Selection cuts on the objects included in the events, \ie, partons,
hadrons or high-level objects, according to the level of sophistication, are
then
implemented and applied to both the signal and background samples. This allows
for the creation of histograms of various quantities in order to be able to
extract (or not, in the worst case scenario) information about a signal in
general swamped by backgrounds.

This procedure is most of the time based on home-made and non-public programs,
especially due to the lack of a dedicated framework. As a consequence, this can
lead to various problems in the validation and the traceability of the analyses
as well as possibly in the interpretation of the results. In this work, we are
alleviating this issue by proposing a single efficient framework for
phenomenological analyses at any level of sophistication. We introduce the
package \madanalysis\ 5, an open source program based on a
multi-purpose \cpp\ kernel, denoted \sampleanalyzer, which uses the \rooot\
platform \cite{Brun:1997pa}.

Using the strengths of a \python\ interface, the user can define his own physics
analysis in an efficient, flexible and straightforward way. Similarly to the
older \fortran\ and \perl\ version of \madanalysis\ which was linked to 
version 4 of the \madgraph\ program, \madanalysis\ 5 can either be run within
\madgraph\ 5 or as a standalone package. It includes a complete
reorganization of the code and the implementation of many novel functionalities
such as an efficient method to implement cuts as well as to generate
histograms and cut-flow charts in an automated fashion. Moreover, care has been
taken in developing a fast and optimized code.

Consequently, the procedure for performing a phenomenological analysis has been
drastically simplified since the only task left to the user is to define the
corresponding selection cuts and the distributions to be computed. Sometimes,
these embedded features might however not be sufficient according to the needs
of the 
user. In order to overcome this limitation, \madanalysis\ 5 offers an expert
mode of running with unlimited possibilities. It is unlimited in the sense that
the user directly implements, within the \cpp\ kernel, his own analysis. 

Finally, in order to release a single framework for the analysis of
parton-level, hadron-level or reconstructed-level based event simulations,
several event file formats are supported as input, from the \lhe\
files which could describe partonic or hadronic events to the more complex
\stdhep, \hepmc\ and \lhco\ file formats. Let us note that, according to the
needs of the users, interfacing additional event formats, such as, \eg, the \exroot\
format \cite{exroot}, can be easily achieved. 

This paper documents the fifth version of \madanalysis\ and consists in its
user guide. An up-to-date version of this document, together with the program 
can be found on the web page

\maweb
\medskip

\noindent Moreover, an
extended, more pedestrian, version of this manual is also available at the same
Internet address.
The outline of
this paper is as follows. In Section \ref{sec:ma5overview}, we give a general
overview of the program, its philosophy and its structure. Section
\ref{sec:firststeps} illustrates the capabilities of \madanalysis\ 5 without
entering into the details, the latter being given in the next sections. Hence, 
Section \ref{sec:normalmode} provides the guidelines for implementing basic
(but still professional) physics
analyses, \ie, implementing simple selection cuts and creating histograms for
various kinematical distributions, whilst Section \ref{sec:expert} is dedicated
to a more
expert usage of the program which allows us to design more sophisticated analyses.
Our conclusions are presented in Section \ref{sec:conclusions}. Finally, a
technical Appendix on the installation of the program and its dependencies
follows.
%

\section{Overview of \madanalysis\ 5}\label{sec:ma5overview}
\subsection{\madanalysis\ 5 in a nutshell}
The \madanalysis\ 5 package is an open source program allowing us to perform
physics analyses of Monte Carlo event samples in an efficient, flexible and 
straightforward way. It relies on a \cpp\ kernel, named \sampleanalyzer, 
which uses the \rooot\ platform
and interacts with the user by means of a \python\ command line interface.

The distribution of the program includes a home-made reader of event files
created by Monte Carlo event generators. The reader is compliant with Monte Carlo 
samples containing events either at the parton-level, at the hadron-level
or at the reconstructed-level. Hence, a unique framework can be used for
analyzing events embedded equivalently 
into simplified \lhe\ files or into more complex \stdhep, \hepmc\ and \lhco\
files. 

From the information included in the event files, the user can ask 
\madanalysis\ 5 to
generate histograms illustrating various properties of the generated physics
processes. These
properties range from observables related to the whole content of the events,
such as the multiplicity of a given particle species or the missing
transverse energy distribution, to observables associated to a specific particle
such as, for instance,
the transverse-momentum distribution of the final-state muon with the
highest energy or the angular separation between the final-state jets. The user
has also the possibility to compare event samples related to different
physical processes in a straightforward fashion. On the one hand, the investigated 
distributions can be 
stacked on the same histogram, after automatic normalization of the samples 
to an integrated luminosity which the user can specify. 
On the other hand, the distributions can be normalized to unity in order
to facilitate the design of event selection cuts allowing for an efficient
background rejection based on the shape of the distribution. In this case, the
histograms could be superimposed rather than stacked in the aim of improving 
readability. 

Moreover, if available, the whole event history, \ie, all the
event information from the hard-scattering process to the final-state hadrons,
including the mother-particle to daughter-particle relations among the different
stages yielding the final state particle content, is stored. One can emphasize
that this feature is important for sanity and consistency checks of the
generated event samples.

Event selection can be easily performed by means of a series of intuitive
\python\ commands yielding the application of selection cuts to the  
samples under consideration. For
instance, it is possible to only consider, in the current analysis,
events which contain at least a certain amount of missing transverse energy
or a given number of final-state leptons. In the same way, one can access a
specific particle in the event, such as the jet with the hardest
transverse momentum, and require that it fulfills a given geometrical or
kinematical criterion. As a last example, the user could decide to only select
events containing at least two muons whose  
invariant mass is compatible with the $Z$-boson mass. After applying a given
cut, \madanalysis\ 5 proceeds with an automatic computation of the 
associated cut efficiency. 

The ultimate goal of a physics analysis in particle physics is to extract some
signal from a usually swamping background so that one could study its
properties. The \madanalysis\ 5 framework offers the
possibility to tag a specific event sample as a background or signal sample. 
This tagging allows
for an automatic treatment of the signal over background ratio, or of any other
similar observable which can be specified by the user, together with the
associated uncertainty. This quantity is recomputed after each of the different
selection cuts implemented by the user. With these pieces of information
available, optimizing selection cuts consequently becomes easier, which
allows us to investigate in a fast and efficient way whether a given signature
could be observable at colliders.

To display the results in a human-readable form, \madanalysis\ 5 can collect
them either into a \latex\ document to be further compiled or under the form of
an \html\ webpage.

\subsection{Basic concepts}\label{sec:basic_concepts}
The \madanalysis\ 5 program provides to the user a platform including a wide
class of functionalities allowing one to perform sophisticated physics analyses.
Even if the existing possibilities are rather large, see
Section \ref{sec:normalmode} and Section \ref{sec:expert}, implementing an
analysis within the \madanalysis\ 5 framework always follows the same steps, 
each of these steps being linked to one or several of the key features of the
program. These key features are briefly described in this Section, whilst
additional and more detailed information can be found in the rest of this
manual.\\

\noindent \textbf{Sample declaration - datasets}.\\
\indent 
When implementing an analysis within the framework of \madanalysis\ 5, the first
task which is asked of the user is to indicate the Monte Carlo event files to be
processed on run-time. In general, a full Monte Carlo event generation requires
the simulation of several physical processes, such as the signal under
consideration and the associated sources of background. Furthermore, for the
processes with the highest cross sections, more than one event sample may have
to be generated in order to get enough statistical significance. The user then
requires these samples, describing the same physics, to be treated on the same 
footing. In \madanalysis\ 5, this can be performed in a very natural way. The
event files to be merged are gathered under the form of an object dubbed a
\textit{dataset}. When the analysis is effectively executed by the \cpp\ core
of the program, datasets, \ie, collections of event files, are processed rather
than the event files individually.\\

\noindent \textbf{The particle content of the events - particles and
multiparticles}.\\ 
\indent According to the conventions of the different event formats introduced
above, the particles
described in the events are defined, in an unambiguous way, through their
Particle Data Group identifier (PDG-id) \cite{Nakamura:2010zzi}. Similarly, the
algorithms generated by \madanalysis\ 5 are widely based on this 
concept to distinguish the various particle species. The
drawback is obviously that this identifier is most of the time non-intuitive for
the user and can even become unreadable when the set of particles included in
the events becomes large. For instance, the particle content of hadron-level
events consists in hundreds of different particles, each of them being identified
by a different PDG-id. 

In the spirit of the \madgraph\ program, \madanalysis\ 5
alleviates this issue by allowing us to associate a \textit{particle label} to a
given PDG-id. Hence, instead of representing an electron and a positron by the
integer numbers 11 and -11, one can create the (more intuitive) labels
\texttt{e-} and \texttt{e+} and associate them to the PDG-ids 11 and -11,
respectively. It is also possible to collect labels together
through the concept of \textit{multiparticles}. Hence, one could define a label 
\texttt{e} referring to both the electron and the positron. 

In order to implement an analysis in an efficient way, 
it is recommended to the user to define, in a first step, a series of particle
and multiparticle labels facilitating the readability (and then the validation)
of his analysis.\\

\noindent \textbf{Definition of the analysis - selections}.\\ 
\indent Implementing the analysis is the core task of the user. It consists in 
defining  the histograms that have to be generated and the selection cuts that
need to be applied. These two types of objects, \ie, histograms and cuts, are 
uniquely dubbed \textit{selections}. It can be noted that even if
asking for the generation of several histograms is a commutative operation, the 
ordering of the selection cuts is in contrast important. Applying the cuts in a
different order indeed leads to the production of different (intermediate)
histograms and efficiency tables.\\

\noindent \textbf{Running the analysis - jobs}.\\ 
\indent After having created a series of dataset objects and defined the
analysis to be performed, \ie, having implemented the histograms and selection
cuts of interest, the user needs to execute the analysis on the datasets. 
Contrary to the previous steps which rely on the \python\ module of
\madanalysis\ 5, this task is built, for efficiency reasons, upon a \cpp\ code
which is generated by \madanalysis\ 5 and denoted as a \textit{job}. After
\madanalysis\ 5 creates the job, it has to be 
executed by the \sampleanalyzer\ kernel for each of the predefined
datasets. Once the analysis has been performed, the results are subsequently
imported in the \python\ module and re-interpreted by \madanalysis\ 5.\\

\noindent \textbf{Display of the results - reports}.\\ 
\indent The generated results can be collected and displayed in a synthetic
\textit{report}. This report is given either under the \html\ format, as a
webpage, or as a \latex\ document that has to be further compiled. The report 
contains the 
histograms which have been generated and the efficiency tables related to the
implemented selection cuts. Furthermore, a signal over background table is
generated provided the datasets defined by the user have been tagged as signal 
and background samples. 

\subsection{Logical architecture of the program}
\begin{figure}[t]
  \center \includegraphics[width=0.90\textwidth]{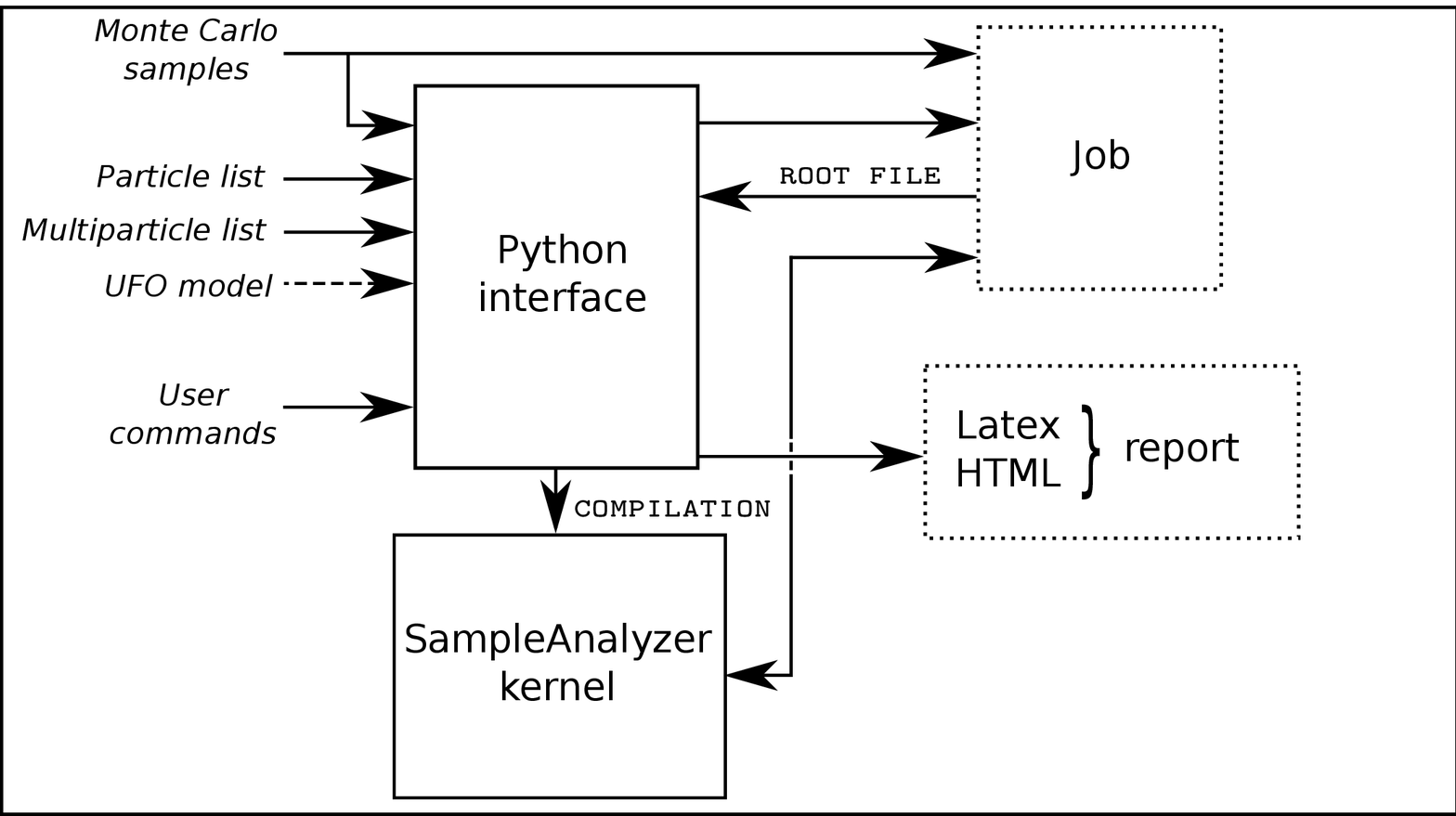}
  \caption{\label{fig:flowchart}The \madanalysis\ 5 flowchart.}
\end{figure}

The \madanalysis\ 5 program has two components, a \python\ command line
interface and a \cpp\ kernel dubbed \sampleanalyzer. In the normal mode of
running of the program, these two modules interact at different stages, as
illustrated in Figure \ref{fig:flowchart}. The way to perform physics analyses in
this mode is presented in detail in Section \ref{sec:firststeps} and
Section \ref{sec:normalmode}. For more sophisticated analyses, users with
advanced skills in programming can bypass the \python\ interface and directly
implement their analysis within the \sampleanalyzer\ framework. This expert mode
of running the code is described in detail in Section \ref{sec:expert}.

The \python\ command line interface of \madanalysis\ 5 consists in a command 
prompt where the user
accesses all the functionalities of the program through a set of commands. This
allows to implement an analysis in a very user-friendly way. 
Each command entered by the user is first checked from the
point of view of the syntax and if necessary, an error message is printed to the
screen. The issued command is then stored in the memory of the computer.

The interface can import different types of external information. Hence, 
a predefined list of particle and multiparticle labels could be imported if
given as a text file or as a UFO model \cite{Degrande:2011ua}. In a similar way,
the list of the Monte Carlo event files to be analyzed can be easily loaded in the
current session of \madanalysis\ 5. 

After the user has typed in all the
commands defining his analysis (definition of the datasets, histograms
and selection cuts), the \python\ interface creates a \cpp\ code using the
\sampleanalyzer\ framework, the previously introduced \madanalysis\ 5 job. This
\cpp\ code comes with a {\tt Makefile} and is kept available to the user for 
further modifications or improvements of the analysis without having to be
regenerated. After the compilation and execution of the job, the \python\
interface loads the results and uses the
\rooot\ library of functions \cite{Brun:1997pa} to normalize and draw the
histograms according to the requirements of the user. The \html\ and \latex\
reports are eventually generated.

The \sampleanalyzer\ \cpp\ kernel of \madanalysis\ 5 is, strictly speaking,
the part of the program dedicated to the analysis of the Monte Carlo event
samples itself. It consists in a framework built upon an adaptive data
format common for all types of Monte Carlo event samples and which contains a
series of well-suited functionalities. In addition, it includes a 
reader compliant with the \lhe, \stdhep, \hepmc\ and \lhco\ event formats
and a library of specific functions facilitating particle physics analyses.
Among the latter, one finds, \eg, methods allowing for boosting four-momenta or
testing if a particle is a final-state particle or not. Let us finally note that
the storage of information within the context of the \sampleanalyzer\ platform
is widely based on the functions implemented within the \rooot\ library.

\section{First steps with \madanalysis\ 5}\label{sec:firststeps}

In this Section, we present the philosophy, main features and the
user-friendliness of \madanalysis\ 5 by means of a simple example. In
contrast, the full list of capabilities of \madanalysis\ 5, which are much
broader than what is shown in this Section, are described in Section
\ref{sec:normalmode}.

For the sake of
the illustration, we decide to perform a toy analysis at the parton-level. We
consider several samples of 1000 events each describing various Standard
Model hard-scattering processes at the LHC running at a center-of-mass
energy of 7 TeV. Events are generated with the Monte
Carlo generator \madgraph\ 5 \cite{Alwall:2011uj}. Neglecting all quark masses
(but the top mass), we employ the leading order set of the CTEQ6 parton density
fit \cite{Pumplin:2002vw} and identify both the renormalization and
factorization scales as the transverse mass of the produced particles.
 
We consider four different event samples describing three different physical
processes. They are stored into the directory \texttt{samples} of \madanalysis\
5. If this directory is not present on the system of the user or if the event
files are not there for any reason, one can type, once the command line
interface of \madanalysis\ 5 has been started (see Section \ref{sec:fs_start}),
\begin{verbatim}
  install samples
\end{verbatim}
This leads to the creation of the directory \texttt{samples}, if relevant, and
to the download from the Internet of the four samples 
\begin{verbatim}
  ttbar_sl_1.lhe.gz
  ttbar_sl_2.lhe.gz
  ttbar_fh.lhe.gz
  zz.lhe.gz
\end{verbatim} 
The first two samples are related to the production of a semi-leptonically
decaying
top-antitop pair where the lepton is an electron or a muon, and the third one is
related to the production of a fully
hadronically decaying top-antitop pair. The last sample describes the
production of a $Z$-boson pair, including also diagrams with virtual
photons, where each of the bosons is decaying either to an electron pair or to a
muon pair.
At the time of event generation, we demand that the produced parton-level jets
have a transverse momentum $p_T > 20$ GeV, a pseudorapidity $|\eta| < 2.5$ and a 
relative distance $\Delta R > 0.4$. Leptons are required to have a
pseudorapidity $|\eta| < 2.5$ and we ask the invariant mass of a pair of two
leptons of the same flavor to be higher than 20 GeV.

\subsection{Starting the command interface}\label{sec:fs_start}
Once downloaded from the web and unpacked, the \madanalysis\ 5 package does not
require any
compilation\footnote{The \cpp\ core of \madanalysis\ 5 has in fact to be
compiled but this task is performed automatically, behind the scenes, without
requiring any interaction from the user.} or configuration and its command
interface, consisting in a command prompt \texttt{ma5>}, can immediately be
launched by issuing
\begin{verbatim}
  bin/ma5
\end{verbatim}
from the directory where \madanalysis\ 5 has been installed. For the
installation procedure of the program and all its dependencies, we refer to
\ref{sec:install}. The user is now able to access all the functionalities of the
tool and can start implementing an analysis.
 
When launched, the program firstly checks that all the
required dependencies, such as the \rooot\ header files and libraries and a
\cpp\ compiler, are present on the system and correctly installed. If not, an
error message is printed, so that the user has enough information to solve the
issue, and the program exits. On its first
run, \madanalysis\ 5 also compiles a static library which is stored into
the directory \texttt{lib} of the distribution. This library is further
used by the \sampleanalyzer\ kernel when analyses are executed.

Secondly, two lists of labels corresponding to standard definitions of particles
and multiparticles are loaded into the memory of the current session of the
program. By default, when \madanalysis\ 5 is used as a standalone package, they
are imported 
from the corresponding files stored in the \texttt{input} directory of the
distribution of the program. In contrast, if \madanalysis\ 5 has been installed
in the directory where \madgraph\ 5 has been unpacked, the lists of particle and
multiparticle labels are directly imported from \madgraph\ 5. Let us note that
several multiparticles, such as \texttt{hadronic} or \texttt{invisible}, are
essential for a correct running of \madanalysis\ 5. Therefore, if not included
in the imported files, they are automatically created.

\subsection{Particles and multiparticles}\label{sec:fs_multip}
As soon as all the particle and multiparticle labels have been imported, 
\madanalysis\ 5 creates links pointing to lists containing all the declared
labels. This allows us to access them in an easy way at any time of the analysis,
by issuing in the command interface
\begin{verbatim}
  display_particles 
  display_multiparticles 
\end{verbatim}
New labels can be created on run-time with the command \texttt{define}, as, \eg, 
\begin{verbatim}
  define mu = mu+ mu-
\end{verbatim} 
This command creates a multiparticle label \texttt{mu} which is associated to
both the muon and the antimuon. Moreover, new labels are automatically
added to the relevant list of (multi)particles. 

The definition of a specific
label can be retrieved with the help of the \texttt{display} command which
outputs the PDG-id of
the associated particle(s). For example, after invoking the two commands 
\begin{verbatim}
  display b
  display l+
\end{verbatim} 
the command interface outputs information about the (multi)particle labels
\texttt{b} and \texttt{l+},
\begin{verbatim}
  The particle 'b' is defined by the PDG-id 5.
  The multiparticle 'l+' is defined by the PDG-ids -11 -13.
\end{verbatim}
As can be seen from the screen output above, the two symbols \texttt{b} and
\texttt{l+} define a $b$-quark and a positively charged lepton different from a
tau, respectively.

\subsection{Importing event samples}\label{sec:fs_import}

Preliminary to performing any analysis, event files under consideration must be
parsed and
loaded into the memory. The command \texttt{import} has been designed for that
purpose. As a mandatory (single) argument, it requires the name of the Monte
Carlo sample(s) to be parsed. In order to import several files at one time, the
wildcard characters \texttt{*} and \texttt{?} are allowed when typing-in the
argument of the function. Hence, the four samples introduced above can be loaded
simultaneously by issuing the command
\begin{verbatim}
  import samples/*.lhe.gz
\end{verbatim}
which is equivalent to the set of four commands
\begin{verbatim}
  import samples/ttbar_sl_1.lhe.gz
  import samples/ttbar_sl_2.lhe.gz
  import samples/ttbar_fh.lhe.gz
  import samples/zz.lhe.gz
\end{verbatim}
The result is the creation of a unique event sample denoted by
\texttt{defaultset}
containing all the imported files. We are now ready to define selection cuts
which
the \sampleanalyzer\ kernel will apply to the events
included in the dataset \texttt{defaultset}. The name as well as the way how to
merge the samples can be tuned according to the needs of the user, but this goes
beyond the scope of this Section and will be addressed in Section
\ref{sec:normalmode}.

\subsection{Selection cuts and creation of histograms}\label{sec:fs_selection}

Creating a histogram representing a specific kinematical distribution 
has been made very efficient in the framework of \madanalysis\ 5 through the
command \texttt{plot}.
This command requires one mandatory argument, the observable to be computed,
and a set of optional arguments containing, among others,
the number of bins of the histogram to be created and the lower and upper bounds
of its $x$-axis. In the setup of the bounds of a histogram,
one must note that the standard unit of energy used in \madanalysis\ 5 is the
GeV.

In the toy analysis which is implemented in the following, we focus on the
missing transverse-energy distribution as well
as on the transverse-momentum distribution of the final state muons. We recall
that we are considering a unique event sample resulting from the merging of 
three $t \bar t$ and one diboson event files, as explained in Section
\ref{sec:fs_import}. In order to create the associated histograms, it is
sufficient to issue the two commands
\begin{verbatim}
  plot MET 
  plot PT(mu) 20 0 100
\end{verbatim} 
where we recall that the multiparticle \texttt{mu} has been defined in Section
\ref{sec:fs_multip} and represents both the muon and the antimuon. The symbol
\texttt{MET} is associated to the missing transverse energy whilst the function
\texttt{PT} stands for the transverse momentum of a given (multi)particle
provided as its argument. The next pieces of information to be passed to the
command \texttt{plot} are optional and related to the binning of the
histograms. By default, \ie, in the case the binning information is not
specified as in 
the first example above, \madanalysis\ 5 uses hard-coded values which depend on
the observable under consideration. In contrast, as in the second example
above, the user can provide, at the time of typing-in the command, the number of
bins 
(20 here) together with the values of the lowest and highest bins (which are
chosen equal to 0 GeV and 100 GeV, respectively, in our example).

We now turn to illustrating the implementation of the two types of selection
cuts which is possible to employ in \madanalysis\ 5. These cuts will be further
applied, by the \sampleanalyzer\ kernel, to the events contained in the 
\texttt{defaultset} dataset. We recall that the program contains a set of
possibilities much broader than what is presented in this
Section and we refer to Section
\ref{sec:normalmode} for more information. In a first step, we decide to
select events where the missing transverse energy ($\slashed{E}_T$) is not too
large, \ie, 
$\slashed{E}_T < 100$ GeV. Rejecting events not fulfilling this criterion can be
very efficiently and easily performed by issuing the command
\begin{verbatim}
  reject MET > 100
\end{verbatim}
The function \texttt{reject} tells \madanalysis\ 5 to remove
from the event selection any event where the missing energy (\texttt{MET}) is
larger than 100 GeV.

As a second illustration about the implementation of selection cuts in
\madanalysis\ 5, we focus on the kinematical properties
of the muons contained in the selected events. In particular, a part of the
events
contain a muon$-$antimuon pair and we would like to represent the corresponding
invariant-mass distribution through a histogram. However, many events do not
exactly contain one muon and one antimuon. This is taken into account by 
\madanalysis\ 5 at the time of the generation of the histogram, where 
one entry is included for each different muon$-$antimuon
pair which can be formed from the event particle content. For a given event, the
number of entries can hence be zero, one or bigger than one according to the
(anti)muon multiplicity of the final state. 

Realistic detectors are in
general not capable of correctly reconstructing too soft particles. Even at the
parton-level, this effect can (and should) be implemented. In our case,
we could consider as (anti)muon candidate only final state (anti)muons with,
\eg, a transverse momentum $p_T > 20$ GeV. This cut can be implemented in
\madanalysis\ 5 through the command \texttt{reject}, but following a syntax
different from above,
\begin{verbatim}
  reject (mu) PT < 20
\end{verbatim}
The effect of this command is to consider, for each of the selected events, as
muon and antimuon candidates only muons and antimuons with 
a transverse momentum (\texttt{PT}) harder than 20 GeV. For all the histograms
which are generated after having typed the line above in the command interface, 
only the selected (anti)muon candidate's are considered. As stated above, we
choose, as an example, to represent the muon-pair invariant mass distribution
computed from the analyzed event samples included in the dataset
\texttt{defaultset}. 
The command to create this histogram is similar to those presented in the
beginning of this Section,
\begin{verbatim}
  plot M(mu+ mu-) 20 0 100
\end{verbatim}
where the arguments of the function \texttt{M}, associated to the invariant
mass, are multiple. This
denotes that we are summing the four-momenta of the muon and the antimuon to
derive the invariant mass to be represented. As it can be seen from the optional
arguments of the command plot, we once again ask for a histogram of 20 bins
with its lower and upper bounds being fixed to 0 GeV and 100
GeV, respectively. We recall that for a specific event, the number 
of entries which are included in the histogram corresponds to the number of
different muon$-$antimuon pairs that can be formed from the final-state particle
content. This can then be any integer number.

Once the selection is defined, it has to be executed through a job to
be run by
\sampleanalyzer. This is done through the command \texttt{submit} which takes as
an argument the name of a directory which will be created,
\begin{verbatim}
  submit <dirname>
\end{verbatim}
The created directory \texttt{<dirname>} contains a series of \cpp\ source and
header files that are necessary for \sampleanalyzer\ to properly run. The
compilation, linking to the external
static library of \madanalysis\ 5 (see Section \ref{sec:fs_start}) and the
execution of the resulting code is handled by \madanalysis\ 5. The screen
output indicates the status of these different tasks and various information
such as the detected format of the event samples or the number of processed
events. In the case anything is not going as smoothly as it should,
\madanalysis\ 5 also prints warning and/or error messages to the user and the
program exits in the worst case scenario.

\subsection{Displaying the results}
After having performed the analysis as indicated in the previous Section,
\madanalysis\ 5 offers several ways to present the results under a
human-readable report. This report can be either generated under the \html\
format or under a \TeX\ format that could be compiled with a \latex\
or a \pdflatex\ compiler. The histograms are saved under a Portable Network
Graphics file (\texttt{.png}) for \html\ and \pdflatex\ reports, and under an
Encapsulated PostScript file (\texttt{.eps}) for \latex\ reports. The
\madanalysis\ 5 commands generating the reports read 
\begin{verbatim}
  generate_latex <dirname>
  generate_pdflatex <dirname>
  generate_html <dirname>
\end{verbatim}
where \texttt{<dirname>} stands for the directory in which the report is 
generated. 

The structure of the report is similar whatever the adopted format is. It 
starts with a table of contents and firstly
displays all the information necessary to ensure the reproducibility
of the analysis. One hence finds the list of commands which have been issued,
in the current session of the program, before the generation of the report,
followed by the employed version of the code as well as the values of all the
setup parameters. In this last category, one has, \eg, 
the list of the event samples which have been analyzed, given together with the
corresponding cross sections and the number of events contained in the event
files. 

\begin{figure}[t]
  \center \includegraphics[width=0.90\textwidth]{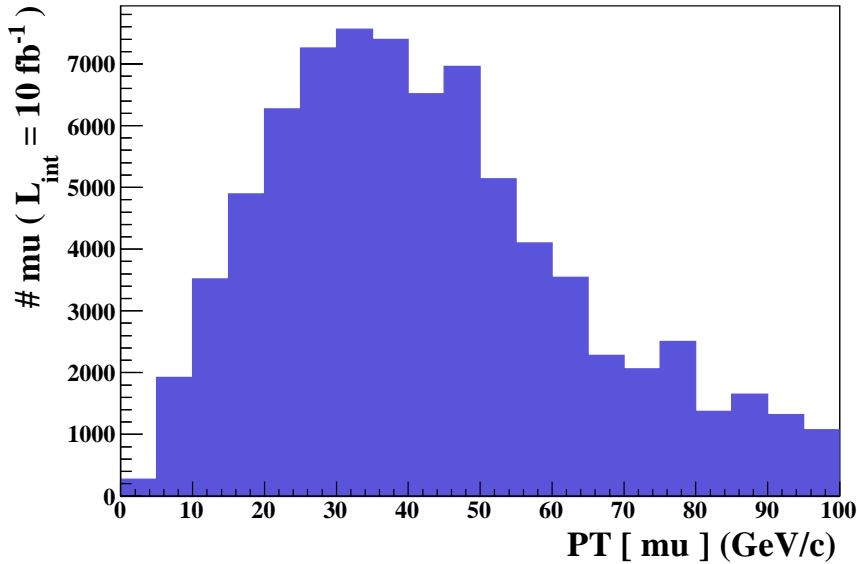}
  \caption{\label{fig:fs1}Transverse-momentum distribution of the muons for a
    dataset consisting of the four event samples introduced in Section 
    \ref{sec:firststeps}.}
\end{figure}

The core of the report contains the results of all the commands related to
histogram creation and to the application of a selection cut. These
commands have been treated one by one by \sampleanalyzer\ and the report follows
this pattern. By default, histograms are normalized to an integrated luminosity
of 10 fb$^{-1}$ and comes together with a summary table containing
information on the mean value of the computed distribution and the associated
root mean square (RMS), 
as well as on the presence of possible underflow and overflow bins in the
histogram. The latter are given as a ratio of the value of the integral of the
represented distribution between its lower and upper bounds. A
green$-$orange$-$red
color code indicates their relative importance, an orange or red color
suggesting more or less strongly to the user  to modify the bounds of the
histograms, if relevant. We refer to
Section \ref{sec:normalmode} for the description of all the properties of the
histograms that can be modified by the user, such as the way to modify the
luminosity or the binning of a given histogram.

In Figure \ref{fig:fs1}, we take the example of the transverse-momentum
distribution of the muons implemented in Section \ref{sec:fs_selection} and
present the histogram generated by \madanalysis\ 5. As stated
above, if several muons are contained in one specific event, they each
correspond to a different entry in the histogram with the same weight. 
The summary table generated
together with the histograms is given in Table \ref{tab:fs1}.  
\begin{table}[t]\center
\begin{tabular}{|c|c|m{1.3cm}|c|c|c|c|}
\hline
\cellcolor{yellow}\textcolor{black}{Dataset}&
\cellcolor{yellow}\textcolor{black}{Integral}&
\cellcolor{yellow}\textcolor{black}{Entries / event}&
\cellcolor{yellow}\textcolor{black}{Mean}&
\cellcolor{yellow}\textcolor{black}{RMS}&
\cellcolor{yellow}\textcolor{black}{Underflow}&
\cellcolor{yellow}\textcolor{black}{Overflow}\\
\hline
\cellcolor{white}\textcolor{black}{defaultset}& \cellcolor{white}\textcolor{black}{82747}& \cellcolor{white}\textcolor{black}{0.752}& \cellcolor{white}\textcolor{black}{42.8177}& \cellcolor{white}\textcolor{black}{21.36}& \cellcolor{orange}\textcolor{black}{0.0}& \cellcolor{orange}\textcolor{black}{6.181}\\
\hline
\end{tabular}
\caption{\label{tab:fs1} Statistics associated to the histogram of Figure
\ref{fig:fs1}.}
\end{table}
In the first column of the table, the name of the dataset is printed. In the
second column, one finds the number of events normalized to an
integrated luminosity of 10 fb$^{-1}$, since this is the default value which has
not been modified. In the third column, the average number of histogram entries
for each event is provided, with the mean and the root mean square of the
distribution under consideration shown in the fourth and fifth columns.
From the orange cells in the table, one immediately notices that the number of
underflow and overflow entries given in the last two columns are only under a
roughly reasonable control.

 \begin{table}[t]
\center
\begin{tabular}{|c|c|c|c|}
\hline
\cellcolor{yellow}\textcolor{black}{Cuts}&
\cellcolor{yellow}\textcolor{black}{Signal (S)}&
\cellcolor{yellow}\textcolor{black}{Background (B)}& \cellcolor{yellow}\textcolor{black}{S vs B}\\
\hline
\cellcolor{white}\textcolor{black}{Initial (no cut)}& \cellcolor{white}\textcolor{black}{109999 +/\-- 789}& \cellcolor{white}\textcolor{black}{}& \cellcolor{white}\textcolor{black}{}\\
\hline
\cellcolor{white}\textcolor{black}{cut 1}& \cellcolor{white}\textcolor{black}{82994 +/\-- 375443}& \cellcolor{white}\textcolor{black}{}& \cellcolor{white}\textcolor{black}{}\\
\hline
\cellcolor{white}\textcolor{black}{cut 2}& \cellcolor{white}\textcolor{black}{82994 +/\-- 612}& \cellcolor{white}\textcolor{black}{}& \cellcolor{white}\textcolor{black}{}\\
\hline

\hline
\end{tabular}
\caption{\label{tab:fs2} Efficiencies of the cuts applied in Section
\ref{sec:fs_selection}.}
\end{table}

Let us finally note that for each selection cut, a summary table of the cut
efficiency is also given, together with the corresponding effect on the signal
over background ratio. In our toy example, we have not tagged any sample as
signal or background. Therefore, this feature is irrelevant here and we refer to
Section \ref{sec:normalmode} for more information. The generated summary of all
cuts is provided in Table \ref{tab:fs2}, where all the events are (by default) 
considered as signal events.

\section{Implementing analyses in an efficient and user-friendly way}
\label{sec:normalmode}

\subsection{Starting a \madanalysis\ 5 session}\label{sec:uf_starting}
The \madanalysis\ 5 package is able to handle several types of Monte Carlo
samples. Supported event files can describe partonic, hadronic or
reconstructed events. The key difference between these three types of events
lies in the basic objects in which the event content is expressed in terms of. 
In the first case, an event consists in a set of
partons (quark, gluons, charged leptons, neutrinos, new states, \etc),
whilst in the second case, it consists in 
hadrons (pions, kaons, baryons, \etc). Finally, a reconstructed event contains a
set of high-level reconstructed objects such as electrons, muons, jets,
\etc. For each of these classes of events, \madanalysis\ has a specific running
mode which the user is required to specify when launching the code,
\begin{verbatim}
  bin/ma5 [level] 
\end{verbatim}
Typing in a shell \texttt{bin/ma5 --partonlevel} or \texttt{bin/ma5 -P} allows
us to
run \madanalysis\ 5 in the mode required to analyze parton-level events, whilst
issuing \texttt{bin/ma5 --hadronlevel} or \texttt{bin/ma5 -H} makes the program
running in the hadron-level mode. In a similar fashion, the two shell commands
\texttt{bin/ma5 --recolevel} and \texttt{bin/ma5 -R} allow us to start 
\madanalysis\ 5 in a ready way to analyze reconstructed-level events. 
In the case no argument is provided, as in the 
example of Section \ref{sec:firststeps}, the parton-level mode is
automatically selected. Of course, the different modes cannot be combined, since
the levels of sophistication are self-excluding. 
Once one of the previously introduced commands is issued, the command line
interface of
\madanalysis\ 5 is started. It consists in a command prompt \texttt{ma5>} where
the user can access all the functionalities of the program and directly
implement a physics analysis by issuing a set of commands. We refer to Section
\ref{sec:uf_cli} for more information about the whole set of existing commands.

The list of commands to be typed can also be provided under the form of a
script, \ie, a simple text file. In order to execute the script,
its path has to be provided as the last argument when typing the
\texttt{bin/ma5} command in a shell. Hence, 
\begin{verbatim}
  bin/ma5 --recolevel <filename>
\end{verbatim}
starts the command interface of \madanalysis\ 5 in the reconstructed-level
mode. All the commands included in the file \texttt{<filename>} are then 
sequentially applied, one after the other. When having such a script executed by
\madanalysis\ 5, it can be useful to bypass all the confirmation questions
usually asked by the program. This can be done by issuing in a shell one of the
two commands
\begin{verbatim}
  bin/ma5 --script --recolevel <filename>
  bin/ma5 -s --recolevel <filename>
\end{verbatim}
This \textit{script mode} also enforces the automatic exit of \madanalysis\ 5 
after the completion of the
script \texttt{<filename>}, in contrast to the \textit{forced mode} described
below. It is also possible to execute several scripts
by providing various filenames, 
\begin{verbatim}
  bin/ma5 -s -R <filename1> <filename2> <filename3>
\end{verbatim}
The different files are handled as concatenated in one single file, \ie, all the
commands included in \texttt{filename1} are processed sequentially, followed by
the commands included in \texttt{filename2}, \etc. Let us note that the
character `\texttt{\#}' can be included when writing down script files.
Everything standing to
the right of this character is considered as a comment by \madanalysis\ 5 and
ignored by the command line interpreter.

\begin{table}\bgfb 
  \bcen \textbf{Table \ref{tab:startingma5}: Syntax to launch the \madanalysis\
     5 program:}\\ \vspace{.3cm}
  \textbf{\texttt{bin/ma5 -h   }  or  \texttt{   bin/ma5 --help}} \ecen
\vspace{-.3truecm}
This allows to display to in a shell a summary of the content of this Table.\\
\vspace{-.2truecm}
  \bcen
  \textbf{\texttt{bin/ma5 -v   }, \texttt{   bin/ma5 --version   }
  or \texttt{   bin/ma5 --release}}  \ecen
\vspace{-.4truecm}
This allows to display in a shell the number of the version of \madanalysis\ 5
present on the system.\\
\vspace{-.2truecm}
  \bcen
  \textbf{\texttt{bin/ma5 -f   }, \texttt{   bin/ma5 --forced}} \ecen
\vspace{-.4truecm}
This allows to run \madanalysis\ 5 in a mode where
confirmation questions are ignored. \\
\vspace{-.2truecm}
  \bcen
  \textbf{\texttt{bin/ma5 -e   }, \texttt{   bin/ma5 -E   } or \texttt{bin/ma5
--expert}} \ecen
\vspace{-.4truecm}
This indicates that \madanalysis\ 5 has to run in expert mode.
In this case, any other option is ignored.\\
\vspace{-.2truecm}
  \bcen \textbf{\texttt{bin/ma5 [level] [files]}} \ecen
\vspace{-.4truecm}
\textbf{\texttt{[level]}}\\
When provided, this optional argument allows to select the type of event files
to analyze. If absent, parton-level events are assumed. The user can choose one
of the following options:\\
\btaby
\texttt{-P} or \texttt{--partonlevel} &  $\quad$Parton-level events.\\ 
\texttt{-H} or \texttt{--hadronlevel} &  $\quad$Hadron-level events.\\
\texttt{-R} or \texttt{--recolevel} &  $\quad$Reconstructed events. \\
\etaby\\ $~$\\
\texttt{[files]}\\
When provided, this optional argument consists in a sequence of filenames,
separated by blank characters, containing each a set of \madanalysis\
5 commands. The files are handled as concatenated and the commands are
applied sequentially. If the option pattern \texttt{-s} or \texttt{--script} is
included, \madanalysis\ 5 exits after having executed the script and does
not ask any confirmation question.
\egfb \textcolor{white}{\caption{\label{tab:startingma5}}}
\end{table}

For highly sophisticated analyses, the features which are presented in the
rest of this section
might not be sufficient. Therefore, \madanalysis\ 5 can be run in an expert
mode by issuing one of the three commands
\begin{verbatim}
  bin/ma5 --expert      bin/ma5 -e      bin/ma5 -E
\end{verbatim}
The possibilities to implement any analysis are here unlimited. More information
about the expert mode is provided in Section \ref{sec:expert}. Let us however
note that in this case, any other option which could be passed when starting the
interface is ignored.

Among the other options which could be provided when starting the interpreter,
one can note that the version of the distribution installed on the
system of the user can be accessed through one of the following commands,
\begin{verbatim}
  bin/ma5 -v      bin/ma5 --version      bin/ma5 --release
\end{verbatim}
and that typing-in
\begin{verbatim}
  bin/ma5 -f      bin/ma5 --forced
\end{verbatim}
ensures a running mode of \madanalysis\ 5 where confirmation questions, such as,
\eg, those printed to the screen when a directory is about to be removed, are
not asked of the user.

The different running modes of \madanalysis\ 5 and the way to cast them are
summarized in Table \ref{tab:startingma5}. They can also be displayed by 
the program when the user types in a shell one of the following commands,
\begin{verbatim}
  bin/ma5 --help     bin/ma5 -h
\end{verbatim}

\subsection{The command line user interface of \madanalysis\ 5}
\label{sec:uf_cli}
The command line interface of \madanalysis\ 5 is built upon the \python\ module
\texttt{cmd} which allows for a flexible processing of commands issued by the
user. It features, among others, text help, tab completion and shell access. In
addition, the history of commands issued by the user can be obtained by means of
the up and down keys of the keyboard. As presented in Section
\ref{sec:uf_starting}, the command interface can be started by issuing in a
shell the command 
\begin{verbatim}
  bin/ma5 [options]
\end{verbatim}
from the directory where \madanalysis\ 5 has been installed. However,
\madanalysis\ 5 could also be started from any location on the computer of the
user, the command above needing only to be modified accordingly.

The user interface consists in a command prompt \texttt{ma5>} where the user can
implement a physics analysis very efficiently and easily by means of a series
of (case sensitive) commands. The related syntax is inspired by the \python\
programming language and a command therefore always starts with an
\textit{action}. Several actions can be typed together in a single command line
after separating them with a semicolumn `\texttt{;}'.
The number of different implemented actions has been made as small as possible
in order to guarantee a quick handling of the code by any physicist. Their list
has been collected in Table \ref{tab:actions}, with basic instructions about
how to use each of these actions. The information included in this table can
also be displayed to the screen at run-time by issuing
\begin{verbatim}
  help [<action>]
\end{verbatim}
where \texttt{<action>} is the action under consideration. Moreover, if
\texttt{help} is typed-in without any argument, the list of all the available
actions is printed out. Let us emphasize that tab completion could also be
used with the aim of obtaining that list. We now dedicate the rest of this Section
\ref{sec:normalmode} to a detailed description of each of those actions.

As stated above, the history of all the actions undertaken by the user can be
accessed through the up and down keys of the
keyboard, as for standard shells. They are read from the file
\texttt{.ma5history} stored in the directory where \madanalysis\ 5 has been
unpacked. This file is updated every time that the user types a command, and 
the list of the last
100 commands executed by the interpreter is saved. In addition, if one
restricts ourselves to the current session of \madanalysis\ 5, the list
of typed commands can be accessed by issuing in the command interface
\begin{verbatim}
  history
\end{verbatim}

Besides actions, the language handled by the interpreter contains
\textit{objects} that actions are acting on. There exist various types of objects
representing particles, multiparticles, datasets, selection cuts
or even the configuration panel of the
current session of the program, denoted by \texttt{main}. In addition, any
object is described by several attributes related to its properties, which are
generically denoted 
as \textit{options}. These attributes can be accessed through the syntax
\textit{object.option}.

Finally, the language of \madanalysis\ 5 contains a set 
of reserved keywords, such as \texttt{all}, \texttt{or} or \texttt{as}, whose
usage is described in the following. Moreover, it is important to keep in mind
that both predefined and user-defined objects have a name
which has to be unique. Therefore, in an analysis, any object which is created,
such as, \eg,  a new instance of the multiparticle class, must have a name different
from those of the existing actions (see Table \ref{tab:actions}) or those of the
objects already defined and/or used in the current analysis.

The syntax introduced above has been developed with a focus on uniformization
and intuition. For the sake of the example, we now focus on the three commands
\texttt{display}, \texttt{set} and \texttt{remove}. 
To display to the screen an object denoted by \texttt{<object>} 
together with its properties, it is enough to type the command 
\begin{verbatim}
  display <object> 
\end{verbatim}
Similarly, the display of a specific property, denoted by \texttt{<option>}, of
the object under consideration can be performed by issuing 
\begin{verbatim}
  display <object>.<option>
\end{verbatim}
On the same footing, the value of the attribute
\texttt{<object>.<option>} can be easily modified via the action \texttt{set}
\begin{verbatim}
  set <object>.<option> = x
\end{verbatim}
where in the example above, the option under consideration is set to the value
\texttt{x}. Finally, an object can be deleted from the computer memory with the
help of the command \texttt{remove}, 
\begin{verbatim}
  remove <object>
\end{verbatim}
However, three objects, the objects \texttt{main}, \texttt{hadronic} and
\texttt{invisible}, as well as all the particles and multiparticles currently used 
in the analysis, \ie, being related to a histogram or to a selection cut,
cannot be deleted.

As already stated in the beginning of this Section, the \texttt{cmd} module of
\python\ allows for efficient access to shell commands directly from inside
the command line interface of \madanalysis\ 5. This requires starting the
command line either with an exclamation mark
`\texttt{!}' or with the command \texttt{shell}. Therefore, the syntax 
\begin{verbatim}
  !<command>
\end{verbatim}
or equivalently
 \begin{verbatim}
  shell <command>
\end{verbatim}
has to be followed in order to execute the command \texttt{<command>} from an
external shell.

\begin{table}
\bgfbn
\multicolumn{2}{c}{\textbf{Table \ref{tab:actions}: Actions available from the
command}}\\
\multicolumn{2}{c}{\textbf{line user interface of \madanalysis\ 5}}\\$~$\\
{\tt define <name> = <def>} & 
   This creates a new (multi)particle label \texttt{<name>} defined through the
   (list of) PDG-id(s) \texttt{<def>}.\\ 
{\tt display <var>} & 
   This displays to the screen either the properties of an object or the value
of one of its options, both generically denoted by \texttt{<var>}.\\ 
{\tt display\textunderscore datasets} & 
   This displays to the screen the list of the imported datasets.\\ 
{\tt display\textunderscore multiparticles} & 
   This displays to the screen the list of defined multiparticle labels.\\ 
{\tt display\textunderscore particles} & 
   This displays to the screen the list of defined particle labels.\\ 
{\tt EOF} & 
   This closes the current session of \madanalysis\ 5.\\ 
{\tt exit} & 
   This closes the current session of \madanalysis\ 5.\\ 
{\tt generate\textunderscore html <dir>} & 
   The report of the current analysis is generated, under the \html\ format, and
     stored in the directory \texttt{<dir>}.\\ 
{\tt generate\textunderscore latex <dir>} & 
   The report of the current analysis is generated, under the \TeX\ format, and
   stored in the directory \texttt{<dir>}. Compilation requires
   a \latex\ compiler.\\ 
{\tt generate\textunderscore pdflatex <dir>} & 
   The report of the current analysis is generated, under the \TeX\ format, and 
   stored in the directory \texttt{<dir>}. Compilation requires a
   \pdflatex\ compiler.\\
{\tt help} & 
   This displays to the screen the list of all the actions implemented in
   \madanalysis\ 5.\\
{\tt history} & 
   This displays to the screen the history of the commands typed in the current
   session of the program.
\egfbn
\textcolor{white}{\caption{\label{tab:actions}}}
\end{table}

\begin{table}
\bgfbn
\multicolumn{2}{c}{\textbf{Table \ref{tab:actions} (continued): Actions
  available from the command}}\\
\multicolumn{2}{c}{\textbf{line user interface of \madanalysis\ 5}}\\$~$\\ 
{\tt import <obj>} & 
   This allows to import in the current session of \madanalysis\ 5 external
    information, generically denoted by \texttt{<obj>}, such as  
   Monte Carlo samples, a configuration used by \sampleanalyzer\ in a
   previous analysis or a UFO model.\\ 
{\tt install <obj> } & 
   This allows to install the external object \texttt{<obj>} from the Internet.
   Currently, the only allowed choice for the value of the variable
   \texttt{<obj>} consists in the keyword \texttt{samples} which yields the
   download of the four example samples presented in Section
   \ref{sec:firststeps}. \\
{\tt plot <obs> [options]} & 
   This creates, in an analysis, an histogram associated to the
   distribution of the observable \texttt{<obs>}. We refer to Section
   \ref{sec:uf_histo} for the list of available options.\\ 
{\tt open <rep>} & 
   This opens the report \texttt{<rep>} of a given analysis, with the
   appropriate program.\\ 
{\tt preview <hist>} & 
   This displays an histogram \texttt{<hist>} in a graphical window.\\ 
{\tt quit} & 
   This closes the current session of \madanalysis\ 5.\\ 
{\tt reject (<prt>) <cond>} & 
   This adds a selection cut. In an analysis, a candidate to a given particle
   species \texttt{<prt>} is ignored if the condition \texttt{<cond>} is
   fulfilled.\\
{\tt reject <cond>} & 
   This adds a selection cut. An event is rejected if the
   condition \texttt{<cond>} is fulfilled.\\ 
{\tt remove <obj>} & 
   This removes the object \texttt{<obj>} from the memory.\egfbn
\end{table}

\begin{table}
\bgfbn
\multicolumn{2}{c}{\textbf{Table \ref{tab:actions} (continued): Actions
  available from the command}}\\
\multicolumn{2}{c}{\textbf{line user interface of \madanalysis\ 5}}\\$~$\\  
{\tt reset} & 
   This reinitializes \madanalysis\ 5 as when a new session starts.\\
{\tt resubmit} & 
   This allows, for an analysis which has already been run by \sampleanalyzer\
    and further modified, to be re-executed.\\
{\tt select (<prt>) <cond>} & 
   This adds a selection cut. In order to consider, in an analysis, a candidate
   to a given particle species \texttt{<prt>}, the condition \texttt{<cond>} has
   to be fulfilled.\\ 
{\tt select <cond>} & 
   This adds a selection cut. Events are selected only if the
   condition \texttt{<cond>} is fulfilled.\\ 
{\tt set <obj>.<opt> = <val>}  & 
   This allows to set the attribute \texttt{<opt>} of the object \texttt{<obj>}
   to the value \texttt{<val>}. \\
{\tt shell <com>} & 
   This allows to run the shell command \texttt{<com>} from the command
   interface of \madanalysis\ 5. The action \texttt{shell} can be replaced by an
   exclamation mark `\texttt{!}'.\\ 
{\tt submit <dir>} & 
   This allows to execute an analysis as a job run by \sampleanalyzer. The \cpp\
   source and header files related to the job are created in the directory
   \texttt{<dir>}.\\ 
{\tt swap <sel1> <sel2>} & 
   This allows to permute the sequence of two instances of the class
   \texttt{selection}, \texttt{<sel1>} and \texttt{<sel2>}. We refer to Section 
   \ref{sec:uf_histo} for more information.
\egfbn
\end{table}

Finally, the configuration of the command line interpreter can be restored as
for a new session of \madanalysis\ 5 by issuing the command
\begin{verbatim}
  reset
\end{verbatim}
Consequently, all the user-defined particles and multiparticles as well as all
the implemented histograms and selection cuts are removed from the computer 
memory.

Before moving on, some remarks on tab completion are in order. This feature
allows
for an easy typing of commands, attributes of the objects, \etc. Typing on the
`\texttt{tab}' key has the effect of printing to the screen all the allowed 
possibilities for completing the command about to be written. It works
equivalently for actions, objects and attributes. Let us emphasize
that this makes tab completion in \madanalysis\ 5 much more advanced than
its counterpart in standard command shells.

\subsection{Datasets}\label{sec:uf_datasets}
In Section \ref{sec:firststeps}, we have shown that four example Monte Carlo
samples can be downloaded from the Internet by employing the command
\texttt{install},
\begin{verbatim}
  install samples
\end{verbatim}
and stored into a directory \texttt{samples} of the current distribution of
\madanalysis\ 5. Actions, such as their gathering into datasets, the creation
of histograms and 
the definition of selection cuts, have been subsequently executed on those
samples. This Section is dedicated to the dataset class of objects, whilst
Section \ref{sec:uf_histo} and Section \ref{sec:uf_cuts} concerns histograms
and cuts, respectively.

In order to load Monte Carlo samples into the computer memory, they have to be
imported from outside the current session of \madanalysis\ 5. This requires the
use of the command \texttt{import}, as briefly presented in both Section
\ref{sec:firststeps} and Table \ref{tab:actions},
\begin{verbatim}
  import <path-to-sample>
\end{verbatim}
The syntax above allows us to import a Monte Carlo sample stored at a location 
\texttt{<path-to-sample>} on the computer of the user. Several samples can be
imported at one time by employing the wildcard characters \texttt{*} and
\texttt{?}. As a generic example, the command
\begin{verbatim}
  import <directory>/*
\end{verbatim}
imports all the samples stored in the directory \texttt{<directory>}.
Moreover, the tilde character \texttt{\~} can be used when typing the path to a
sample. As in standard {\sc Linux} shells, it points to the location of the home
directory of the user.

According to their event format, \madanalysis\ 5 handles differently the 
imported event samples. The key to the procedure lies in the extension of the
event files, up to a possible packing with \textsc{gzip}. The formats currently
supported
are the \lhe\ event file format, whose corresponding file extensions are
\texttt{.lhe} or \texttt{.lhe.gz}, the \stdhep\ event file format, whose 
corresponding file extensions are \texttt{.hep} or \texttt{.hep.gz}, the 
\hepmc\ event file format, whose corresponding defining extensions
are \texttt{.hepmc} or \texttt{.hepmc.gz} and the \lhco\ event file
format, whose corresponding event file extensions are \texttt{.lhco} or
\texttt{.lhco.gz}. 

Of course, all the different formats are not appropriate for any level of
sophistication of the analysis. For instance, using the parton-level mode of
\madanalysis\ 5 to analyze an event file compliant with the \lhco\ format 
is clearly inadequate. In more detail, events stored in files compliant with
the \lhe\ format can be imported whatever is the level of sophistication of
the analysis, \ie, equally for parton-level, hadron-level or
reconstructed-level analyses. A remark is in order here. 
In principle, the \lhe\ format is not supposed to describe reconstructed events.
However, there exist routines external to \madanalysis\ 5, such as the
\textsc{Hep2LHE} converter included in \madgraph, which allow us to
efficiently convert events as produced by hadronization algorithms or a fast
detector simulation tool into
reconstructed events. Those routines use a jet algorithm in order to reconstruct
light and $b$-tagged jets, and the total missing transverse energy of the
event is eventually computed. In contrast, \stdhep\ and \hepmc\ event files
can only store parton-level and hadron-level events, whilst complementary
\lhco\ files can only be used after high-level objects have been
reconstructed, \ie\ at the reconstructed-level after detector simulation.

The effect of the action \texttt{import} is to unify all the
imported samples into one single object dubbed \textit{dataset}. If nothing is
specified by the user, events are
gathered together into a dataset called \texttt{defaultset}. The possibility to
import events and collect them into different datasets
allows us, for instance, to differentiate background event samples from signal
event samples, as with the following commands
\begin{verbatim}
  import <path-to-signal-events>/* as signalset 
  import <path-to-background-events>/* as backgroundset
\end{verbatim}
The first command imports all the event samples present in the directory 
\texttt{<path-to-signal-events>} and collects them into one single
dataset labeled as \texttt{signalset}. Similarly, the second command loads into
the current session of \madanalysis\ 5 all the event files located in the
directory \texttt{<path-to-background-events>} and gathers them together into a
dataset labeled as \texttt{backgroundset}.
On the same footing, this feature can be used to collect efficiently
event files corresponding to different sources of background,
\begin{verbatim}
  import <path-to-events>/ttbar* as ttbar 
  import <path-to-events>/zz* as zz 
  import <path-to-events>/drellyan* as drellyan 
\end{verbatim}
These three commands import into \madanalysis\ 5 three series of samples, related
respectively to top-antitop pair, $Z$-boson pair and Drell-Yan events. Three
different datasets, \texttt{ttbar}, \texttt{zz} and \texttt{drellyan}, are 
created so that these events can be treated separately when performing the
analysis.

A dataset has several properties which are implemented as options of the
\texttt{dataset} class and whose value can be 
modified with the command \texttt{set}. These attributes can be grouped into
three categories of properties which are summarized in Table
\ref{tab:dataset}.

The first category of options consists in fact in one single property which is
related to the attribute \texttt{type}. It
allows us to tag the corresponding event sample(s) as a part of the signal or
the background. Taking the example of a generic dataset \texttt{<dataset>}, the
two commands
\begin{verbatim}
  set <dataset>.type = background    
  set <dataset>.type = signal
\end{verbatim}
tag it as a dataset belonging to the series of background or signal event
samples, respectively. By default, a dataset is always considered as of the type 
\texttt{signal}. In an analysis, this distinction between
signal and background is mandatory for a correct (automated) computation of the
cut efficiencies by \madanalysis\ 5, as well as for the corresponding derivation
of the signal over background ratio.

When creating histograms, their overall normalization is related to both the 
luminosity, which can be specified by the user (see Section \ref{sec:uf_submit})
and the cross sections
associated to the different datasets included in the histograms. These cross
sections are included in \lhe\ event files and are directly read and imported
into
the current session of the program, in the case \lhe\ samples are analyzed.
In contrast, the numerical value of the cross sections are 
absent from \stdhep, \hepmc \ and \lhco\ files. Therefore, in most of the
cases, the user has to indicate the cross section manually. Moreover, in the
case of \lhe\ files, one could also want to modify the values which have been
read. For instance, the 
cross section associated to an event sample which has been generated with a
leading order Monte
Carlo tool could be modified by the user in order to account for
next-to-leading-order normalization effects.

The second category of options
related to the \texttt{dataset} class offers two ways to
perform the above-mentioned task, either through the \texttt{weight} attribute
or through the \texttt{xsection} attribute. 
A weight different from one can be assigned to each event of a
dataset by modifying the value of the attribute \texttt{weight} of
the \texttt{dataset} class, 
\begin{verbatim}
  set <dataset>.weight = <weight>
\end{verbatim}
The command line above leads to the assignment of a weight \texttt{<weight>} to
each event included in a generic dataset 
which has been defined as \texttt{<dataset>}. Consequently, the default value of
one has been superseded by the value \texttt{<weight>}. In contrast, the
user can also decide to leave the weight of each event unchanged and equal to
unity, and modify instead the value of the cross section. The new value of the
cross section has to be stored through the attribute \texttt{xsection}
of the \texttt{dataset} class,
\begin{verbatim}
  set <dataset>.xsection = <value>
\end{verbatim}
If the value \texttt{<value>} is different from the default choice of zero, 
\madanalysis\ 5 uses it at the time of the creation of the histogram when
calculating its normalization,
ignoring the value of the cross section possibly included in the event file. 

The last class of attributes associated to \texttt{dataset} objects concerns
the layout of the histograms generated by \madanalysis\ 5, and in
particular the style of the curves associated to the datasets which can be
customized. 
On the one hand, styles and colors can be modified via 
the value of the attributes \texttt{linestyle}, \texttt{linewidth},
\texttt{linecolor}, \texttt{backstyle} and \texttt{backcolor}. On the other
hand, the text used in histogram legends can be set by the user through the
attribute \texttt{title}. The first three attributes above 
are associated to the type of lines (solid, dashed or
dotted) used in the histograms, their width and their color (blue, green, none,
purple, white, black, cyan, gray, orange, red or yellow). The two
attributes \texttt{backstyle} and \texttt{backcolor} refer to the surface
under a histogram and the style (solid, dotted, or hatched lines) and color (blue,
green, none, purple, white, black, cyan, gray, orange, red or yellow) employed
when it is drawn. 

\begin{table}
\bgfbyy
\multicolumn{2}{c}{\textbf{Table \ref{tab:dataset}: List of the attributes of
the \texttt{dataset} class}}\\
\multicolumn{2}{c}{\textbf{\texttt{set <dataset>.<option> = <value>}}}\\$~$\\
{\tt backcolor} & 
   In an histogram, this changes the color filling the surface under the curve
   associated to the dataset under consideration. The allowed choices are
   \texttt{auto} (default), \texttt{blue}, \texttt{green}, \texttt{none},
   \texttt{purple},
   \texttt{white}, \texttt{black}, \texttt{cyan}, \texttt{gray},
   \texttt{orange}, \texttt{red} and \texttt{yellow}. Integer numbers between
   one and four can be added or subtracted to these values (see Figure
   \ref{fig:color}).\\
{\tt backstyle} & 
   In an histogram, this changes the style employed when filling the surface under
   the curve associated to the dataset under consideration. The allowed 
   choices are \texttt{auto}, 
   \texttt{solid}, \texttt{dotted}, \texttt{hline},
   \texttt{dline} and \texttt{vline} (see Figure \ref{fig:back}).\\ 
{\tt linecolor} & 
   In an histogram, this changes the color of the line of the curve associated
   to the dataset under consideration. For the allowed choices, we refer to the
   attribute \texttt{backcolor}.\\
{\tt linestyle} & 
   In an histogram, this changes the style of the line of the curve associated
   to the dataset under consideration. The allowed choices are
   \texttt{solid}, \texttt{dashed}, \texttt{dotted} and
   \texttt{dash-dotted}.\\ 
{\tt linewidth} & 
   In an histogram, this changes the width of the line of the curve associated
   to the dataset under consideration. It takes an integer value between one and
   ten.\\ 
{\tt title} & 
   This change the string used in histogram legends for the dataset
   under consideration. The possible
   choices are either \texttt{auto} (the name of the dataset) or any
   string under a valid \TeX\ form.\\ 
{\tt type} & 
   This modifies the type of a dataset, associating it to the set of 
   either signal (\texttt{signal}) or background (\texttt{background}) samples.\\ 
{\tt weight} & 
   This allows to change the weight of each event included in the dataset. The
   default value is one.\\
{\tt xsection} & 
   This allows to modify the total cross section associated to the events
   included in the dataset under consideration. The value can be any real number. 
\egfbyy
\textcolor{white}{\caption{\label{tab:dataset}}}
\end{table}

\begin{figure}[t]
  \center \includegraphics[width=0.90\textwidth]{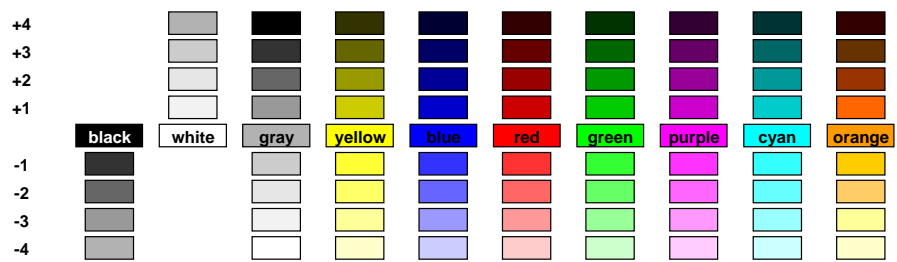}
  \caption{\label{fig:color}Complete list of allowed choices for the values
taken by the attributes
\texttt{backcolor} and \texttt{linecolor} of the \texttt{dataset} class.}
\end{figure}

The default value for the two attributes related to colors in histograms, 
\texttt{linecolor} and \texttt{backcolor}, 
is \texttt{auto}. This means that \madanalysis\ 5 handles the color features
automatically, assigning different colors to the datasets created by the user.
For a specific dataset object denoted by \texttt{<dataset>}, the values of the
two color attributes can be, as usual, superseded by employing the command
\texttt{set},
\begin{verbatim}
  set <dataset>.linecolor = <color>
  set <dataset>.backcolor = <color>
\end{verbatim}
The supported values for the variable \texttt{<color>} are intuitive and
read \texttt{auto}, \texttt{blue}, \texttt{green}, \texttt{none},
\texttt{purple}, \texttt{white}, \texttt{black}, \texttt{cyan}, \texttt{gray},
\texttt{orange}, \texttt{red} or \texttt{yellow}. For histograms employing a
large number of datasets, this panel of colors might however be insufficient.
Shades of these basic colors can be used by adding or subtracting an
integer number between one and four to those colors. For instance, the set of
commands
\begin{verbatim}
  set <dataset1>.backcolor = red+1  
  set <dataset2>.backcolor = red+2 
  set <dataset3>.backcolor = red+3 
  set <dataset4>.backcolor = red+4 
\end{verbatim}
allows us to assign four different shades of red to the four datasets
\texttt{<dataset1>}, \texttt{<dataset2>}, \texttt{<dataset3>} and
\texttt{<dataset4>}. Consequently, when a
stacked histogram containing these four datasets is drawn, the associated
surfaces will all be filled with different colors derived from the basic red.
The complete list of colors integrated in \madanalysis\ 5 is illustrated in
Figure \ref{fig:color}.

\begin{figure}[t]
  \center \includegraphics[width=0.90\textwidth]{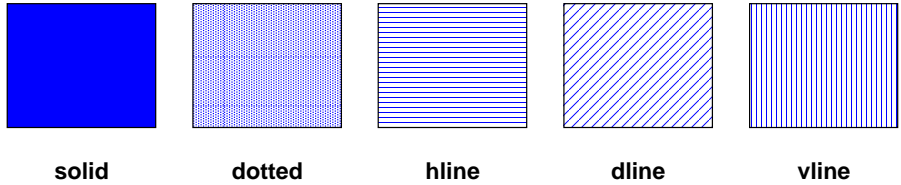}
  \caption{\label{fig:back}Complete list of styles allowed for the values
possibly taken by the attribute \texttt{backstyle} of the \texttt{dataset}
class.}
\end{figure}

The style of the lines of the curves drawn in the histograms can be modified
through the attribute \texttt{linestyle} of the dataset class,
\begin{verbatim}
  set <dataset>.linestyle = <value>
\end{verbatim}
This command changes the default employed solid style to the value
\texttt{<value>}, which can be either
\texttt{solid}, \texttt{dashed}, \texttt{dotted} or \texttt{dash-dotted}.
Similarly, the attribute \texttt{linewidth} allows us to change
the width of the drawn lines,
\begin{verbatim}
  set <dataset>.linewidth = <value>
\end{verbatim}
where \texttt{<value>} is an integer number between one and ten, 
the default value being unity. The option 
\texttt{backstyle} of the \texttt{dataset} class is linked to
the style employed to fill the surface under a histogram and can be set to a
new value by issuing in the command line interface
\begin{verbatim}
  set <dataset>.backstyle = <value>
\end{verbatim}
The allowed choices for the parameter \texttt{<value>} are either
the default value \texttt{auto} or \texttt{solid}, \texttt{dotted}, \texttt{hline},
\texttt{dline} or \texttt{vline}, as presented in Figure \ref{fig:back}. The
last three choices, less intuitive, stand for horizontal, diagonal and vertical
lines, respectively.

Finally, when creating a histogram where the curves related to several datasets
are stacked or superimposed,
\madanalysis\ 5 includes a legend with the explanation of the color/style code
employed to distinguish the datasets. By default, the name of the dataset is
used in this legend, but this can be modified by the user once setting the
attribute \texttt{title} to a string consisting of a valid \TeX\ expression,
\begin{verbatim}
  set <dataset>.title = <string>
\end{verbatim}
where \texttt{<string>} could stand, \eg, in the case of a dataset describing
events related to the production of a pair of $W$-bosons, for 
\texttt{"W\^{}\{+\} W\^{}\{-\}"} (including the quotation marks).

The effects of the customization of the histograms are only taken into account
when the
reports containing the results are generated. Therefore, the user does not
have to issue the command \texttt{submit} each time he modifies the style of
a curve or how the area under a curve is filled inside a histogram. 

Before moving on, let us recall that the command \texttt{display\textunderscore
datasets} introduced in Table \ref{tab:actions} allows us to display to the screen
the list of all the instances of the \texttt{dataset} class which have been
created in the current session of \madanalysis\ 5. Furthermore, the action
\texttt{display} can also be used on datasets. This leads to the printing to the
screen of the current values of all the attributes of a given dataset. Hence,
for a dataset labeled by \texttt{ttbar} where no attribute has been
modified by the user, the effect of the command
\begin{verbatim}
   display ttbar
\end{verbatim}
is to print to the screen the following pieces of information,
\begin{verbatim}
   Name of the dataset = ttbar (signal)
   Title = 'ttbar'
   User-imposed cross section = 0.0
   User-imposed weight of the set = 1.0
   Line color in histograms = auto
   Line style in histograms = solid
   Line width in histograms = 1
   Background color in histograms = auto
   Background style in histograms = solid
   List of event files included in this dataset:
    - ttbar.lhe.gz
\end{verbatim}
This indeed summarizes the status of the instance of the \texttt{dataset} class
labeled \texttt{ttbar} and displays the value of all the options.

Finally, a dataset can be deleted from the memory of the computer by employing
the command \texttt{remove}. If several event files are
included in a single dataset, it is however not possible to remove a particular
file. In this case, the whole dataset has to be removed and then recreated
without including this file. 

\subsection{Particles and multiparticles}\label{sec:uf_labels}
When starting a session of \madanalysis\ 5, lists of predefined particle and
multiparticle
labels are loaded into the memory, as it is mentioned in the example of Section
\ref{sec:firststeps}. These lists are taken from the content of the directory
\texttt{madanalysis/input} which contains several files with label definitions.
Not all the files present in this directory are loaded when a new session of
\madanalysis\ 5 starts. Indeed, according to a running with the aim of analyzing
parton-level, hadron-level or reconstructed-level events, only some of the files
are appropriate, and only these are imported when \madanalysis\ 5 is
launched.

For parton-level analyses, the particle labels are
imported from the file \texttt{particles\textunderscore name\textunderscore
default.txt} whilst the multiparticle labels are read from the file 
\texttt{multiparticles\textunderscore default.txt}. These two files contains
standard names for the Standard Model and the Minimal Supersymmetric Standard
Model particles (\eg, an antimuon is denoted by \texttt{mu+} and the lightest
neutralino by \texttt{n1}) as well as appropriate definitions for the
multiparticles \texttt{hadronic} and \texttt{invisible} (see below). For
hadron-level analyses, only a single file with a list of 415
hadrons (see Ref.\ \cite{Nakamura:2010zzi}),  
\texttt{hadron\textunderscore default.txt}, is imported. Finally, for analyses
at the reconstructed-level, the two lists \texttt{reco\textunderscore
default.txt} and \texttt{multiparticles\textunderscore reco\textunderscore
default.txt} are imported. They contain labels for high-level
reconstructed objects necessary for such a level of sophistication of the Monte
Carlo simulations. The set of predefined labels is rather short and consists of
\texttt{e+}, \texttt{e-},
\texttt{mu+}, \texttt{mu-}, \texttt{ta+}, \texttt{ta-} for charged
leptons, and \texttt{j} and \texttt{b} and \texttt{nb} for jets, $b$-tagged jets
and non-$b$-tagged jets. In addition, several (intuitive) multiparticle labels
are included, \texttt{e}, \texttt{mu}, \texttt{ta}, \texttt{l+}, \texttt{l-}
and \texttt{l} for various combinations of charged leptons, as well as two special
labels dedicated to muons and antimuons,
\texttt{mu-\textunderscore isol} and \texttt{mu+\textunderscore isol}, in
case they are isolated. 

Let us note that if \madanalysis\ 5 has been
installed in the directory where \madgraph\ 5 is unpacked, 
the predefined particle and multiparticle labels 
are directly imported from \madgraph\ 5 rather than from the directory
\texttt{madanalysis/input} of \madanalysis\ 5. 

Two multiparticles, denoted by the labels \texttt{invisible} and
\texttt{hadronic},
have a special role and are essential in any analysis. They
are respectively related to the computation of observables related to the
missing energy and to the hadronic activity. 
Therefore, if they are not included
in the imported files, they are automatically created by \madanalysis\ 5 when
a new session starts. Moreover, their special role prevents them from being deleted 
with the command \texttt{remove}.

Full support is also offered to load entire UFO models
\cite{Degrande:2011ua}. The UFO format is automatically detected by
\madanalysis\ 5 so that, in order to load the model, it is enough to issue in
the interpreter 
\begin{verbatim}
  import <path-to-UFO-files>
\end{verbatim}
where \texttt{<path-to-UFO-files>} is the location of the UFO model under
consideration. The list of particle labels
is derived from the \texttt{particles.py} file containing the UFO implementation
of the particles of the model.
Moreover, any stable, electrically and color neutral particle, with the
exception of the photon, is
automatically added to the \texttt{invisible} multiparticle object.

As already presented in Section \ref{sec:firststeps} and in Table
\ref{tab:actions},
the command interpreter of \madanalysis\ 5 contains two actions for
printing to the screen the list of all the particle and multiparticle labels
which have been defined in the current session,
\begin{verbatim}
  display_particles      display_multiparticles
\end{verbatim}
Moreover, as for any instance of any class, the properties of a given particle
or multiparticle object can be obtained with the action \texttt{display}, 
\begin{verbatim}
  display <label>
\end{verbatim}
where \texttt{<label>} is the label of the (multi)particle under consideration.
The effect of the
command above is to print to the screen, in the case of a particle, the PDG-id
linked to the label \texttt{<label>}. Similarly, for multiparticle labels, the
whole list of associated PDG-ids is displayed. 

As already shown in Section \ref{sec:firststeps} where we have taken the example
of muons and antimuons, new particle and multiparticle labels can
be created, according to the needs of the user, by means of the command
\texttt{define},
\begin{verbatim}
  define <label> = <identifiers>
\end{verbatim}
The command above creates a label denoted by \texttt{<label>} and
associates to
it the content of the parameter \texttt{<identifiers>}. If the value of 
\texttt{<identifiers>} consists in one single integer number, a new particle
label
related to the corresponding PDG-id is created. In the case we have either 
several integer numbers separated by spaces or several other particle and
multiparticle labels separated by spaces, a new multiparticle label is created
and linked to the list of corresponding PDG-ids. The \texttt{define}
command also allows for redefining existing labels or to add extra
(multi)particles to a definition. In this last case, the user is allowed to
issue commands such as, \eg, 
\begin{verbatim}
  define <label> = <label> <identifiers>
\end{verbatim}
The effect of the command above is to add the particle(s) contained in the
variable \texttt{<identifiers>} to the definition of the label
\texttt{<label>}. 

Even if, in principle, the user is allowed
to choose any name for the labels to be created, \madanalysis\ 5 contains a set
of reserved keywords that cannot be used as labels, such as the symbols
\texttt{all}, \texttt{or} and \texttt{and} (see below). In addition, the names of
the different actions (see Table \ref{tab:actions}) cannot be used for
(multi)particle labels.

At any time, a label not already used for the definition of a histogram or of a
cut can be deleted from the memory of the computer by issuing
\begin{verbatim}
  remove <label>
\end{verbatim}
where the label to be deleted is denoted by \texttt{<label>}. 
Due to their particular nature, the labels \texttt{hadronic} and
\texttt{invisible} however can never be removed.

\subsection{Creating histograms}\label{sec:uf_histo}
There are two types of observables that can be represented as histograms by
\madanalysis\ 5, global observables (see Table \ref{tab:g_observables}), related
to the entire event content, and observables related to a given type of
particle (see Table \ref{tab:p_observables}), thus specific to only a part of 
the event. In addition, a few additional observables are dedicated to 
analyses at the reconstructed-level and are collected in Table
\ref{tab:r_observables}.

We first address the description of the global observables. Two of them are
related to the missing
energy. The corresponding symbols implemented in \madanalysis\ 5 are
denoted by \texttt{MET} and \texttt{MHT} and correspond to the missing transverse
energy $\slashed{E}_T$ and the missing hadronic transverse energy
$\slashed{H}_T$, respectively. The definitions of these two quantities are
not unique in the literature, and we decide to adopt the choice
\bed
  \slashed{E}_T = \bigg|\bigg| \sum_{\text{visible particles}} \vec p_T \bigg|
    \bigg| \qquad\text{and}\qquad
  \slashed{H}_T = \bigg|\bigg| \sum_{\text{hadronic particles}} \vec p_T
    \bigg| \bigg| \ , 
\label{eq:miss}\eed
where $\vec{p}_T$ stands for the particle transverse momentum. 
At the parton-level, the generic name \textit{hadronic particles} stands for
gluons and light and $b$-quarks, but not for top quarks (decaying
most of the time to a $W$-boson and a $b$-quark). 
At the hadronic-level and reconstructed-level, \textit{hadronic particles} are 
trivially hadrons and jets, respectively.
These definitions are related to the default values of the multiparticle labels 
\texttt{hadronic} and \texttt{invisible}. We remind that the latter can be
modified by the user by
means of the command \texttt{define} (see Section \ref{sec:uf_labels}).

Even if particle objects are not explicitly tagged as 
visible, any non-invisible object, \ie, an object whose PDG-id is not 
included in the definition
of the label \texttt{invisible}, is considered by \madanalysis\ 5 as visible.
Similarly, any object which is not tagged as hadronic is considered as
non-hadronic.

Concerning the missing transverse energy $\slashed{E}_T$, a remark is in order.
Detector simulation tools are in general internally computing the missing
energy, following a built-in definition which might be different
from the one of Eq.\ \eqref{eq:miss}. This value is stored under a special tag
in the event files, compliant with the \lhco\ format relevant for
analyses at the reconstructed-level. When this is  
read by \madanalysis\ 5, the value of the missing transverse energy
is subsequently imported and always
supersedes the one which would have been computed by employing Eq.\
\eqref{eq:miss}.

Parallel to those observables related to the invisible sector of the events, 
two important global observables are related to the visible
objects, the total transverse energy $E_T$ and the total transverse
hadronic energy $H_T$. In \madanalysis\ 5, these kinematical quantities are
defined by
\bed 
  E_T = \sum_{\text{visible particles}} \big|\big| \vec p_T \big|
    \big| \qquad\text{and}\qquad
  H_T = \sum_{\text{hadronic particles}} \big|\big| \vec p_T \big|
    \big| \ ,
\label{eq:vis}\eed
and are associated to the symbols \texttt{TET} and \texttt{THT}, respectively.

Creating histograms associated to the four variables introduced above follows
the standard syntax of the command \texttt{plot},
\begin{verbatim}
  plot <observable> <nbins> <min> <max>
\end{verbatim} 
where the value of the variable \texttt{<observable>} corresponds here either to
\texttt{MET}, \texttt{MHT}, \texttt{TET} or \texttt{THT}. The number of bins,
the value of the lowest bin and the one of the highest bin are optional
information which can be passed through the 
symbols \texttt{<nbins>}, \texttt{<min>} and \texttt{<max>}, respectively.
If left unspecified, \madanalysis\ 5 uses built-in values.

The effect of the command \texttt{plot} is to create a new instance of the class
\texttt{selection} which has been designed to handle histograms (and cuts, as it
is shown in Section \ref{sec:uf_cuts}). 
The labeling of the \texttt{selection} objects is internally handled by
\madanalysis\ 5.The first time that the command \texttt{plot} is issued
in the current session of \madanalysis\ 5, the label \texttt{selection[1]} is 
assigned to the corresponding histogram. The second occurrence of the command
\texttt{plot} leads to the creation of the object \texttt{selection[2]}, and so
on. The full list of instances of the \texttt{selection} class which have been
created can be displayed to the screen with the \texttt{display} command,
\begin{verbatim}
  display selection
\end{verbatim}
As for particle and multiparticle labels, typing in the interpreter
\begin{verbatim}
  display selection[<i>]
\end{verbatim}
allows us to display the properties of the \texttt{<i>}$^{\text{th}}$ created
\texttt{selection} object. This \texttt{<i>}$^{\text{th}}$ selection can always
be deleted from the memory of the computer with the action \texttt{remove},
\begin{verbatim}
  remove selection[<i>]
\end{verbatim}
After the removal of a specific selection, the numbering of the different
created selections is
automatically adapted by \madanalysis\ 5 so that the sequence of the integer 
numbers
is respected, without any hole. Even if handled automatically, the ordering of
the selections can be modified by the user by means of the command
\texttt{swap}. Issuing
\begin{verbatim}
  swap selection[<i>] selection[<j>]
\end{verbatim}
results then in the exchange of the \texttt{<i>}$^{\text{th}}$ and
\texttt{<j>}$^{\text{th}}$ instances of the \texttt{selection} class. 

Objects of the \texttt{selection} class have several attributes, as shown in
Table \ref{tab:selection}. In this section, we however only focus on
\texttt{selection} objects related to histograms. In the case of cuts, we refer
to Section \ref{sec:uf_cuts}.
The number of bins, the value of the lowest bin of a histogram and the one of
its highest bin are stored in the attributes \texttt{nbins}, \texttt{xmin} and
\texttt{xmax}, respectively. Their values are fixed at the time the command
\texttt{plot} is issued in the interpreter. They can however be further
modified by the use of the command \texttt{set}, as for any other attribute of
an object, by typing in the command interface, \eg,
\begin{verbatim}
  set selection[<i>].xmax = 100
\end{verbatim}
This command allows us to set the value of the highest bin of the histogram
associated to the object \texttt{selection[<i>]} to 100.

\begin{table}
\bgfbyy
\multicolumn{2}{c}{\textbf{Table \ref{tab:selection}: List of the attributes of
the \texttt{selection} class}}\\
{\tt logX}  & If set to \texttt{true}, this enforces a logarithmic scale for the
$x$-axis. If set to \texttt{false} (default), a linear scale is used. \\
{\tt logY}  & If set to \texttt{true}, this enforces a logarithmic scale for the
$y$-axis. If set to \texttt{false} (default), a linear scale is used.\\
{\tt nbins} & Number of bins of the histogram. The value taken by this attribute
   must be an integer.\\
{\tt rank}  & This refers to the observable to be used for ordering particles
  of the same type. The value of this option has to be anything among
  \texttt{ETAordering} (pseudorapidity ordering),
  \texttt{ETordering} (transverse-energy ordering), \texttt{Eordering} (energy
  ordering), \texttt{Pordering} (momentum ordering), \texttt{PTordering}
  (transverse-momentum ordering, the default
  choice), \texttt{PXordering} (ordering according to the $x$-component of the
  momentum), \texttt{PYordering} (ordering according to the $y$-component of the
  momentum) or \texttt{PZordering} (ordering according to the $z$-component of
  the momentum). \\
{\tt stacking\textunderscore method} & .\\
& When several datasets are represented on an histogram, the different curves can
be stacked (\texttt{stack}), superimposed (\texttt{superimpose}) or
normalized to unity (\texttt{normalize2one}), including superimposing. The
default value is \texttt{auto} and then refers to the properties of the
attribute \texttt{stacking\textunderscore method} of the object
\texttt{main} (see Section \ref{sec:uf_submit}).\\ 
{\tt statuscode}  & By default, only final state particles are considered in
  histograms. This attribute allows us to consider instead initial or intermediate
  particles, or all the particle content. The (intuitive)
  allowed values are \texttt{allstate}, \texttt{initialstate}, 
  \texttt{finalstate} (default) and
  \texttt{interstate}.\\
{\tt titleX} & The title of the $x$-axis of the histogram, given as a string.\\
{\tt titleY} & The title of the $y$-axis of the histogram, given as a string.\\
{\tt xmin}  & Central value of the lowest bin of the histogram. This must be a
   real number. \\
{\tt xmax}  & Central value of the highest bin of the histogram. This must be a
   real number. \\
\egfbyy
\textcolor{white}{\caption{\label{tab:selection}}}
\end{table}

Two of the other
attributes of the \texttt{selection} class are related to the scales used when
drawing the axes of the histograms. By default, a linear scale is employed.
Logarithmic scales can be enforced
through the boolean attributes \texttt{logX} and \texttt{logY} which can then
take the values \texttt{true} or \texttt{false} (default choice). For instance,
issuing
\begin{verbatim}
  set selection[<i>].logY = true
  set selection[<j>].logX = false
\end{verbatim}
ensures the usage of a logarithmic scale for the $y$-axis of the histogram
related to the object \texttt{selection[<i>]} and the one of a linear scale for
the $x$-axis of the histogram associated to the object \texttt{selection[<j>]}. 

Another important attribute related to the layout of the histograms concerns the
stacking method used for the curves related to the different datasets created
by the user. By default, the curves are drawn as stacked one above each other,
but this can be
modified through the attribute \texttt{stacking\textunderscore method} of the
class \texttt{selection}. For instance, focusing on the object
\texttt{selection[<i>]}, the command
\begin{verbatim}
  set selection[<i>].stacking_method = <value>
\end{verbatim}
allows us to change the employed stacking method to the value \texttt{<value>}. By
default, this attribute is set to the value \texttt{auto}. This means that the
value of the attribute \texttt{stacking\textunderscore method} of the object
\texttt{main} (see Section \ref{sec:uf_submit}) is employed. The other
allowed choices are \texttt{stack}, \texttt{superimpose} and
\texttt{normalize2one}. In the first case, the curves are all stacked, whilst in
the second case, they are superimposed. The last possibility for the attribute
\texttt{stacking\textunderscore method}, \ie, \texttt{normalize2one}, has been 
designed for comparing the
shapes of the curves related to the different datasets. Here, the 
normalization of each curve is set to one and they are drawn superimposed. Let
us note that this consists in a helpful feature when optimizing selection cuts.

To achieve the description of the functions allowing us to tune the layout of a
histogram, the user is allowed to change the titles of the $x$-axis and $y$-axis
by means of the attributes \texttt{titleX} and \texttt{titleY} of the
\texttt{selection} class,
\begin{verbatim}
  set selection[<i>].titleX = <string>
  set selection[<i>].titleY = <string'>
\end{verbatim}
The effect of the commands above leads to the use of the strings
\texttt{<string>} and \texttt{<string'>} as titles for the $x$-axis and $y$-axis
of the histogram represented by the object \texttt{selection[<i>]}, respectively.

The attributes of an instance of the \texttt{selection} class related to a
histogram can also be directly set when issuing the command \texttt{plot},
\begin{verbatim}
  plot <observable> <nbins> <min> <max> [options]
\end{verbatim} 
The optional pattern \texttt{[options]} stands for \texttt{logX}, \texttt{logY},
\texttt{stack}, \texttt{superimpose}, \texttt{normalize2one}
or the value of any of the other attributes presented below,
\texttt{statuscode} and \texttt{rank}. Multiple keywords are allowed 
(however only one for each attribute). 
In this way, the values of several options can be passed at one time, each 
separated
by a space character. For instance, creating a histogram representing the
visible transverse hadronic energy with a $y$-axis represented with a logarithmic
scale and where all the curves are drawn superimposed can be done by issuing the
command
\begin{verbatim}
  plot THT [logY superimpose]
\end{verbatim} 
The binning is then automatically handled by \madanalysis\ 5, since it is not
provided by the user.

\begin{table}
\bgfbyyy
\multicolumn{2}{c}{\textbf{Table \ref{tab:g_observables}: List of the global 
observables that can}}\\
\multicolumn{2}{c}{\textbf{be represented by a histogram}}\\
{\tt NAPID} & 
   The particle multiplicity of the events after mapping particles and
   antiparticles.\\
{\tt NPID} & 
   The particle multiplicity of the events. \\ 
{\tt MET} & 
   The missing transverse energy as defined in Eq.\ \eqref{eq:miss}. At the
   reconstructed-level, the missing energy is however directly read from the
   \lhco\ file. \\
{\tt MHT} & 
   The missing transverse hadronic energy as defined in Eq.\ \eqref{eq:miss}. \\
{\tt SQRTS} & Partonic center of mass energy. Only available for parton-level
              and hadron-level event samples.\\
{\tt TET} & 
   The visible transverse energy as defined in Eq.\ \eqref{eq:vis}. \\
{\tt THT} & 
   The visible transverse hadronic energy as defined in Eq.\ \eqref{eq:vis}. \\
\egfbyyy
\textcolor{white}{\caption{\label{tab:g_observables}}}
\end{table}

In addition to the kinematical global observables $E_T$, $\slashed{E}_T$, $H_T$
and $\slashed{H}_T$ introduced in the beginning of this Section, the partonic
center-of-mass energy can be represented by a histogram by typing the command
\begin{verbatim}
  plot SQRTS
\end{verbatim}
Finally, two other
global observables are available. The latter are related to the particle
multiplicity of the events. In order to draw the associated histograms, it is
enough to enter in the command interface the two commands
\begin{verbatim}
  plot NPID 
  plot NAPID
\end{verbatim}
The first command, 
\texttt{plot NPID}, generates a histogram where one bin is associated to 
each possible type of final-state particle, the height of the bin being related
to the
multiplicity of the corresponding particle within the whole sample. Hence, if
several particles of the same type are present in one specific event, they 
correspond to several entries in the histogram. The second command above,
\texttt{plot NAPID}, produces a similar histogram after mapping
antiparticles and particles. For both types of histograms, the labels of the
$x$-axis correspond to the particle label imported in \madanalysis\ 5. If
non-existing, the PDG-ids are used instead. We emphasize that these histograms
are special histograms dedicated to the task of getting an idea about the
particles present in the input sample. To compute the multiplicity of a given
particle species, we refer to the observable \texttt{N} described below.

The second large class of observables that can be represented by histograms in
\madanalysis\ 5 refers to the kinematical properties 
of the particles contained
in the events. Hence, distributions such as the invariant mass or the
transverse momentum  of given particle species can be computed. The complete
list of implemented observables can be found in Table \ref{tab:p_observables}.

Creating histograms associated to a given property \texttt{<observable>} of a
specific particle represented by the label \texttt{<label>} is also based on the
command \texttt{plot}. In this case, the syntax is however slightly
different as for the global observables. The symbol \texttt{<observable>} has to
be seen as a function which
takes as the argument the label associated to the particle under consideration,
\begin{verbatim}
  plot <observable>(<label>) 
\end{verbatim} 
For instance, if \texttt{mu+} stands for the particle label related to antimuons, 
the command
\begin{verbatim}
  plot PT(mu+)
\end{verbatim} 
results in representing by a histogram the transverse-momentum distribution of
all the antimuons in the sample. As above, if several antimuons are included in
one single
event, they contribute to several entries in the histogram.
In the case of the relative distance between two particles (denoted by
\texttt{DELTAR}), two particle labels \texttt{<label1>} and \texttt{<label2>}
are required, 
\begin{verbatim}
  plot DELTAR(<label1>, <label2>) 
\end{verbatim}

As illustrated above, when studying the kinematical properties of a
given particle species, the normalization of the histograms reflects the total
number of
particles of this type included in the full sample. This can be different from
the total number of events, since one single event could describe a final state
with zero, one or several particles of the 
considered type, which then
corresponds to zero, one or several entries in the produced
histograms\footnote{In this Section, we do not consider the luminosity
which also affects the normalization of the histograms. This feature is
described in more detail in Section \ref{sec:uf_submit}.}.

\begin{table}
\bgfbyy
\multicolumn{2}{c}{\textbf{Table \ref{tab:p_observables}: List of the 
kinematical observables that can}}\\
\multicolumn{2}{c}{\textbf{be represented by histograms}}\\
{\tt BETA}  & Velocity $\beta = v/c$.\\
{\tt DELTAR}& Relative distance between two objects in the $\eta-\phi$ plane.\\
{\tt E}     & Energy.\\
{\tt ET}    & Transverse energy.\\
{\tt ETA}   & Pseudorapidity.\\
{\tt GAMMA} & Lorentz factor.\\
{\tt M}     & Invariant mass.\\ 
{\tt N}     & Particle multiplicity.\\
{\tt MT}    & Transverse mass.\\
{\tt P}     & Norm of the momentum.\\ 
{\tt PHI}   & Azimuthal angle of the momentum.\\
{\tt PT}    & Norm of the transverse momentum.\\
{\tt PX}    & Projection of the momentum on the $x$-axis.\\
{\tt PY}    & Projection of the momentum on the $y$-axis.\\
{\tt PZ}    & Projection of the momentum on the $z$-axis.\\
{\tt R}     & Position of the object in the $\eta-\phi$ plane.\\ 
{\tt THETA} & Angle between the momentum and the beam axis.\\
{\tt Y}     & Rapidity.\\
\egfbyy
\textcolor{white}{\caption{\label{tab:p_observables}}}
\end{table}

The syntax detailed above is also valid for multiparticle objects. In this case,
the label \texttt{<label>} is related to an instance of the multiparticle
class. Histograms are created by treating on the same footing all the particles
linked to the label \texttt{<label>}. For example, the set of commands
\begin{verbatim}
  define <multi> = <particle1> <particle2>
  plot <observable>(<multi>)
\end{verbatim} 
defines, in a first step, a multiparticle label denoted by \texttt{<multi>} and
linked to the particle species \texttt{<particle1>} and \texttt{<particle2>}
(see Section \ref{sec:uf_labels}). In
a second step, a property represented by the observable \texttt{<observable>} is
investigated. For a specific event, each particle of the type
\texttt{<particle1>} or \texttt{<particle2>} is associated to one 
entry in the histogram.

As shown above, if several particles of the considered type appear in a
specific event, they always correspond to several entries in the histograms.
In phenomenological analyses, it is however often more relevant to only consider
the leading particle, \ie, the one which has the highest value of a kinematical
variable such as the transverse momentum or the energy. The squared brackets
`\texttt{[ ]}' allow us to access leading, next-to-leading, \etc,
particles. 
For instance, the histogram resulting from the command
\begin{verbatim}
  plot <observable>(<label>[<i>]) 
\end{verbatim} 
represents the 
distribution of the observable \texttt{<observable>} for the
particle of the type \texttt{<label>} with the \texttt{<i>}$^\text{th}$ largest
transverse momentum, \texttt{<i>} being a positive integer. Similarly, a 
negative value of the parameter \texttt{<i>} corresponds to a pointer to 
the particle with
the \texttt{<i>}$^\text{th}$ smallest transverse momentum. Events where the
number of particles of the type under consideration is smaller than
the absolute value of \texttt{<i>} are ignored at the time of the creation of 
the histogram.

By default, the ordering variable employed in \madanalysis\ 5 is the transverse
momentum. Other possible choices exist and the information is passed, for a
given histogram, through the value of the attribute \texttt{rank} of the
\texttt{selection} class. Hence, considering the instance of the class
\texttt{selection[<i>]}, one can modify the ordering variable by issuing 
\begin{verbatim}
   set selection[<i>].rank = <value>
\end{verbatim}
The parameter \texttt{<value>} can take any value among \texttt{ETAordering}
(pseudorapidity ordering), \texttt{ETordering} (transverse-energy ordering),
\texttt{Eordering} (energy ordering), \texttt{Pordering} (momentum ordering),
\texttt{PTordering} (transverse-momentum ordering), \texttt{PXordering}
(ordering according to the $x$-component of the momentum), \texttt{PYordering}
(ordering according to the $y$-component of the momentum) and
\texttt{PZordering} (ordering according to the $z$-component of the momentum).

For all the histograms produced so far, \madanalysis\ 5 only uses information
related to the final-state particles. However, for phenomenological purposes, it
is often interesting
to investigate the properties of the intermediate particles or those of
final-state particles issued from the decays of a specific type of intermediate
particle. This requires, of course, that the relevant information is available.
This is not the case for event samples at the reconstructed-level since these
features are totally absent from event files under the \lhco\ format. In
contrast, the user can
access, for hadron-level or parton-level event files, to the
whole particle history by a set of mother-to-daughter relations. To benefit from
those relations, there exist
two special functions in \madanalysis\ 5.

Firstly, the symbol `\texttt{<}' links one particle or set of particles to their
direct mother. The command line
\begin{verbatim}
  plot <observable>(<type1> < <type2>)
\end{verbatim}
allows us hence to study a given property, which is represented by the symbol
\texttt{<observable>}, of the particles of type \texttt{<type1>} included in the
final state of the events under consideration. However, in order
to correspond to an entry in the histogram, the particles of type \texttt{type1}
must be 
issued from the direct decay of a particle of type \texttt{<type2>}.
Doubling the symbol `\texttt{<}', \ie, replacing it by `\texttt{<<}', allows us to
remove the restriction of a \textit{direct} decay. Finally, the usage of these
symbols recursively
\begin{verbatim}
  plot <observable>(<type1> < <type2> << <type3>) 
\end{verbatim}
allows us to focus on entire decay chains. Here, we investigate the properties of
the particle species \texttt{<type1>}, but only for particles of type
\texttt{type1} issued from a direct decay of a particle of type
\texttt{<type2>}, which is itself issued from a decay of a particle represented
by the label \texttt{<type3>}, this last decay possibly occurring in several
steps. 

Secondly, kinematical properties of
the intermediate particles can be directly investigated through the option
\texttt{statuscode} of the \texttt{selection} class. As suggested above, by
default, only final-state particles are considered. This corresponds to the
value \texttt{finalstate} of the attribute \texttt{statuscode}. 
Issuing in the interpreter 
\begin{verbatim}
  set selection[<i>].statuscode = interstate
\end{verbatim}
indicates that, for the selection \texttt{selection[<i>]}, we are not
considering final-state particles anymore, but only intermediate states. On
the same footing, setting \texttt{statuscode} to the value
\texttt{initialstate} allows us to focus on the initial-state particles only,
whilst
setting it to \texttt{allstate} allows us to consider equivalently initial-state,
final-state and intermediate-state particles.

Key observables to design efficient selection cuts in an analysis are
in general related to more than one single particle. For
instance, highlighting a new $Z'$ gauge boson in Drell-Yan events and estimating
its mass with a good accuracy rely on the invariant-mass distribution of the
produced lepton pair. All the functions of Table \ref{tab:p_observables}, but
the \texttt{DELTAR} observable which requires exactly two arguments, 
can take an arbitrary number of arguments. This allows us to
combine particles before computing the kinematical distribution to be
represented. Hence, the two equivalent command lines 
\begin{verbatim}
  plot <observable>(<prtcl1> <prtcl2>) 
  plot v<observable>(<prtcl1> <prtcl2>) 
\end{verbatim} 
lead to the creation of a histogram showing the distribution of the
observable \texttt{<observable>}. The observable is computed on the basis of the
combined 
four-vector built from the sum of the four-momentum of the particles
\texttt{<prtcl1>} and \texttt{<prtcl2>}. The optional prefix `\texttt{v}'
indicates that the four-momenta are combined vectorially (other options are
shown below). If, for a given event, several pairs of particles
\texttt{<prtcl1>} and \texttt{<prtcl2>} can be formed, each possible pair 
leads to one different entry in the histogram. Hence, the histogram describing
the invariant mass of a muon pair can be created by issuing
\begin{verbatim}
  plot M(mu+ mu-) 
\end{verbatim} 
where, as before, the binning is automatically handled by \madanalysis\ 5. The
observable that is computed corresponds to the norm of the sum of the
four-momentum of the muon and the one of the antimuon.
This syntax can straightforwardly
be generalized to multiparticles or to combinations of more than two particles. 
A remark is however in order here. If \texttt{<multi>} denotes the label
associated to an instance of the multiparticle class, issuing
\begin{verbatim}
  plot <observable>(<multi> <multi>) 
\end{verbatim}
generates a histogram where each entry corresponds to one \textit{different}
combination of the particles represented by the multiparticle \texttt{<multi>}.
\madanalysis\ 5 indeed forbids double-counting any combination.

By default, the particles are combined by adding vectorially their
four-momentum, \ie,
\be
  p_\mu = \sum_i p^i_\mu \ ,
\ee
where $p_\mu$ is the resulting four-momentum and $p_\mu^i$ are the four-momenta
of the particles to be combined. The four-vector $p^\mu$ is the one which is 
used when computing the value of the observable to be represented on the
corresponding histogram. \madanalysis\ 5 offers additional ways to perform this
combination. The sum could be, in contrast, done scalarly, \ie, by firstly
computing the considered observable for each of the particles to be combined
and secondly adding the results. To select this option, the user must add a
prefix `\texttt{s}' in front of the name of the observable when typing the
command in the interpreter,
\begin{verbatim}
  plot s<observable>(<prtcl1> ...)
\end{verbatim} 
where the dots stand for the list of particles to be combined. In the case the
user wants to combine all particles of a given type \texttt{<prtcl>} in an
event, the reserved keyword \texttt{all} can be used in order to simplify the
syntax,
\begin{verbatim}
  plot <observable(all <prtcl>) 
\end{verbatim}

If two (and only two) particles are combined, differences can also be computed,
rather than sums. To allow for this option, it is enough to include one of the
the prefixes `\texttt{dv}', `\texttt{vd}', `\texttt{d}', `\texttt{ds}',
`\texttt{sd}' or
`\texttt{r}' in front of the symbol of the observable to be computed. For the
options `\texttt{dv}', `\texttt{vd}' and `\texttt{d}', vectorial differences are
considered. Hence, the result of one of the equivalent commands
\begin{verbatim}
  plot dv<observable>(<prtcl1> <prtcl2>) 
  plot vd<observable>(<prtcl1> <prtcl2>)
  plot  d<observable>(<prtcl1> <prtcl2>)
\end{verbatim} 
is to subtract the two four-momenta from each other,
\be
  p_\mu = p^1_\mu - p^2_\mu \ ,
\ee
$p^1_\mu$ being the four-momentum of the particle associated to the label
\texttt{<prtcl1>} and $p^2_\mu$ the one of the particle associated to the label
\texttt{<prtcl2>} and then compute the observable \texttt{<observable>} from
the resulting four-vector $p^\mu$. In the case of the prefixes `\texttt{ds}' and
`\texttt{sd}', 
\begin{verbatim}
  plot ds<observable>(<prtcl1> <prtcl2>) 
  plot sd<observable>(<prtcl1> <prtcl2>) 
\end{verbatim} 
scalar differences are computed, \ie, the observable \texttt{<observable>} is
computed for each of the particles \texttt{<prtcl1>} and \texttt{<prtcl2>} and
the results are subtracted from each other.
Relative differences can also be computed by means of the prefix
`\texttt{r}',
\begin{verbatim}
  plot r<observable>(<prtcl1> <prtcl2>) 
\end{verbatim} 
In this case, the results consist in the scalar difference of the values of the
observable computed for the two particles \texttt{<prtcl1>} and
\texttt{<prtcl2>} taken individually, which is however given relative to the value
of the observable for the first particle represented by the label
\texttt{<prtcl1>}. This corresponds then to the quantity
\be
  \frac{\texttt{<observable>}(\texttt{<prtcl1>}) -
    \texttt{<observable>}(\texttt{<prtcl2>})}{
    \texttt{<observable>}(\texttt{<prtcl1>})}
\ . 
\ee
If the considered observable vanishes for the particle labeled by
\texttt{<prtcl1>}, the value zero is returned.

\begin{table}
\bgfbyy
\multicolumn{2}{c}{\textbf{Table \ref{tab:r_observables}: List of the 
additional observables that can}}\\
\multicolumn{2}{c}{\textbf{be represented by histograms for reconstructed
  events}}\\
{\tt EE\textunderscore HE} & Ratio of the electromagnetic and hadronic energy
   for a given object.\\
{\tt HE\textunderscore EE} & Ratio of the hadronic and electromagnetic energy
   for a given object.\\
{\tt NTRACKS}  & Number of tracks in a jet. This returns zero for non-jet
  objects.\\
\egfbyy
\textcolor{white}{\caption{\label{tab:r_observables}}}
\end{table}

A final way for combining observables is related to the reserved word
\texttt{and}. When the considered observable has to be evaluated for several
possible combinations of particles, the user can use the keyword \texttt{and} to
efficiently implement the corresponding command in \madanalysis\ 5, 
\begin{verbatim}
  plot <observable>(<p1> <p2> and <p3> <p4>) 
\end{verbatim} 
The command line above computes the observable represented by the symbol
\texttt{<observable>}, firstly, for the vectorial combination of the particles 
\texttt{<p1>} and \texttt{<p2>} and secondly, for the vectorial combination of
the particles \texttt{<p3>} and \texttt{<p4>}. The two distributions are
eventually summed.

Three additional observables can be represented by histograms in the case of
fully-reconstructed objects. To allow for generating histograms for these
observables, \madanalysis\ 5 must be run in the
reconstructed-level mode. These observables consist of the ratio between the
hadronic and the
electromagnetic energy for a given object (the ratio between the energy
deposited in the electromagnetic and hadronic calorimeters of a detector), the
inverse ratio and the number of tracks within a
(reconstructed) jet. For objects different from a jet, this last observable 
always returns 
the number zero. The associated symbols are \texttt{HE\textunderscore EE},
\texttt{EE\textunderscore HE} 
and \texttt{NTRACKS} and their definitions are collected in
Table \ref{tab:r_observables}. The corresponding histograms can be created by
following the usual syntax,
\begin{verbatim}
  plot <observable>(<label>) 
\end{verbatim} 
Let us note that for these three observables, combining objects is not
supported, \ie, only one single (multi)particle label can be passed as an
argument.

\subsection{Selection cuts}\label{sec:uf_cuts}
In \madanalysis\ 5, the process of event selection is based on two equivalent
classes of kinematical cuts which can be applied to the imported datasets. The
program offers to the user the two choices of either selecting or
rejecting events in the case a certain condition is fulfilled. This task can be 
performed at the level of the command line interpreter of the program by
means of the two actions \texttt{select} and \texttt{reject}. The associated
syntax is very intuitive and reads
\begin{verbatim}
  select <condition> [<options>]
  reject <condition> [<options>]
\end{verbatim}
For the first (second) command, events are selected (rejected) if the
condition \texttt{<condition>} is satisfied. Hence, the command
\begin{verbatim}
  reject PT(mu) > 50
\end{verbatim}
leads to the rejection of all the events where at least one muon with a transverse
momentum $p_T > 50$ GeV is found, whilst issuing
\begin{verbatim}
  select M(e+ e-) > 100
\end{verbatim} 
allows for the selection of all the events where we have a electron-positron
pair with an invariant mass larger than 100 GeV. Internally, the effects of the
commands above are to create instances of the \texttt{selection} class with
special properties. Contrary to the command \texttt{plot} which is related to
the creation of histograms, the commands \texttt{select} and \texttt{reject}
lead to the production of tables of cut efficiencies. 
Consequently, the only attributes of the \texttt{selection} class which are
relevant are \texttt{rank} and \texttt{statuscode} (see Table
\ref{tab:selection}). They can be either passed directly at the time of
typing-in
the commands, by including the desired values as the optional parameter
\texttt{<options>} above, or at a
later stage by means of the command \texttt{set} (see Section
\ref{sec:uf_histo}). 

Modifying the \texttt{rank} attribute only plays a role if
the ordering of the particles is necessary information for a good application
of the condition, as \eg, if we are constraining some observable related to the
leading or next-to-leading particles. Furthermore, setting the option
\texttt{statuscode} to a non-default value allows us to apply, if needed by the
user, cuts on initial or intermediate states rather than on final states only.

The condition \texttt{<condition>} must be given according to the pattern
\begin{verbatim}
  <observable> <logical-operator> <value>
\end{verbatim}
where the observable \texttt{<observable>} can be any observable from Tables
\ref{tab:g_observables}, \ref{tab:p_observables} and \ref{tab:r_observables},
computed for a given particle or for any combination of particles, 
with the exception of the global variables related to the symbols \texttt{NPID}
and \texttt{NAPID}\footnote{To implement selection cuts on the multiplicity of a
given particle species, cuts on the observable \texttt{N} have to be performed,
rather than on the global observable \texttt{NPID} and \texttt{NAPID}.}. 
The supported logical operators are 
`\texttt{>}' (greater
than), `\texttt{>=}' (greater than or equal to), `\texttt{<}' (smaller
than), `\texttt{<=}' (smaller than or equal to), `\texttt{==}' (equal to)
and `\texttt{!=}' (different from).

Conditions can also be combined by using one or several of the connecting
keywords \texttt{and} (logical \textit{and}) and \texttt{or} (logical
\textit{inclusive or}), such as
in
\begin{verbatim}
  <cond1> <connector1> <cond2> <connector2> <cond3>
\end{verbatim}
The condition above consists of the combination of the three conditions
\texttt{<cond1>}, \texttt{<cond2>} and \texttt{<cond3>} by employing the two 
connecting keywords \texttt{<connector1>} and \texttt{<connector2>} being
\texttt{and} or \texttt{or}. Moreover, brackets are also
authorized by the syntax for handling more complex conditions.
In the special case both upper and lower limits are imposed on a given
observable, the condition can be easily implemented by means of the keyword
\texttt{and}. However, there exists a more compact syntax  
\begin{verbatim}
  <value1> <logical-operator> <obs> <logical-operator> <value2>
\end{verbatim}
which could be employed. The logical operator \texttt{<logical-operator>} is
here used twice. The parameter \texttt{<obs>} denotes the observable and
\texttt{<value1>} and \texttt{<value2>} the imposed bounds. 

So far, we have supposed that all the objects present in each event can be used
for applying selection cuts. However, this is barely the case in most
realistic phenomenological analyses, since, for example, too soft particles are
in general omitted. This feature can also be implemented in analyses performed
with \madanalysis\ 5 by means of the commands \texttt{select} and
\texttt{reject}, but following a slightly different syntax as the one shown 
before,
\begin{verbatim}
  select (<particle>) <condition> [<options>]
  reject (<particle>) <condition> [<options>]
\end{verbatim}
The commands above allow us to respectively select and reject any particle
associated to
the label \texttt{<particle>}, if the condition \texttt{<condition>} is
fulfilled. The syntax for typing-in the condition is similar to the one employed
for implementing conditions associated to the selection or rejection of 
events, with the difference that the observable entering the
condition cannot here take any argument and must simply be one of the symbols
presented in Tables \ref{tab:p_observables} and \ref{tab:r_observables}. 

In the case the particle candidate is rejected, the event is
considered as without containing this particle. 
Let us note that whilst selecting and rejecting events have a
direct influence on the signal over background ratio, selection or rejecting
candidates to a particle type only affects the number of entries for one event 
in the histograms.

\subsection{Executing an analysis and displaying the results} \label{sec:uf_submit}
Once an analysis has been implemented, it must be passed to \sampleanalyzer,
the \cpp\ kernel of \madanalysis\ 5, for execution by means of the command
\begin{verbatim}
  submit <dirname>
\end{verbatim}
A directory named \texttt{dirname} is created and all the
files necessary for \sampleanalyzer\ to properly run are generated and included
in this directory. The code is further compiled and linked to the external
static library of \madanalysis\ (see Section \ref{sec:fs_start}), and the
execution of the resulting program is eventually managed by \madanalysis\ 5.

This execution starts with the reading of the event samples under consideration
and their
storing in the memory of the computer according to a format internal to
\madanalysis\ 5. All the histograms required by the user are then sequentially
created, including the application to all the events of the defined selection
cuts. As an output, a \rooot\ file \cite{Brun:1997pa} is generated so that the
analysis can be accessed later, as, \eg, directly in the \rooot\ framework or 
in a new session of \madanalysis\ 5. In
this last case, the \sampleanalyzer\ (executed) job can be imported by means of
the command \texttt{import}
\begin{verbatim}
  import <dirname>
\end{verbatim}
where the directory \texttt{<dirname>} contains the analysis
previously performed.

After modifications such as asking for the creation of a
new histogram or the application of a new selection cut, the user does not have
to submit the \sampleanalyzer\ job entirely again. A much faster option, saving
a sensible amount of computing time, consists in the command
\begin{verbatim}
  resubmit 
\end{verbatim}
This updates the already existing \sampleanalyzer\ directory and only the
differences with respect to the original analysis are executed.

Once the \rooot\ file has been created by \sampleanalyzer, the command 
\texttt{preview} allows for
the display of a single histogram,
\begin{verbatim}
  preview selection[<i>]
\end{verbatim} 
where in the generic example above, the \texttt{<i>}$^\text{th}$ histogram is
asked to be displayed to the screen, in a \rooot\ popup window. The command
\texttt{preview} only works for displaying histograms. Consequently,
previewing the efficiency table associated to a selection cut is not possible.

A more complete report, with all the selection cuts, efficiency tables and
histograms can be generated by issuing in the command line interface one of the
three commands
\begin{verbatim}
  generate_html <html-dirname>
  generate_latex <tex-dirname>
  generate_pdflatex <pdftex-dirname>
\end{verbatim}
The first command, \texttt{generate\textunderscore html}, generates the report
under the \html\ format and stores the 
files in the directory \texttt{<html-dirname>}. 
The second and third commands, \texttt{generate\textunderscore latex} and
\texttt{generate\textunderscore pdflatex}, are related to the creation of \TeX\
files to be compiled with
the help of the shell commands \texttt{latex} and \texttt{pdflatex},
respectively. In these two cases, the \TeX\ files are already compiled by
\madanalysis\
5, if the \texttt{latex} and \texttt{pdflatex} commands are available on the
system of the user. 

Histograms are exported to (non-\rooot) figures
under either the Encapsulated PostScript (\texttt{.eps}) format, required for a
proper compilation with
\texttt{latex}, or to the  Portable Network
Graphics (\texttt{.png}) format for \html\ files or \TeX\ files to be
compiled with \texttt{pdflatex}. In addition, each figure is also saved as a
\rooot\ macro which can be modified by the user for further processing.
Once generated, the report can be immediately displayed by typing-in the command
\begin{verbatim}
  open <report-dirname>
\end{verbatim}
which opens the report in a web browser.

The analysis, and thus the generation of the histograms and the efficiency
tables, is so far based on the default configuration of \madanalysis\ 5. This
configuration can
be modified by superseding the default values of the attributes of the object
\texttt{main}, which are listed in Table \ref{tab:main} and Table
\ref{tab:main2}.

\begin{table}
\bgfbyy
\multicolumn{2}{c}{\textbf{Table \ref{tab:main}: List of the attributes of
the object \texttt{main}}}\\
\multicolumn{2}{c}{\textbf{\texttt{set main.<option> = <value>}}}\\$~$\\
{\tt currentdir} & Current directory in which any directory created by
\madanalysis\ 5 is stored.\\
{\tt lumi} & 
   This allows us to modify the value of the integrated luminosity, in fb$^{-1}$,
   used for the normalization of the histograms. The default value is 10
   fb$^{-1}$.\\
{\tt normalize} & 
   This defines the way the histograms are normalized. The allowed choices are
   the number of events included in the different datasets (\texttt{none}), the
   integrated luminosity without (\texttt{lumi}) or accounting for the event
   weights associated to each dataset (the default choice,
   \texttt{lumi\textunderscore weight}).\\ 
{\tt SBerror} & This fixes the way the uncertainty on the signal over background
ratios is computed by \madanalysis\ 5. The attribute \texttt{<value>} is the
corresponding  analytical formula passed as a valid \python\ expression given as
a string. It has to depend on \texttt{S} (number of signal events), \texttt{B}
(number of background events),
\texttt{ES} (uncertainty on the number of signal events) and \texttt{EB}
(uncertainty on the number of 
background events). It is automatically handled for the most simple cases (see Eq.\
\eqref{eq:sb}).\\
{\tt SBratio} & This fixes the way signal over background ratios are
computed by \madanalysis\ 5. The attribute \texttt{<value>} is the corresponding 
analytical
formula passed as a valid \python\ expression given as a string. It has to depend on
\texttt{S} (number of signal events) and \texttt{B} (number of background events).\\
{\tt stacking\textunderscore method} &  \\
 & When several datasets are represented on histograms, the different curves can
be stacked (\texttt{stack}, default), superimposed (\texttt{superimpose}) or
normalized to unity (\texttt{normalize2one}), including superimposing. When the
\texttt{stacking\textunderscore method} attribute of an instance of the class
\texttt{selection} is set to \texttt{auto}, the value of
\texttt{main.stacking\textunderscore method} is employed.\\ 
\egfbyy
\textcolor{white}{\caption{\label{tab:main}}}
\end{table}

Two options allow us to control the normalization of the histogram, \texttt{lumi}
and \texttt{normalize}. This last attribute defines the way the histograms are
normalized. The allowed values are \texttt{none}, \texttt{lumi}
and \texttt{lumi\textunderscore weight} and can be set as for any other
attribute of any class,
\begin{verbatim}
  set main.normalize = <value>
\end{verbatim}
For the first choice 
(\texttt{<value>} $=$ \texttt{none}), the total number of events included in
each dataset is kept. The two other options imply that the histograms are
normalized with respect to the integrated luminosity. The difference 
lies in the weight which can be possibly associated to each event (see
Section \ref{sec:uf_datasets}), which can be ignored 
(\texttt{<value>} $=$ \texttt{lumi}) or included (\texttt{<value>} $=$
\texttt{lumi\textunderscore weight}). This last possibility consists in the
default choice. 

By default, all the histograms are normalized to an
integrated luminosity of 10 fb$^{-1}$. This value can be updated by
modifying the attribute \texttt{lumi} of the object \texttt{main},
\begin{verbatim}
  set main.lumi = <new-value>
\end{verbatim}
where the new value of the integrated luminosity, \texttt{<new-value>}, is given
in fb$^{-1}$.

Once all the datasets have been defined as part of the signal or background
samples, 
\madanalysis\ 5 can compute automatically the signal ($S$) over background ($B$)
ratio, and thus the efficiency of each selection cut. This feature is related to
an attribute of the object \texttt{main} denoted by \texttt{SBratio}. It refers 
to an analytical formula, expressed as a valid \python\ expression, which
indicates how the signal over background ratio must be calculated\footnote{The
formula is internally stored as an instance of the \texttt{TFormula} class. This
structure is defined in the \rooot\ library linked to \madanalysis\ 5, and all 
the associated attributes can therefore be used. We
refer to the \rooot\ manual for more information \cite{Brun:1997pa}.}. When
implementing this formula, the symbols related to the signal and background
number of events are \texttt{S} and \texttt{B}, respectively. By default,
the signal over background ratio is computed according to \texttt{S/B}. 
This can be modified through the command \texttt{set}, by issuing in the
interpreter,
\begin{verbatim}
  set main.SBratio = '<formula>'
\end{verbatim}
where \texttt{<formula>} is, as sketched above, a valid \python\ expression
depending on the two variables \texttt{S} and \texttt{B}. For instance, the
command 
\begin{verbatim}
  set main.SBratio = 'S/sqrt(S+B)'
\end{verbatim}
enforces the signal over background ratio $r$ to be computed according to 
\be
  r = \frac{S}{\sqrt{S+B}} \ .
\ee
In the case the signal over background ratio is undefined due, \eg, to the
evaluation of the squared root of a negative number or to a division by zero,
the value zero is returned. Let us emphasize that it is safer for the user 
to use real numbers 
when typing-in the analytical expression \texttt{<formula>}, rather than
integer numbers. We indeed recall that $1/2$ is evaluated as $0$ whilst $1./2.$
returns $0.5$.

It is fundamental to associate an uncertainty to the signal over background
ratio. The way to compute this quantity is related to the \texttt{SBerror}
attribute of the object \texttt{main}. As for \texttt{SBratio}, it refers to an
analytical formula, given as a valid \python\ expression, which indicates how
the uncertainty on the signal over background ratio must be computed. For the
three choices
\bed
  r_1 = \frac{S}{B} \ , \qquad
  r_2 = \frac{S}{S+B} \qquad\text{and}\qquad
  r_3 = \frac{S}{\sqrt{S+B}} \ , 
\label{eq:sb}\eed
as well as for the three additional cases obtained when $S$ and $B$ are exchanged, 
\madanalysis\ 5 automatically detects the formula stored in the \texttt{SBratio}
attribute. It then updates accordingly the uncertainty, setting the
\texttt{SBerror} attribute to the values $\Delta r_i$ given by
\be\bsp
  \Delta r_1 =&\ \frac{\sqrt{B^2 (\Delta S)^2 + S^2 (\Delta B)^2}}{B^2} \ , \\
  \Delta r_2 =&\ \frac{\sqrt{B^2 (\Delta S)^2 + S^2 (\Delta B)^2}}{(S+B)^2} \ , \\
  \Delta r_3 =&\ \frac{\sqrt{(S+2B)^2 (\Delta S)^2 + S^2 (\Delta B)^2}}{2
    (S+B)^{3/2}} \ ,
\esp\ee 
where $\Delta S$ and $\Delta B$ are the uncertainties on the signal and on the
background number of events. If the user wants to use another formula or to set
himself the \texttt{SBerror} attribute, the command \texttt{set} has to be
employed,
\begin{verbatim}
  set main.SBerror = '<formula>'
\end{verbatim}
where \texttt{<formula>} depends this time on the number of signal and
background events \texttt{S} and \texttt{B} as well as on the associated
uncertainties \texttt{ES} ($\equiv \Delta S$) and \texttt{EB} ($\equiv \Delta
B$). For instance, implementing by hand the error $\Delta r_1$ above would give
\begin{verbatim}
  set main.SBerror = 'sqrt(B**2*ES**2+S**2*EB**2)/B**2'
\end{verbatim}

\begin{table}
\bgfbyy
\multicolumn{2}{c}{\textbf{Table \ref{tab:main2}: List of the attributes of
the object \texttt{main}}}\\
\multicolumn{2}{c}{\textbf{associated to the isolation of the muons.}}\\
\multicolumn{2}{c}{\textbf{\texttt{set main.isolation.<option> = <value>}}}\\$~$\\
{\tt algo} & This specifies the algorithm to be employed for muon
isolation. The allowed choices are \texttt{DELTAR} (default) and \texttt{SUMPT}.
In the first case, we require that no tracks lies inside a cone around the
muon. In the second case, the sum of the transverse-momentum
of all the tracks inside the cone and the
ratio of the summed transverse energy of these tracks over their summed
transverse momentum must be lower than values specified by the user. For 
the second algorithm, the size of the cone is fixed by the detector simulation
tool.\\
{\tt deltaR} & This specifies the radius of the isolation cone to be used by the
\texttt{DELTAR} isolation algorithm.\\
{\tt ET\textunderscore PT} & This specifies the value of the ratio of the sum
of the transverse energy of the tracks lying in the isolation cone over the sum 
of their transverse-momentum to be used by the \texttt{SUMPT} algorithm.\\
{\tt sumPT} & This specifies the transverse-momentum threshold to be used by the
\texttt{SUMPT} algorithm.\\
\egfbyy
\textcolor{white}{\caption{\label{tab:main2}}}
\end{table}

Four specific attributes of the object \texttt{main} concern
muon isolation when \madanalysis\ 5 is run in order to analyze
reconstructed-level events. The user has the possibility to
choose the algorithm which is employed by \madanalysis\ 5 when defining an
isolated muon. Two choices are implemented and can be adopted by issuing one of
the two commands
\begin{verbatim}
  set main.isolation.algo = DELTAR
  set main.isolation.algo = SUMPT
\end{verbatim}
In the first case, a muon is tagged as isolated when no track lies inside a cone
around the muon. The size of this cone can be specified by the user by typing in the 
command interface
\begin{verbatim}
  set main.isolation.deltaR = <value>
\end{verbatim}
where \texttt{<value>} is a floating-point number. For the second algorithm, a 
muon is
considered as isolated when the sum of the transverse momentum of all the tracks
lying in a cone around the muon is lower than a value \texttt{<value>}. The
latter can be specified by typing
\begin{verbatim}
  set main.isolation.sumPT = <value>
\end{verbatim}
In addition, the ratio of the sum of the transverse energy of these tracks over
the sum of their transverse momentum has also to be lower than a value
\texttt{<value'>} to be
specified. This value is provided by means of the command 
\begin{verbatim}
  set main.isolation.ET_PT = <value'>
\end{verbatim}

On a fairly different line, the last attribute of the object \texttt{main} which
can be modified by the user is denoted by \texttt{currentdir}
\begin{verbatim}
  set main.currentdir = <dirname>
\end{verbatim}
It fixes the path to the directory in which any file and/or directory created by
\madanalysis\ 5 is stored.

\section{\madanalysis\ 5 for expert users}\label{sec:expert}

Besides its user-friendliness, the way of using \madanalysis\ 5 described in
Section \ref{sec:normalmode} is (obviously) restricted by the set of 
functionalities that have been implemented. 
The latter allow, in general, to perform rather traditional and standard
analyses but might not be sufficient for more sophisticated and/or exotic
investigations. For example, the normal running mode of the program does not
allow us to create a histogram related to either the distribution of a new
observable or to the one of an existing observable that needs to be computed 
in a reference frame different from the laboratory reference frame. 
Along the same lines, selection cuts must match the pattern presented in
Section \ref{sec:uf_cuts}, which forbids any other selection than cutting on the
implemented kinematical variables by means of assigning a lower and/or upper
bound on 
the result of the computation of the associated observable. Finally, as a
last example, two- or three-dimensional histograms, necessary for correlation
studies, are not included.

In order to overcome the above-mentioned restrictions as well as any type of,
even unforeseen, limitations, \madanalysis\ 5 comes with an expert
mode of running. In this case, the possibilities are only limited by the
programming skills of the user and his originality in designing the analysis. 
The user is asked to implement the entire analysis himself, knowing that he is 
able to benefit from all the strengths
of the \sampleanalyzer\ framework. The latter comes indeed with its own set of
reading routines for the event samples, its own data format and a large class of
functions and methods ready to be employed. In addition, an automated
compilation of the analysis \cpp\ files as well as a programming error 
management system are included.

As already presented in Section \ref{sec:uf_starting}, the expert mode of
\madanalysis\ 5 can be set by issuing in a shell one of the three commands
\begin{verbatim}
  bin/ma5 --expert      bin/ma5 -e      bin/ma5 -E
\end{verbatim}

\subsection{The \sampleanalyzer\ framework}\label{sec:em_sampleana}
In the expert mode, there are two possibilities in order to create an analysis,
\ie, to implement the \cpp\ source and header files which are automatically
generated by \madanalysis\ 5 in its normal mode of running. Either one can
start from and extend an existing analysis already generated by \madanalysis\
5, or one can design a new analysis from scratch.

In the first case, a working directory has already been created through the
issue, in the command line interface of \madanalysis\ 5, of the 
command \texttt{submit} (see Section \ref{sec:uf_submit}). Consequently, at
least one dataset has been imported (see Section \ref{sec:uf_datasets}) and at
least one histogram has been created (see Section \ref{sec:uf_histo}). 
The created working directory contains three sub-directories, 
\texttt{SampleAnalyzer}, \texttt{lists}
and \texttt{root}. The first of these directories, 
\texttt{SampleAnalyzer}, includes all
the files necessary for having the \cpp\ kernel of \madanalysis\ 5 properly
running and executing the analysis implemented by the user. When issuing the
command \texttt{submit} in the command interface, all the \python\ commands
entered by the user are translated into a set of three \cpp\ files,
\texttt{user.cpp}, \texttt{user.h} and \texttt{analysisList.cpp}, stored in the
sub-directory \texttt{Analysis} of \texttt{SampleAnalyzer}.

The two other sub-directories included in the working directory, 
\texttt{lists} and \texttt{root}, are dedicated to the input and output files,
respectively. As mentioned in Section \ref{sec:uf_datasets}, the imported events
are gathered into different datasets. For each dataset, a single list,
containing the paths to the various associated event files, is stored in
the directory \texttt{lists}. After being executed, \sampleanalyzer\ creates a
\rooot\ file for each of the defined datasets. These files, stored into the
directory \texttt{root}, contain the necessary information to
create the histograms requested by the user. 

When the user starts an analysis from scratch, the situation is similar to
the one described in the paragraphs above, with the exception that the working
directory is initialized directly from the shell, when \madanalysis\ 5 is
launched. When typing in a shell the command
\begin{verbatim}
  bin/ma5 --expert 
\end{verbatim}
or equivalently any of the three commands recalled in the introduction of this
section, \madanalysis\ 5 indeed asks a series of questions to the user
such as the name of the working directory to be created or the chosen label for
the analysis to be created (see below).

As a result, a working directory is created,
together with the sub-directory \texttt{SampleAnalyzer} which contains
a blank analysis. The implementation of this analysis, as in any
analysis, is divided into the implementation of three core functions. 
The latter have to be provided
by the user in the file
\texttt{user.cpp} which comes together with the corresponding header file
\texttt{user.h}. Both files are stored in the sub-directory
\texttt{SampleAnalyzer/Analysis}, together with an additional file,
\texttt{analysisList.cpp}, that contains
the list of all the analyses which are/will be implemented in the 
working directory under consideration. After the
creation of a fresh working directory, 
this list only contains a single analysis, the blank analysis 
pre-implemented in the file \texttt{user.cpp}. However, there is no limitation
on the number of analyses which can be included 
and nothing prevents this list from becoming very large. 

Once the three \cpp\ files
introduced above have been created, \madanalysis\ 5 exits and the user can then
start implementing his analysis by modifying these files.
In this Section \ref{sec:em_sampleana}, we do not address the way in which 
an analysis has to be implemented, since this is the scope of Section
\ref{sec:em_dataformat}, Section \ref{sec:em_sampleformat} and Section
\ref{sec:em_tools}, 
but we only describe the steps leading to the execution of the
analysis and the generation of the output files. 

In order to properly compile and execute the implemented analysis, 
the linking to the
external dependencies, such as the \rooot\ header files and libraries, must be
performed appropriately. This is allowed by a correct setting of the 
environment variables
\texttt{LD\textunderscore LIBRARY\textunderscore PATH} (or rather
\texttt{DYLD\textunderscore LIBRARY\textunderscore PATH} for {\sc MacOS}
systems),
\texttt{LIBRARY\textunderscore PATH} and 
\texttt{CPLUS\textunderscore INCLUDE\textunderscore PATH} prior to the
compilation.
This task has been rendered automatic by means of the
scripts \texttt{setup.sh} and \texttt{setup.csh} included in the
\texttt{SampleAnalyzer} sub-directory. All the above-mentioned environment
variables can be appropriately set at once by issuing 
\begin{verbatim}
  source setup.sh
\end{verbatim}
in a \texttt{bash} shell or
\begin{verbatim}
  source setup.csh
\end{verbatim}
in a \texttt{tcsh} shell before compiling and/or executing  \sampleanalyzer.

After this step, the analysis can be compiled with the help of the
\texttt{Makefile} present in the \texttt{SampleAnalyzer} directory, assuming
that the {\sc GNU Make} package has been installed on the computer of the user. 
As for any program to be compiled with {\sc GNU Make}, it is enough to issue in a
shell the command
\begin{verbatim}
  make
\end{verbatim}
which also takes care of linking the external libraries to the
\sampleanalyzer\ program. In addition, the command
\begin{verbatim}
  make clean
\end{verbatim}
has also been implemented and leads to the cleaning of the various
sub-directories included in the current working directory. 

Once compiled, the \sampleanalyzer\ core, containing the user's analysis, is
ready to be executed. The only input left to be provided consists in the
location paths of the event samples. They have, for a given dataset, to be 
collected into a single text file, denoted in the following by
\texttt{<datasetlist>}. The \sampleanalyzer\ package is then simply run by
issuing in a shell
\begin{verbatim}
  SampleAnalyzer [ options ] <datasetlist>
\end{verbatim}
If the event samples are collected into several datasets, \sampleanalyzer\ has
to be run once for each of the datasets to be included in the analysis, with a
different text file with the paths to the relevant event samples.

The only option supported by \sampleanalyzer\ consists in
specifying the label of the analysis to be performed. In particular, this allows
us
to include several analyses, each of them specified by a unique label, in one
single working directory of \sampleanalyzer. The user can hence indicate which
analysis to perform on run-time. If the option pattern is not provided, 
\sampleanalyzer\ lists to the screen all the analyses included in the file
\texttt{analysisList.cpp}, together with the corresponding labels, and asks to
the user to make his choice. 

The label of an analysis is defined in the corresponding header file. In the
case of a fresh working directory just created by \madanalysis\ 5, the chosen
name for the label is asked by the program and exported to the file \texttt{user.h}.
We refer to Section \ref{sec:em_template} for more information
about the way to declare labels independently from \madanalysis\ 5. 
Assuming that the label under
consideration is denoted by \texttt{<label>}, \sampleanalyzer\ is launched by
issuing in a shell
\begin{verbatim}
  SampleAnalyzer --analysis=<label> <datasetlist>
\end{verbatim}
As a result, the analysis defined by the label \texttt{<label>} 
is executed by \sampleanalyzer\ and the output files are generated
according to what the user has implemented in the \cpp\ files related to the 
corresponding analysis.

\subsection{Implementing new analyses using the analysis template}
\label{sec:em_template}
As mentioned in Section \ref{sec:em_sampleana}, the implementation of
an analysis within the \sampleanalyzer\ framework consists in the
writing of three files. Two of them are related to the analysis itself, \ie, 
one \cpp\ source file together with the associated header
file. Since a given working directory of \sampleanalyzer\ can
include many analyses, their list must be provided. This is the aim of the third
file, \texttt{analysisList.cpp}, which is common to all the present analyses.

As presented in the previous Section, a pair of such analysis source/header
files is automatically created when \madanalysis\ 5 is run in expert mode.
These files are denoted by \texttt{user.cpp} and \texttt{user.h}. However, we
adopt, in the following, the generic names \texttt{name.cpp} and
\texttt{name.h} since the user has in fact the freedom to choose the name
of the files. The only requirement is that all the filenames are different. 
Moreover, all these files have to be stored, together with the list of the
implemented
analyses included in the file \texttt{analysisList.cpp}, in the sub-directory
\texttt{SampleAnalyzer/Analysis}.

The pair of (generic) header and source analysis files \texttt{name.cpp}
and \texttt{name.h} contains the declaration of a class denoted by 
\texttt{name}, \ie, having the same name as the files. 
The class \texttt{name} is a daughter class
inheriting from the base class \texttt{AnalysisBase} that 
contains (empty) analysis methods. These methods are then specified at the level of
the definition of the daughter classes.

The structure of the header file \texttt{name.h} follows, for
any of the analyses included in the working directory,
\begin{verbatim}
  #ifndef analysis_name_h
  #define analysis_name_h
  
  #include "Core/AnalysisBase.h"
  
  class name: public AnalysisBase
  {
    INIT_ANALYSIS(name,label)
  
    public:
      virtual void Initialize();
      virtual void Finalize(const SampleFormat& summary, 
                 const std::vector<SampleFormat>& files);
      virtual void Execute(const SampleFormat& sample, 
                           const EventFormat& event);
  
    private:
  };
  #endif
\end{verbatim}
The only pieces among these predefined lines to be modified by the user are
the name tag \texttt{name},
related to the filename, and the label of the analysis \texttt{label} which is
a string. This label is the one that can be provided as an option 
when running the \sampleanalyzer\ code (see Section \ref{sec:em_sampleana}). 

In the \cpp\ code above, the \texttt{INIT\textunderscore ANALYSIS}
macro automatically creates the constructor and destructor methods 
associated to 
the class \texttt{name} and consistently links the filename to the name
of the analysis, which must be the same. As for any class
derived from the mother class \texttt{AnalysisBase}, the class \texttt{name}
must contain the three methods \texttt{Initialize}, \texttt{Execute} and
\texttt{Finalize}. 
Apart from this, the user is free to include his own set of additional
functions and variables as required to perform his analysis. 

The role of the three functions above is intuitive and related to their
names. When \sampleanalyzer\ is
executed in a shell, the program starts by calling (once) the function
\texttt{Initialize} 
\begin{verbatim}
  void Initialize()
\end{verbatim}
Its aim is to initialize all the internal variables, as well as
the entire set of user-defined variables such as the histograms that have been 
requested (and that must be properly declared in the header file
\texttt{name.h}).

After initialization, \sampleanalyzer\ loops over all the events which are
passed to the code. As shown in Section \ref{sec:em_sampleana}, 
the list of event files is collected into a single file which is passed as an
argument when executing the program from a shell. For each event, the function
\texttt{Execute} is called in order to perform the analysis. Among others, the
histograms are filled event by event and the cuts are applied. The method,
\begin{verbatim}
  void Execute(const SampleFormat& sample, 
               const EventFormat& event)
\end{verbatim}
takes as arguments, on the one hand, the event \texttt{event} under
consideration, which is passed as a set of particles with specific properties,
and on the other hand, general information on the entire event sample \texttt{sample}
currently analyzed, such as the
corresponding integrated cross section necessary for a proper normalization of
the histograms to be produced.

Once all the events have been processed, the function \texttt{Finalize} is
eventually called (once) by \sampleanalyzer
\begin{verbatim}
  void Finalize(const SampleFormat& summary, 
                const std::vector<SampleFormat>& files)
\end{verbatim}
Its aim is twofold. Firstly, it allows
for the creation of the histograms and cut-flow charts requested by the user.
To implement the latter in an easy way, the user can employ all the functionalities
included in the \rooot\ library and we therefore refer to the \rooot\ manual for
more information \cite{Brun:1997pa}.
Secondly, the method \texttt{Finalize} stores the results of the analysis 
as an output \rooot\ files which can be further processed or modified. 
The method takes as arguments general information
related to the event samples, \texttt{summary}, such as the
total cross section and the associated uncertainty. However, this time, instead
of having this information linked to a specific event sample, an 
average over the whole list of samples included in the dataset under
consideration has been performed. 
In the case the user wants to compute additional quantities when
implementing the method \texttt{Finalize}, 
the same information, related to each of the samples individually, is passed as
the second argument which is denoted, in the example above, as  \texttt{files}.  

These three predefined functions have to be implemented by the user in the 
corresponding source file \texttt{name.cpp}. To this aim, we refer to the
description of the internal data format used
by \sampleanalyzer\ for event processing (see Section \ref{sec:em_dataformat}
and Section \ref{sec:em_sampleformat})
and to the one of the methods included in the mother class \texttt{AnalysisBase}
(see Section \ref{sec:em_tools}).

As mentioned above, several analyses can be included in the same working
directory. The only requirement consists in implementing them in different files
and classes since each analysis is unambiguously defined by its
(file)name, which must therefore be unique. In order for
\sampleanalyzer\ to correctly handle them, the user must refer to the various
classes and files in the \cpp\ source file 
\texttt{analysisList.cpp}, located in the sub-directory
\texttt{SampleAnalyzer/Analysis}. This file contains a link to each of the
header files associated to the analyses to be included. In addition, one
instance of each analysis class is created. The architecture of this file
follows the structure 
\begin{verbatim}
  #include "Analysis/name1.h"
  #include "Analysis/name2.h"
  ...
  #include "Core/AnalysisManager.h"
  #include "Core/logger.h"
  
  void AnalysisManager::BuildTable()
  {
    Add(new name1);
    Add(new name2);
    ....
  }
\end{verbatim}
At least two analyses, denoted by \texttt{name1} and \texttt{name2} are
implemented and 
the corresponding header files have been included in \sampleanalyzer. 
In the example above, the dots stand for possible additional analyses
that the user might want to embed too. In the case the user has only
implemented one single analysis, the file \texttt{analysisList.cpp} must still
be present. It however then only includes the header file of this analysis and
creates an instance of the related class.

In order to facilitate the implementation of new analyses, \sampleanalyzer\
comes with a \python\ script \texttt{newAnalysis.py} located in the directory
\texttt{SampleAnalyzer}. As shown above, creating a new analysis with the name
\texttt{newname} requires the implementation of the two \cpp\ files 
\texttt{newname.h} and \texttt{newname.cpp} and then update the file
\texttt{analysisList.cpp} in order to include the new analysis. 
The analysis-independent part of this task has been
automated through this \python\ script. The user can use it from a shell, by
typing 
\begin{verbatim}
  newAnalysis.py newname
\end{verbatim}
The script starts by asking the user to type-in the label of the new
analysis.
As a result, the two files containing the declaration of the new analysis
class are created (with a blank analysis included) and the list of the existing
analyses in \texttt{analysisList.cpp} is updated. The user must now 
start implementing the analysis itself. 

To this aim, he has to declare the
variables necessary for the analysis and implement the three methods
\texttt{Initialize}, \texttt{Execute} and \texttt{Finalize}, together with
possible additional user-defined functions. This step requires a knowledge of
the methods already included in the base class \texttt{AnalysisBase} and
the one of the data format used internally by \sampleanalyzer. This is the scope of
Section \ref{sec:em_dataformat}, Section \ref{sec:em_sampleformat}  and Section
\ref{sec:em_tools}.

\subsection{The data format used by \sampleanalyzer} \label{sec:em_dataformat}
The function \texttt{Execute} introduced in the previous Section
takes as a second argument an object of the type \texttt{EventFormat}, which
points to the current event being analyzed denoted by \texttt{event} in the
following. We dedicate this section to the description of the \texttt{EventFormat}
class, which can also be found as a \doxy\ documentation on the \madanalysis\ 5
website,

\maweb
\medskip

In order to implement his analysis and to apply it to the event under
consideration,
the user generally needs to access various pieces of information related to this
event.  
For example, the momentum of a specific type of object could be needed to
implement some sophisticated cuts. To this aim, the \sampleanalyzer\
framework contains many built-in methods. They are split into two categories,
the first one being related to hadron-level and parton-level events, which are
generically named, in the following, as Monte Carlo level events, and the second
one to reconstructed events.

When processing the event \texttt{event}, \sampleanalyzer\ automatically
creates, when reading the event, an object
with a structure appropriate to store the information included in the event 
under consideration,
\begin{verbatim}
  event.mc()            event.rec()
\end{verbatim} 
These two objects are \cpp\ pointers to all the methods implemented 
to facilitate the design of an analysis at the Monte Carlo level 
and at the reconstructed-level, respectively. These methods are extensively
described in the rest of this Section.  

\subsubsection{The data format for parton-level or hadron-level
events}\label{sec:mcformat}

The pointer \texttt{event.mc()} introduced above allows us to access general
information related to a (Monte Carlo) event denoted by \texttt{event} 
such as the weight of the
event or the employed value of the strong coupling constant. In addition,
the whole set of initial-, intermediate- and final-state particles, together 
with their kinematical properties, is available. These properties form
the so-called data format of \sampleanalyzer\ for Monte Carlo level events and
are summarized in Table \ref{tab:mcformat}.

\begin{table}
\bgfbnn
\multicolumn{2}{c}{\textbf{Table \ref{tab:mcformat}: Methods
related to the parton-level}}\\
\multicolumn{2}{c}{\textbf{and hadron-level event format}}\\$~$\\
\multicolumn{2}{l}{Let \texttt{ev} be an \texttt{EventFormat} object, \ie, an
instance of the class related}\\
\multicolumn{2}{l}{to events issued from a partonic or hadronic Monte Carlo sample.}\\$~$\\
{\tt ev.mc()->alphaQCD()} & This returns, as a floating-point number, the value of the
strong coupling constant used in the event.\\
{\tt ev.mc()->alphaQED()} & This returns, as a floating-point number, the value of the
electromagnetic coupling constant used in the event.\\
{\tt ev.mc()->particles()} & This returns a vector of \texttt{MCParticleFormat}
objects whose each entry consists in one of the particles present in the
event, together with its properties. All initial, intermediate and final
state particles are included.\\
{\tt ev.mc()->processId()} & This returns an unsigned integer number related to
the tag of the physical process the event is originating from. It is especially
useful when several physical processes are merged into one event sample.\\
{\tt ev.mc()->scale()} & This returns, as a floating-point number, the factorization
   scale.\\
{\tt ev.mc()->weight()} & This returns, as a floating-point number, the event weight.\\
\egfbnn
\textcolor{white}{\caption{\label{tab:mcformat}}}
\end{table}

For Monte Carlo samples which include events related to several physical
processes, matrix-element generators usually assign to each event a tag, \ie, an
unsigned integer number, that allows us to identify the physical process which the
event is originating from. If the user needs to access this tag in the analysis,
it is available through the function 
\begin{verbatim}
  event.mc()->processId()
\end{verbatim} 
which returns an unsigned integer. With a similar syntax, the weight associated 
to the event \texttt{event} can be used in the \cpp\ source file of the analysis
through,
\begin{verbatim}
  event.mc()->weight()
\end{verbatim} 
which returns a floating-point number. If the implementation of the analysis 
requires the evaluation of the factorization scale, it can be obtained
from the method 
\begin{verbatim}
  event.mc()->scale()
\end{verbatim} 
which gives the results as a floating-point number. In contrast, the
renormalization scale is not available but the values of the strong and
electromagnetic coupling constants (including a possible
running) can also be employed in the analysis, 
\begin{verbatim}
  event.mc()->alphaQCD()
  event.mc()->alphaQED()
\end{verbatim} 
These last two methods also return floating-point numbers. 

More importantly, implementing histograms or selection cuts requires us
to investigate the particle content of the event, as well as the properties of
one or several of these particles. All the initial-, intermediate- and final-state
particles included in the event \texttt{event} are stored in a vector of
\texttt{MCParticleFormat} objects, which can be called in the analysis 
through
\begin{verbatim}
  event.mc()->particles()
\end{verbatim} 
The syntax above returns a vector that each entry consists in a particle
present in the event, together with its properties, given as an instance of the
\texttt{MCParticleFormat} class.
 
In order to implement a loop
over all the particle content of the event \texttt{event}, in, \eg, the
function \texttt{Execute} of the analysis source file, it is sufficient to
program
\begin{verbatim}
  unsigned int n = event.mc()->particles().size();
  for (unsigned int i=0; i<n; i++)  { ... }
\end{verbatim}
where we recall that at this stage, all the initial-, intermediate- and
final-state particles are considered equivalently. We will show in Section
\ref{sec:em_tools} how to implement loops over, \eg, the final-state
particles only. Similarly, denoting by the object
\texttt{prt} an instance of the \texttt{MCParticleFormat} class, the
\texttt{i}$^{\text{th}}$ particle is given by
\begin{verbatim}
  MCParticleFormat* prt =  &event.mc()->particles()[i];
\end{verbatim}
In the rest of this Section, we focus on the attributes of the object
\texttt{prt}.

\begin{table}
\bgfbalign
\multicolumn{2}{c}{\textbf{Table \ref{tab:partformat}: Methods
related to the \texttt{MCParticleFormat} class}}\\$~$\\
\multicolumn{2}{l}{Let \texttt{prt} be a \texttt{MCParticleFormat} object.}\\$~$\\
{\tt prt.ctau()} & This returns, as a floating-point number, the decay length of
the particle, assuming that it moves at the speed of light.\\
{\tt prt.momentum()} & This returns a \texttt{TLorentzVector} containing the
four-momentum of the particle \texttt{prt}.\\
{\tt prt.mother1()} & This returns a \texttt{MCParticleFormat} object pointing
to the mother particle of the particle \texttt{prt}, \ie, either the particle which
decays into \texttt{prt} (plus other particles) or a particle which interacts
with another particle (see \texttt{mother2()}) to produce the particle
\texttt{prt}.\\
{\tt prt.mother2()} & This returns a \texttt{MCParticleFormat} object pointing
to the mother particle of the particle \texttt{prt}, but only in the case 
\texttt{prt} is produced from the interaction of two particles. These particles
can be accessed through the two methods \texttt{mother1()} and
\texttt{mother2()}.\\
{\tt prt.pdgid()} & This returns an integer number standing for the PDG-id of the
particle.\\
{\tt prt.spin()} & This returns a floating-point number, the cosine of the angle
between the momentum of the particle \texttt{prt} and its spin vector, computed
in the laboratory reference frame.\\
{\tt prt.statuscode()} & This returns an integer depending on the initial-state,
intermediate-state or final-state nature of the particle. The conventions
on this integer number are taken from Ref.\ \cite{Boos:2001cv}.
\egfbalign
\textcolor{white}{\caption{\label{tab:partformat}}}
\end{table}

The class \texttt{MCParticleFormat} contains seven methods which allow
to extract and use the properties of a given particle within the analysis. These
methods are summarized in Table
\ref{tab:partformat}.

The PDG-id of the particle \texttt{prt} can be accessed through
\begin{verbatim}
  prt->pdgid()
\end{verbatim}
which returns the corresponding signed integer number. The
\texttt{statuscode} of the particle, \ie, its tag as an initial-state,
intermediate-state or final-state particle, can be obtained through the function
\begin{verbatim}
  prt->statuscode()
\end{verbatim}
which returns an integer associated to the initial-state, intermediate-state or
final-state nature of the particle. For the conventions on this integer number
we refer to Ref.\ \cite{Boos:2001cv}. 

For non-initial-state particles, information on 
mother particle(s) can be extracted (and used in the analysis) through the two 
methods
\begin{verbatim}
  prt->mother1()
  prt->mother2()
\end{verbatim}
which return the \texttt{MCParticleFormat} objects related to the particle(s)
from which the current particle \texttt{prt} is issued. In the case of a decay
chain, only the \texttt{mother1} method has to be used. In contrast, when the 
particle \texttt{prt}
is produced from the interaction of two particles, both the
\texttt{mother1} and \texttt{mother2} give results, each of them
pointing to one of the initial particles.

The most important property of the \texttt{MCParticleFormat} class to be used 
in physics
analyses consists in the particle four-momentum, accessible from 
\begin{verbatim}
  prt->momentum()
\end{verbatim}
The result of this function is given as a \texttt{TLorentzVector}, a
\rooot\ class appropriate to store four-momentum. In addition, this class 
contains various 
methods to compute a large set of observables which can be derived from the
knowledge of the four-momentum, such as the energy or the transverse momentum.
The complete list of these observables is given, together with the associated
syntax, in
Table \ref{tab:tlorentz}. 

An additional observable related to the
four-momentum, and more in particular to the three-momentum component of the
four-momentum, which can be employed in an analysis reads
\begin{verbatim}
  prt->spin()
\end{verbatim}
This returns a floating-point number which stands for the cosine of the angle
between the
momentum of the particle \texttt{prt} and its spin vector, evaluated in the
laboratory reference frame.

Finally, the decay length of
the particle (assuming that the particle is moving at the speed of light) can
also be easily obtained through the function 
\begin{verbatim}
  prt->ctau()
\end{verbatim}
which returns a floating-point number too.

\begin{table}
\bgfbyy
\multicolumn{2}{c}{\textbf{Table \ref{tab:tlorentz}: Common methods
related to the \texttt{MCParticleFormat}}}\\
\multicolumn{2}{c}{\textbf{and \texttt{RecParticleFormat} 
classes}}\\$~$\\
\multicolumn{2}{l}{Let \texttt{P} be a \texttt{MCParticleFormat} or
\texttt{RecParticleFormat} object.}\\$~$\\
{\tt P.angle(P2)} & Angle between the momenta of the objects \texttt{P} and
     \texttt{P2}, where \texttt{P2} is an instance of
     the \texttt{MCParticleFormat} or \texttt{RecParticleFormat} class. \\ 
{\tt P.beta()}  & Velocity $\beta = v/c$.\\
{\tt P.dr(P2)} & Relative distance between the objects \texttt{P} and
     \texttt{P2} in the $\eta-\phi$ plane, where \texttt{P2} is an instance of
     the \texttt{MCParticleFormat} or \texttt{RecParticleFormat} class. \\ 
{\tt P.e()}     & Energy.\\
{\tt P.et()}    & Transverse energy.\\
{\tt P.eta()}   & Pseudorapidity.\\
{\tt P.gamma()} & Lorentz factor.\\
{\tt P.m()}     & Invariant mass.\\ 
{\tt P.mt()}    & Transverse mass.\\
{\tt P.p()}     & Norm of the momentum.\\ 
{\tt P.phi()}   & Azimuthal angle of the momentum.\\
{\tt P.pt()}    & Norm of the transverse momentum.\\
{\tt P.px()}    & Projection of the momentum on the $x$-axis.\\
{\tt P.py()}    & Projection of the momentum on the $y$-axis.\\
{\tt P.pz()}    & Projection of the momentum on the $z$-axis.\\
{\tt P.r()}   & Position of the object in the $\eta-\phi$ plane.\\ 
{\tt P.theta()} & Angle between the momentum and the beam axis.\\
{\tt P.y()}  & Rapidity.\\
\egfbyy
\textcolor{white}{\caption{\label{tab:tlorentz}}}
\end{table}

\subsubsection{The data format for reconstructed events}
At the beginning of this Section, we have introduced two types of data format
which are used for event processing by \sampleanalyzer. They consist in the 
two sides of the more general \texttt{EventFormat} class. In
the previous Section, we have focused on the description of 
an event object \texttt{event}
read from a parton-level or hadron-level sample. We have shown that its  
properties are embedded, in the \sampleanalyzer\ framework, into a structure
which can be browsed from the \cpp\ pointer \texttt{event.mc()}.

\begin{table}
\bgfbnn
\multicolumn{2}{c}{\textbf{Table \ref{tab:recformat}: Methods
related to the reconstructed event format}}\\$~$\\
\multicolumn{2}{l}{Let \texttt{ev} be an \texttt{EventFormat} object, \ie, an
instance of the class related}\\
\multicolumn{2}{l}{to events issued from a reconstructed Monte Carlo sample.}\\$~$\\
{\tt ev.rec()->electrons()} & This returns a vector with all the 
electrons of the event, encoded as \texttt{RecLeptonFormat} objects.\\
{\tt ev.rec()->jets()} & This returns a vector with all the 
jets of the event, given as \texttt{RecJetFormat} objects.\\
{\tt ev.rec()->met()} & This points to the missing energy of the
event, encoded as a \texttt{RecMETFormat} object.\\
{\tt ev.rec()->muons()} & This returns a vector with all the
muons of the event, encoded as \texttt{RecLeptonFormat} objects.\\
{\tt ev.rec()->taus()} & This returns a vector with all the 
taus of the event, given as \texttt{RecTauFormat} objects.\\
\egfbnn
\textcolor{white}{\caption{\label{tab:recformat}}}
\end{table}

Since the properties of
reconstructed events are in general very
different from those of Monte Carlo events, \sampleanalyzer\ embeds the latter into
another structure which is, this time, 
linked to the \cpp\ pointer \texttt{event.rec()}. It consists in
five methods, collected in Table \ref{tab:recformat}, associated to the five
types of objects that can be reconstructed,
\ie, electrons, muons, taus, jets and missing energy\footnote{By electrons,
muons and taus, we are considering both the corresponding particles and
antiparticles.}. 
In the \sampleanalyzer\
framework, each reconstructed objects is associated to a specific class
derived from the generic \texttt{RecParticleFormat} mother class and depending
on its species. In the
following, we adopt the choice of describing the daughter classes, more relevant
for the user, rather than the generic class.

Two methods are associated to first and second generation charged leptons, \ie,
to the reconstructed electrons and muons, present in the event \texttt{event},
\begin{verbatim}
  event.rec()->electrons()
  event.rec()->muons()
\end{verbatim}
They return vectors that the entries are the different reconstructed electrons and
muons of the event, respectively. Each reconstructed electron or muon is encoded
as a
\texttt{RecLeptonFormat} object. This last class has six associated methods,
gathered in Table \ref{tab:reclept}, related to the attributes of the
reconstructed leptonic electron and muon objects. 

As for Monte Carlo event samples, the four-momentum of the reconstructed objects 
is one of the most incontrovertible variables to be employed in analyses. For 
a reconstructed lepton \texttt{lep}, its four-momentum can be included and used 
in the analysis source file by calling the function
\begin{verbatim}
  lep.momentum()
\end{verbatim}
the syntax being similar to the one employed for particles included in 
parton-level and hadron-level events.
This method returns a \texttt{TLorentzVector} object containing the
corresponding four-momentum. Therefore, the methods associated to all the 
observables which can be derived from the knowledge of the four-momentum,
collected in Table \ref{tab:tlorentz}, are also methods of the
\texttt{RecLeptonFormat} class.

\begin{table}
\bgfbnn
\multicolumn{2}{c}{\textbf{Table \ref{tab:reclept}: Methods
related to the \texttt{RecLeptonFormat} class}}\\$~$\\
\multicolumn{2}{l}{Let \texttt{lep} be an \texttt{RecLeptonFormat} object, \ie,
an instance of the class}\\
\multicolumn{2}{l}{describing the reconstructed electrons and muons.}\\$~$\\
{\tt lep.charge()} & This returns the electric charge of the reconstructed
object \texttt{lep} as a floating-point number. The returned value consists in 
$-1$ or $+1$.\\
{\tt lep.EEoverHE()} & This returns the ratio of the electromagnetic and hadronic 
energy for the reconstructed object \texttt{lep}, given as a 
floating-point number.\\
{\tt lep.ET\textunderscore PT\textunderscore isol()} & This returns the ratio of
the values of the functions 
{\tt sumET\textunderscore isol()} and {\tt sumPT\textunderscore isol()}.\\
{\tt lep.HEoverEE()} & This returns the ratio of the hadronic and
electromagnetic energy for the reconstructed object \texttt{lep}, given as a 
floating-point number.\\
{\tt lep.momentum()} & This returns a \texttt{TLorentzVector} containing the
four-momentum of the reconstructed object \texttt{lep}.\\
{\tt lep.sumET\textunderscore isol()} & This returns, if the object \texttt{lep}
is a muon, the sum of the transverse energy of all the tracks lying in a cone
around the muon. The size of the cone is fixed by the detector simulation
tool. If \texttt{lep} is an electron, this function returns zero.\\
{\tt lep.sumPT\textunderscore isol()} & This returns, if the object \texttt{lep}
is a muon, the sum of the transverse momentum of all the tracks lying in a cone
around the muon. The size of the cone is fixed by the detector simulation
tool. If \texttt{lep} is an electron, this function returns zero.\\$~$\\
\multicolumn{2}{l}{All the methods presented in Table \ref{tab:tlorentz} can
also be used.}\\
\egfbnn
\textcolor{white}{\caption{\label{tab:reclept}}}
\end{table}

Reconstructed leptons are considered as electrons or muons
regardless of their charge. In the case the user wants to select them
according to the electric charge, the method
\begin{verbatim}
  lep.charge()
\end{verbatim}
of the \texttt{RecLeptonFormat} class can be used. It returns, as a
floating-point number, the electric charge of the reconstructed lepton which
can be equal to either $-1$ or $+1$
according to its antiparticle or particle nature.

Two methods allow us to get information on the splitting of the reconstructed
electron or muon energy between the electromagnetic and the
ha\-dro\-nic parts of the detector,
\begin{verbatim}
  lep.EEoverHE()
  lep.HEoverEE()
\end{verbatim}
These methods compute the ratio between the energy deposited in the
electromagnetic
ca\-lo\-ri\-me\-ter and the one deposited in the hadronic calorimeter, and
\textit{vice-versa}, and return them as floating-point numbers. 

Finally, the \texttt{RecLeptonFormat} class contains three specific
methods related to muon isolation (see also Section \ref{sec:em_tools}). The
algorithms to be employed for deciding if a muon is
considered as isolated or not require in general the evaluation of two
quantities, the sum of the transverse momentum of all tracks lying in a cone
around the muon candidate and the sum of their transverse energy. These two
observables can be accessed by typing
\begin{verbatim}
  lep.sumPT_isol()
  lep.sumET_isol()
\end{verbatim}
which return a zero value in the case the lepton \texttt{lep} is an electron.
In contrast, for muons, the value read from the event file is employed.
The size of the cone is fixed by the fast detector simulation tool and is not
available in the \lhco\ event format. The ratio of the values of these two
functions can be obtained via the function
\begin{verbatim}
  lep.ET_PT_isol()
\end{verbatim}

Tau leptons being unstable, they always decay either into a narrow jet, into a muon
or into an electron, each time in association with missing energy. Therefore, a
specific class, different from the \texttt{RecLeptonFormat} class, exists in
order to embed reconstructed taus. This class is denoted by \texttt{RecTauFormat}.
In the internal data format used by \sampleanalyzer, the pointer to the
reconstructed event, \texttt{event.rec()}, contains a specific method to access
all the taus present in the event
\begin{verbatim}
  event.rec()->taus()
\end{verbatim}
This returns, as a vector of \texttt{RecTauFormat} objects, all the
reconstructed tau leptons of the event (regardless of their electric charge).

\begin{table}[t]
\bgfbnn
\multicolumn{2}{c}{\textbf{Table \ref{tab:rectau}: Methods
related to the \texttt{RecTauFormat} class}}\\$~$\\
\multicolumn{2}{l}{Let \texttt{tau} be an instance of the \texttt{RecTauFormat}
class, \ie, the class}\\
\multicolumn{2}{l}{describing the reconstructed taus.}\\$~$\\
{\tt tau.charge()} & This returns the electric charge of the 
object \texttt{tau} as a floating-point number. The returned value consists in
$-1$ or $+1$.\\
{\tt tau.EEoverHE()} & This returns the ratio of the electromagnetic and hadronic 
energy for the object \texttt{tau}, given as a 
floating-point number.\\
{\tt tau.HEoverEE()} & This returns the ratio of the hadronic and
electromagnetic energy for the object \texttt{tau}, given as a 
floating-point number.\\
{\tt tau.momentum()} & This returns a \texttt{TLorentzVector} containing the
four-momentum of the object \texttt{tau}.\\
{\tt tau.ntracks()} & This returns, as a short integer, the number of charged
tracks contained in the object \texttt{tau}.\\$~$\\
\multicolumn{2}{l}{All the methods presented in Table \ref{tab:tlorentz} can
also be used.}\\
\egfbnn
\textcolor{white}{\caption{\label{tab:rectau}}}
\end{table}

In addition to the four methods \texttt{momentum()}, \texttt{charge()},
\texttt{EEoverHE()}, \texttt{HEoverEE()} and the methods given in Table
\ref{tab:tlorentz} already present for charged leptons of the first and second
generations, the \texttt{RecTauFormat} class includes an additional method
extracting the number of charged tracks issued from the decaying tau and included
in the reconstructed object. Denoting by \texttt{tau} the reconstructed object,
this number, given as a short integer, can be obtained and further used in the
implementation of the analysis by 
\begin{verbatim}
  tau.ntracks()
\end{verbatim}
The entire list of methods associated to the \texttt{RecTauFormat} class can be
found in Table \ref{tab:rectau}.

\begin{table}[!t]
\bgfbnn
\multicolumn{2}{c}{\textbf{Table \ref{tab:recjet}: Methods
related to the \texttt{RecJetFormat} class}}\\$~$\\
\multicolumn{2}{l}{Let \texttt{j} be an \texttt{RecJetFormat} object, \ie, an
instance of the class}\\
\multicolumn{2}{l}{describing the reconstructed jets.}\\$~$\\
{\tt j.btag()} & This returns \texttt{true} or \texttt{false} according
to the $b$-tagging (or not) of the object \texttt{j}.\\
{\tt j.EEoverHE()} & This returns the ratio of the electromagnetic and hadronic 
energy for the object \texttt{j}, given as a 
floating-point number.\\
{\tt j.HEoverEE()} & This returns the ratio of the hadronic and
electromagnetic energy for the object \texttt{j}, given as a 
floating-point number.\\
{\tt j.momentum()} & This returns a \texttt{TLorentzVector} containing the
four-momentum of the object \texttt{j}.\\
{\tt j.ntracks()} & This returns, as a short integer, the number of charged
tracks contained in the object \texttt{j}.\\$~$\\
\multicolumn{2}{l}{All the methods presented in Table \ref{tab:tlorentz} can
also be used.}\\
\egfbnn
\textcolor{white}{\caption{\label{tab:recjet}}}
\end{table}

As for electrons, muons and taus, the entire set of reconstructed jets included
in the event \texttt{event} can be obtained through the intuitive method 
of the \texttt{event.rec()} structure, 
\begin{verbatim}
  event.rec()->jets()
\end{verbatim}
This returns a vector of instances of the \texttt{RecJetFormat} class whose 
properties are collected in Table \ref{tab:recjet}.

Reconstructed jets are characterized by their momentum, the number of charged
tracks induced by the parton showering, fragmentation and ha\-dro\-ni\-za\-tion of the
initial partons and the ratio of the hadronic and
electromagnetic parts of the jet energy. These properties are related to the
methods \texttt{momentum()}, \texttt{ntracks()}, \texttt{EEoverHE()}
and \texttt{HEoverEE()} as well as all those included in Table
\ref{tab:tlorentz}, of the \texttt{RecJetFormat} class. In addition,
an extra method returns \texttt{true} or \texttt{false} according to the fact
that the jet has been tagged as a $b$-jet or not, 
\begin{verbatim}
  j.btag()
\end{verbatim}
where \texttt{j} denotes an instance of the \texttt{RecJetFormat} class. 

The last type of objects which are included in reconstructed events consists
in the associated missing transverse energy. The structure
\texttt{event.rec()} comes with the related method 
\begin{verbatim}
  event.rec()->met()
\end{verbatim}
which returns a two-dimensional vector implemented as a \texttt{TVector2} 
object. It contains then the direction and magnitude of the missing energy in the
transverse plane as shown in Table \ref{tab:recmet}.

\begin{table}
\bgfbnn
\multicolumn{2}{c}{\textbf{Table \ref{tab:recmet}: Methods
related to the \texttt{RecMetFormat} class}}\\$~$\\
\multicolumn{2}{l}{Let \texttt{miss} be a \texttt{RecMetFormat} object, \ie, the
variable containing the}\\
\multicolumn{2}{l}{reconstructed missing energy associated to an event.}\\$~$\\
{\tt miss.mag()} & This returns the magnitude of the missing energy represented
by the object \texttt{miss} as a floating-point number.\\
{\tt miss.phi()} & This returns the azimuthal angle of the missing energy
represented by the object \texttt{miss} as a floating-point number.\\
{\tt miss.x()} & This returns the $x$-component of the missing energy
represented by the object \texttt{miss} as a floating-point number.\\
{\tt miss.y()} & This returns the $y$-component of the missing energy
represented by the object \texttt{miss} as a floating-point number.
\egfbnn
\textcolor{white}{\caption{\label{tab:recmet}}}
\end{table}

The $x$-component and $y$-component of the missing transverse energy can be
used when implementing an analysis through the two methods 
\begin{verbatim}
  miss.x()
  miss.y()
\end{verbatim}
which return floating-point numbers associated to the two components of the
two-dimensional vector describing the missing energy, the object \texttt{miss}
being an instance of the \texttt{RecMETFormat} class. In the case the user
prefers to use polar coordinates instead of Cartesian coordinates when
implementing his analysis, he can use the two methods 
\begin{verbatim}
  miss.mag()
  miss.phi()
\end{verbatim}
which return the magnitude and the azimuthal angle of the two-dimensional
transverse-momentum vector as floating-point numbers.

\subsection{The sample format used by \sampleanalyzer} \label{sec:em_sampleformat}
In Section \ref{sec:em_template}, we have introduced the function
\texttt{Execute} as the main function of the analysis class. We have shown that it 
requires two arguments, the event being currently analyzed, stored as an
\texttt{EventFormat} object, and general information about the sample which 
this event belongs to, passed as a \texttt{SampleFormat} object. This format
is also the one to be used for the arguments of the function \texttt{Finalize},
dedicated to the creation of the output files containing the results of the
analysis. In this case, the user must pass two arguments, an instance of the
\texttt{SampleFormat} class associated to all the event samples included in the
dataset under consideration, \ie, information averaged over all the samples, and
one vector of \texttt{SampleFormat} objects with one entry for each of the
samples included in the dataset, \ie, the same information, but given for each
individual sample.

The \texttt{SampleFormat} class contains several 
methods allowing us to access general information about the sample under
consideration, provided the information is available. If not, the associated
entries in the \texttt{SampleFormat} structure 
are left non-initialized and the corresponding quantity cannot consequently be
used in an analysis. A full \doxy\ documentation is available on the 
\madanalysis\ 5 website,

\maweb
\medskip

Since information supplementing the events is only available within \lhe\ and
\stdhep\ files, the \texttt{SampleFormat} structure is then only relevant
for the analysis of such event files. In particular, \lhco\ files with
reconstructed events do not include anything but the events. 
In this case, it is however still possible to pass additional information
such as cross sections
to \sampleanalyzer\ through the attributes of the \texttt{dataset} class (see
Section \ref{sec:uf_datasets}).

In Section \ref{sec:em_dataformat}, we have shown that \sampleanalyzer\ is
creating an \texttt{EventFormat} structure each time an event is processed,
resulting in a \cpp\ pointer pointing to the whole methods available for the
\texttt{EventFormat} structure. Similarly, when processing a new sample
generically denoted by \texttt{sample},  \sampleanalyzer\ creates a \cpp\
pointer
\begin{verbatim}
  sample.mc()  
\end{verbatim} 
which points to a structure containing all the methods allowing us to access the
global information of the event file. These methods are collected in Table
\ref{tab:sampleformatbeam} and Table \ref{tab:sampleformat}. 

As shown by its name, \texttt{sample.mc()} is related to parton-level or
hadron-level events. The counterpart of this object in the case of a sample
containing
reconstructed events, \texttt{sample.rec()}, has been implemented within the
\sampleanalyzer\ framework. However, the only event file format appropriate for
reconstructed events, the \lhco\ format, does
not leave a possibility for including additional information to the events.
Therefore, the pointer \texttt{sample.rec()} points to a set of null
information. If in the future, a new format for reconstructed
events is designed, with the room for global information about the event sample,
the related structure in the \sampleanalyzer\ framework will be ready for the
new format.

\begin{table}
\bgfbyyy
\multicolumn{2}{c}{\textbf{Table \ref{tab:sampleformatbeam}: Methods
related to the \texttt{SampleFormat} class}}\\
\multicolumn{2}{c}{\textbf{giving information on the colliding beams}}\\$~$\\
\multicolumn{2}{l}{Let \texttt{sample} be a \texttt{SampleFormat} object.}\\$~$\\
{\tt sample.mc()->beamE().first} & \\
& This returns, as a floating-point number, the
energy of the first of the colliding beams.\\
{\tt sample.mc()->beamPDFauthor().first} & \\
& This returns the author group of the parton density set used for the first
of the colliding beams, as an unsigned integer number.\\
{\tt sample.mc()->beamPDFID().first} & \\
& This returns the identifier, as an unsigned integer number, of the parton
density set used for the
first of the colliding beams, within a given author group of parton densities.\\
{\tt sample.mc()->beamPDGID().first} & \\
& This returns the PDG-id of the first of the colliding beams, as an integer
number.\\
{\tt sample.mc()->beamE().second} & \\
& This returns, as a floating-point number, the
energy of the second of the colliding beams.\\
{\tt sample.mc()->beamPDFauthor().second} & \\
& This returns the author group of the parton density set used for the second
of the colliding beams, as an unsigned integer number.\\
{\tt sample.mc()->beamPDFID().second} & \\
& This returns the identifier, as an unsigned integer number, of the parton
density set used for the second of the colliding beams, within a given author
group of parton densities.\\
{\tt sample.mc()->beamPDGID().second} & \\
& This returns the PDG-id of the second
of the colliding beams, as an integer number.\\
\egfbyyy
\textcolor{white}{\caption{\label{tab:sampleformatbeam}}}
\end{table}

The first series of methods available within the \texttt{SampleFormat}
structure, collected in Table \ref{tab:sampleformatbeam}, are related to the
description of the initial colliding beams. 
The two members of the \texttt{SampleFormat} class
\begin{verbatim}
  sample.mc()-> beamPDGID().first
  sample.mc()-> beamPDGID().second
\end{verbatim} 
return, as integer numbers, the PDG-id of the first and second beams,
respectively, whilst the four class members
\begin{verbatim}
  sample.mc()-> beamPDFauthor().first
  sample.mc()-> beamPDFID().first
  sample.mc()-> beamPDFauthor().second
  sample.mc()-> beamPDFID().second
\end{verbatim} 
return, as four unsigned integer numbers, information with respect to the set of
parton density functions used for the beams. These identifying numbers are
exported from the event samples (if available) and correspond to the numbering
scheme employed by matrix element generators to identify the set of parton
densities employed. According to the Les Houches conventions, this scheme is
based on the \pdflib\ \cite{PlothowBesch:1992qj} and \lhapdf\
\cite{Giele:2002hx} packages. Hence, the method
\texttt{beamPDFauthor()} is related to the author group which has released the 
parton density relevant for the sample under consideration and the method
\texttt{beamPDFID()} indicates which specific set of parton densities of the
corresponding group has been used. 

The last pieces of information available with respect to the beams which can be
stored in the event files, and thus exported to the \sampleanalyzer\ framework, 
consist in their energy. This can be used in the implementation of
an analysis by means of the two methods
\begin{verbatim}
  sample.mc()-> beamE().first
  sample.mc()-> beamE().second
\end{verbatim}   
for the first and second beams, respectively.

\begin{table}
\bgfbyyy
\multicolumn{2}{c}{\textbf{Table \ref{tab:sampleformat}: Other methods
related to the \texttt{SampleFormat} class}}\\$~$\\
\multicolumn{2}{l}{Let \texttt{sample} be a \texttt{SampleFormat} object.}\\$~$\\
{\tt sample.mc()->weightingmode()} & \\
& This returns, as an integer number,
information on the type of events present in the sample (\eg, weighted versus
unweighted events).\\
{\tt sample.mc()->processes()} & \\
& This returns a vector where the entries are \texttt{ProcessFormat} objects
containing basic information about each of the physical processes included in
the sample \texttt{sample}.\\
{\tt sample.mc()->xsection()} & \\
& This returns the cross section associated to the event sample \texttt{sample},
as a floating-point number.\\
{\tt sample.mc()->xsection\textunderscore error()} & \\
& This returns the Monte Carlo uncertainty related to the cross section
associated to the event sample \texttt{sample}, as a floating-point number.\\
\egfbyyy
\textcolor{white}{\caption{\label{tab:sampleformat}}}
\end{table}

The second series of methods available from the \texttt{SampleFormat} class are
related to the sample itself. When available,
information about the weighting of the events is passed to \sampleanalyzer\ and
can be accessed through 
\begin{verbatim}
  sample.mc()-> weightingmode()
\end{verbatim}
According to the \lhe\ format \cite{Boos:2001cv,Alwall:2006yp}, this type of
information is stored as an integer number which indicates how event
weights and cross sections have to be interpreted, and the
\texttt{weightingmode()} method returns this integer number. Even if 
presently, \sampleanalyzer\ does not fully support weighted events, the
architecture has already been implemented with respect to future
developments of the program. 
Finally, the most important quantities included in this
second series of methods concern the cross section associated to the sample
\texttt{sample}, together with the corresponding uncertainty. Both can be
called at the level of the analysis by using the methods 
\begin{verbatim}
  sample.mc()->xsection() 
  sample.mc()->xsection_error() 
\end{verbatim}   
which return floating-point numbers.

As mentioned in Section \ref{sec:mcformat}, several physical processes can be
included within the same event sample. In this case, each of the processes is
associated with an identifying tag (see the \texttt{processId} method introduced
in Section \ref{sec:mcformat}). The \texttt{SampleFormat} structure allows us to
easily access detailed information about the processes individually. This
information is stored 
into a new structure, the \texttt{ProcessFormat} class, all of whose 
associated methods are summarized in Table \ref{tab:procformat}. 
The method 
\begin{verbatim}
  sample.mc()->processes() 
\end{verbatim}   
returns a vector that each entry is a \texttt{ProcessFormat} object. For
an instance of this class denoted by \texttt{process}, information on the
identifier of the process can be obtained through
\begin{verbatim}
  process.processId()
\end{verbatim}   
the corresponding cross section together with its associated uncertainty through
the methods
\begin{verbatim}
  process.xsection() 
  process.xsection_error() 
\end{verbatim}   
and finally, the maximum weight carried by a single event originating from the
subprocess \texttt{process} through 
\begin{verbatim}
  process.maxweight()
\end{verbatim}

\begin{table}[t]
\bgfbyyy
\multicolumn{2}{c}{\textbf{Table \ref{tab:procformat}: Methods
related to the \texttt{ProcessFormat} class}}\\$~$\\
\multicolumn{2}{l}{Let \texttt{proc} be an instance of the \texttt{ProcessFormat}
class.}\\$~$\\
{\tt proc.processId()} & \\
& This returns an unsigned integer number related to
the tag of the physical process the event is originating from.\\
{\tt proc.xsection()} & \\
& This returns the cross section associated to the process \texttt{proc},
as a floating-point number.\\
{\tt proc.xsection\textunderscore error()} & \\
& This returns the Monte Carlo uncertainty related to the cross section
associated to the process \texttt{proc}, as a floating-point number.\\
{\tt proc.maxweight()} & \\
& This returns the maximum weight carried by an event associated to the
process \texttt{proc}, as a floating-point number.\\
\egfbyyy
\textcolor{white}{\caption{\label{tab:procformat}}}
\end{table}

\subsection{Framework services} \label{sec:em_tools}
In this Section, we describe the various services included in the
\sampleanalyzer\ framework. These services are components of the
program that are initialized (internally or by the user) at the beginning 
of the execution of \sampleanalyzer\ (within the \texttt{Initialize} method) and
can then be further called within the analysis as many times
as necessary. Two series of services are currently available, message services
and physics services and are described in the rest of this Section.
We recall that full \doxy\ documentation is available on the \madanalysis\ 5
website,

\maweb

\subsubsection{Message services}
This class of functions has been implemented and can be used by the users in
order to print text to the screen during an on-going
analysis in a rather sophisticated fashion.

As for the implementation of any \cpp\ program, the user can, when designing his
analysis, use the standard \cpp\ streamers
\texttt{std::cout} and \texttt{std:cerr} in order to implement messages to be
printed to the screen. However, these methods only include two levels of
messages, \ie,
normal
and error messages. It is hence rather cumbersome to handle multi-level
messages and give information both on the reason leading to the printing of the
message and on its location in the analysis code in a clear and useful way.

Therefore, \sampleanalyzer\ includes its own message
services with four different levels of printing, \texttt{DEBUG}, \texttt{INFO},
\texttt{WARNING} and \texttt{ERROR}. The way to use these four modes mimics the
one of the \cpp\ commands \texttt{std::cout} and \texttt{std::cerr},
\begin{verbatim}
  DEBUG   << "Debug message." << std::endl;
  INFO    << "Information message." << std::endl;
  WARNING << "Warning message." << std::endl;
  ERROR   << "Error message." << std::endl;
\end{verbatim}

Each message level is associated to a different color. Debugging messages are
printed in yellow, information messages in white, warning messages in purple and
error messages in red. However, if the user is not interested in the color of
the messages, this message can be switched off by including, in
the source code of the analysis, the line
\begin{verbatim}
  LEVEL->DisableColor()
\end{verbatim}
The color can be enabled again through the command 
\begin{verbatim}
  LEVEL->EnableColor()
\end{verbatim}
In the command lines above, the keyword \texttt{LEVEL} stands for any of the
four levels of message, \texttt{DEBUG},
\texttt{INFO}, \texttt{WARNING} or \texttt{ERROR}.

Warning and error messages have a special role concerning
the debugging of the analysis code. In addition to the message, the 
line number of the
code having generated the message is also printed in order to facilitate
the debugging. 

Finally, let us note that messages associated to a given level
can be fully switched off, if this is needed by the user, by including in the
analysis code the command
\begin{verbatim}
  LEVEL->Mute()
\end{verbatim}
Messages can be restored by implementing
\begin{verbatim}
  LEVEL->UnMute()
\end{verbatim}
where the keyword \texttt{LEVEL} again stands for any of the four levels of
message.

\subsubsection{Physics services}

Under the name \textit{physics services}, we collect a series of methods and 
functions aiming to facilitate the writing of an analysis by the user. It
includes, among others, functions to compute global observables related to the
entire event, such as the transverse missing energy, and this for any type of
event (parton-level, hadron-level and reconstructed-level). All these functions
can be called through the \cpp\ pointer \texttt{PHYSICS}.

Before moving on to the description of these functions, one must note that in 
the case the user wants to use, within his analysis, one or several of the 
methods related to the invisible or hadronizing particles, the definition of
these invisible and hadronic particles must first be provided.
This task is performed within the 
\texttt{Initialize} function introduced in Section \ref{sec:em_template} with
the help of two different functions, one being associated to the invisible
particles and another one being associated to the particles taking part in the
hadronic activity. However, before declaring a new particle as invisible or
hadronizing, the physics services class must be initialized by means of one of 
the commands
\begin{verbatim}
  PHYSICS->mcConfig()->Reset()
  PHYSICS->recConfig()->Reset()
\end{verbatim}
In the case the event samples which are analyzed consist in parton-level and
hadron-level event files, the method \texttt{mcConfig} is used whilst for
reconstructed-level event files, \texttt{recConfig} is instead used. Then, a
particle whose PDG-id is \texttt{PDGID} can be declared as an 
invisible particle by means of 
\begin{verbatim}
  PHYSICS->mcConfig()->AddInvisibleId(PDGID)
\end{verbatim}
In the non-expert mode of \madanalysis\ 5, the definition of the multiparticle
\texttt{invisible} (see
Section \ref{sec:uf_labels}) corresponds to several calls of the 
function above behind the scenes. Taking the example of an analysis at
the parton-level in the context of the Minimal
Supersymmetric Standard Model, the invisible particles consist of the neutrinos
and the lightest superpartner assumed to be the lightest neutralino. The
corresponding declaration within the \sampleanalyzer\ framework reads 
\begin{verbatim}
  PHYSICS->mcConfig().AddInvisibleId(-16);
  PHYSICS->mcConfig().AddInvisibleId(-14);
  PHYSICS->mcConfig().AddInvisibleId(-12);
  PHYSICS->mcConfig().AddInvisibleId(12);
  PHYSICS->mcConfig().AddInvisibleId(14);
  PHYSICS->mcConfig().AddInvisibleId(16);
  PHYSICS->mcConfig().AddInvisibleId(1000022); 
\end{verbatim}

Similarly, the method 
\begin{verbatim}
  PHYSICS->mcConfig()->AddHadronicId(PDGID)
\end{verbatim}
declares the particle whose PDG-id is given by \texttt{PDGID} as a particle 
related to the hadronic activity in an event. Again, in the non-expert mode of
\madanalysis\ 5, the definition of the multiparticle \texttt{hadronic} (see
Section \ref{sec:uf_labels}) corresponds to a series of calls to this last
function. For the example of a parton-level analysis within the Standard Model,
the related declaration would be given by
\begin{verbatim}
  PHYSICS->mcConfig().AddHadronicId(-5);
  PHYSICS->mcConfig().AddHadronicId(-4);
  PHYSICS->mcConfig().AddHadronicId(-3);
  PHYSICS->mcConfig().AddHadronicId(-2);
  PHYSICS->mcConfig().AddHadronicId(-1);
  PHYSICS->mcConfig().AddHadronicId(1);
  PHYSICS->mcConfig().AddHadronicId(2);
  PHYSICS->mcConfig().AddHadronicId(3);
  PHYSICS->mcConfig().AddHadronicId(4);
  PHYSICS->mcConfig().AddHadronicId(5);
  PHYSICS->mcConfig().AddHadronicId(21);
\end{verbatim}
 
In the case of events at the reconstructed-level, the user may need to specify
the algorithm to be employed when testing the isolation of a muon. This can be
done through the two (self-excluding) methods
\begin{verbatim}
  PHYSICS->recConfig().UseDeltaRIsolation(deltaR=0.5);
  PHYSICS->recConfig().UseSumPTIsolation(sumPT, ET_PT);
\end{verbatim}
In the first case, a muon is tagged as isolated when no track lies inside a cone
of size \texttt{deltaR} around the muon. The default size of this cone is set to
0.5. In the second case, the muon is tagged as isolated when the sum of the
transverse momentum of all the tracks lying in a cone around the muon is lower
than \texttt{sumPT} and in addition, when the sum of the transverse energy of
these tracks over the sum of their transverse momentum is lower than
\texttt{ET\textunderscore PT}. By default, the first algorithm is adopted with
\texttt{deltaR} being set to 0.5.

The usage of all the methods associated to the initialization of the physics
services, \ie, to the \texttt{PHYSICS->mcConfig()} and
\texttt{PHYSICS->recConfig()} objects, is summarized in Table
\ref{tab:physics1}. All the other methods included in the physics services are
in general called
within the function \texttt{Execute}, at the core of the analysis and are
summarized in Table \ref{tab:physics2} and Table \ref{tab:physics3}.

\begin{table}
\bgfbyyyy
\multicolumn{2}{c}{\textbf{Table \ref{tab:physics1}: Methods
related to the initialization}}\\
\multicolumn{2}{c}{\textbf{of the physics services}}\\$~$\\
{\tt PHYSICS->mcConfig()->AddHadronic(PDGID)} & \\
& This adds the particle whose PDG-id is \texttt{PDGID} to the list of the
invisible particles.\\
{\tt PHYSICS->mcConfig()->AddInvisible(PDGID)} & \\
& This adds the particle whose PDG-id is \texttt{PDGID} to the list of the
particles taking part of the hadronic activity of an event.\\
{\tt PHYSICS->mcConfig()->Reset()} & \\
& This initializes physics services when analyzing parton-level or hadron-level
events.\\
{\tt PHYSICS->recConfig()->Reset()} & \\
& This initializes physics services when analyzing 
reconstructed-level events.\\
\egfbyyyy
\textcolor{white}{\caption{\label{tab:physics1}}}
\end{table}

Three boolean functions exist in order to test if a particle \texttt{prt} is
invisible or visible as well as if this particle takes part in the hadronic
activity of the event or not,
\begin{verbatim}
  PHYSICS->IsHadronic(prt)
  PHYSICS->IsInvisible(prt)
  PHYSICS->IsVisible(prt)
\end{verbatim}
When analyzing partonic or hadronic event samples, the particle \texttt{prt} is
passed as a \texttt{MCParticleFormat} object. In contrast, for
analyses at the reconstructed-level, \texttt{prt} is an instance of the
\texttt{RecParticleFormat} class\footnote{The \texttt{RecParticleFormat} class
is the
mother class of all the classes defining reconstructed objects, \ie, the
\texttt{RecLeptonFormat}, \texttt{RecJetFormat} and \texttt{RecTauFormat}
classes.}.
In order for these three methods to correctly work,
it is necessary to declare which particle is invisible and which particle takes
part in the hadronic activity as shown above. In
the reconstructed-level case, an additional method exists to test whether a muon is
isolated within an event \texttt{event},
\begin{verbatim}
  PHYSICS->IsIsolatedMuon(prt,event)
\end{verbatim}
which returns always \texttt{false} for particles which are not muon candidates.
In the case of muons, this method applies the isolation algorithm chosen by the
user when setting the \texttt{PHYSICS->recConfig()} properties. 

\begin{table}[t]
\bgfbyyyy
\multicolumn{2}{c}{\textbf{Table \ref{tab:physics2}: Boolean methods
included in the physics services}}\\$~$\\
\multicolumn{2}{p{.9\textwidth}}{Let \texttt{prt} be an instance of either the
\texttt{MCParticleFormat} or the \texttt{RecParticleFormat} classes.}\\$~$\\
{\tt PHYSICS->IsFinalState(prt)} &\\
&This checks if the particle \texttt{prt} is a final-state particle.\\
{\tt PHYSICS->IsHadronic(prt)} &\\
&This checks if the particle \texttt{prt} takes part to the hadronic
activity in an event.\\
{\tt PHYSICS->IsInitinalState(prt)}& \\
&This checks if the particle \texttt{prt} is an initial-state particle.\\
{\tt PHYSICS->IsInterState(prt)}& \\
&This checks if the particle \texttt{prt} is an intermediate-state 
particle.\\
{\tt PHYSICS->IsInvisible(prt)}& \\
&This checks if the particle \texttt{prt} is an invisible particle.\\
{\tt IsIsolatedMuon(prt,evt)} &\\ 
& This checks if the particle \texttt{prt} is
a muon isolated from the other particles of the event \texttt{evt}.\\
{\tt PHYSICS->IsVisible(prt)}& \\
&This checks if the particle \texttt{prt} is a visible particle.\\
\egfbyyyy
\textcolor{white}{\caption{\label{tab:physics2}}}
\end{table}

Another set of three methods checks whether a given particle \texttt{prt} is a
final-state, initial-state or intermediate-state particle,
\begin{verbatim}
  PHYSICS->IsFinalState(prt)
  PHYSICS->IsInitialState(prt)
  PHYSICS->IsInterState(prt)
\end{verbatim}
These functions all return a boolean value according to the final-state,
initial-state or intermediate-state nature of the particle under consideration.
As
above, those methods work equivalently for analyses at the hadronic, partonic
and reconstructed levels, the particle \texttt{prt} being hence either an instance of
the \texttt{MCParticleFormat} or of the \texttt{RecParticleFormat} classes.
Since status codes are
defined in a different fashion according to the event file format, we have
adopted the choice to include these features within the physics services rather than 
the data format itself, which allows us to have a unified way to probe the
initial-,
intermediate- or final-state nature of the particles included in an event.
These last series of functions allows us to implement efficiently a loop 
over, \eg,  the final-state particles of the event, and not on all the particles, 
in contrast to the example given in Section \ref{sec:mcformat},
\begin{verbatim}
  unsigned int n = event.mc()->particles().size();
  for (unsigned int i=0; i<n; i++) 
  {
    MCParticleFormat* prt = &event.mc()->particles()[i];
    if(PHYSICS->IsFinalState(prt))
      { ... }
  }
\end{verbatim}

Five methods included in the physics services are dedicated to the computation
of global observables related to the entire event content: the missing
transverse energy $\slashed{E}_T$, the missing hadronic energy $\slashed{H}_T$,
the total transverse energy $E_T$ and the total hadronic energy $H_T$ as defined
in Eq.\ \eqref{eq:miss} and Eq.\ \eqref{eq:vis}, as well as the partonic 
center-of-mass energy. The associated functions read, respectively, 
\begin{verbatim}
  PHYSICS->EventMET(evt->mc())   PHYSICS->EventMET(evt->rec())
  PHYSICS->EventMHT(evt->mc())   PHYSICS->EventMHT(evt->rec())
  PHYSICS->EventTET(evt->mc())   PHYSICS->EventTET(evt->rec())
  PHYSICS->EventTHT(evt->mc())   PHYSICS->EventTHT(evt->rec())
  PHYSICS->SqrtS(event->mc())    PHYSICS->SqrtS(event->rec())
\end{verbatim}
where \texttt{evt} is an instance of the \texttt{EventFormat} class.
These five methods work for any level of analysis 
(partonic, hadronic and reconstructed) and then take accordingly as argument
either an  \texttt{event->mc()} or an \texttt{event->rec()} object. 
The value, in GeV, of the corresponding observable is returned as a
floating-point number.

\begin{table}
\bgfbyyyy
\multicolumn{2}{c}{\textbf{Table \ref{tab:physics3}: Other methods
included in the physics services}}\\$~$\\
\multicolumn{2}{p{.9\textwidth}}{Let \texttt{evt} be an
instance of the \texttt{EventFormat} class assuming to point to an
\texttt{evt->mc()} or a \texttt{evt->rec()}
structure.} \\$~$\\
{\tt PHYSICS->ToRestFrame(prt,prt1)} &\\
&This recalculates (and modifies) the four-momentum of the particle \texttt{prt}
after a Lorentz boost to the rest frame of the particle \texttt{prt1}. The
objects \texttt{prt} and \texttt{prt1} are two instances of the
\texttt{MCParticleFormat} or \texttt{RecParticleFormat} classes.\\
{\tt PHYSICS->EventMET(evt)} &\\
& This computes the missing transverse energy $\slashed{E}_T$ associated
to the event \texttt{evt}. \\
{\tt PHYSICS->EventMHT(evt)} &\\
& This computes the missing transverse hadronic energy $\slashed{H}_T$ 
associated to the event \texttt{evt}. \\
{\tt PHYSICS->EventTET(evt)}& \\
&This computes the total transverse energy $E_T$ associated
to the event \texttt{evt}. \\
{\tt PHYSICS->EventTHT(evt)}& \\
&This computes the total transverse hadronic energy $H_T$ 
associated to the event \texttt{evt}. \\
{\tt PHYSICS->sort(prtvector,orderobs)}& \\
&This sorts a vector of \texttt{MCParticleFormat} or \texttt{MCRecFormat}
objects according to the ordering observable \texttt{orderobs}. The allowed
choices for the ordering variable are \texttt{ETAordering}
(pseudorapidity ordering), \texttt{ETordering} (transverse-energy ordering),
\texttt{Eordering} (energy ordering), \texttt{Pordering} (momentum ordering),
\texttt{PTordering} (transverse-momentum ordering), \texttt{PXordering}
(ordering according to the $x$-component of the momentum), \texttt{PYordering}
(ordering according to the $y$-component of the momentum) and
\texttt{PZordering} (ordering according to the $z$-component of the momentum).\\
{\tt PHYSICS->SqrtS(evt)}& \\
&This returns the partonic center-of-mass energy of the event in GeV.\\
\egfbyyyy
\textcolor{white}{\caption{\label{tab:physics3}}}
\end{table}

For most analyses, it is important to be able to order the particles according
to a specific variable. Therefore, in order to avoid requiring the user
to implement his own sorting algorithm, physics services include one
dedicated function, 
\begin{verbatim}
  PHYSICS->sort(prtVect, OrderingObs)
\end{verbatim}
The command above allows us to sort a vector of particles \texttt{prtVect},
that each entry is either a \texttt{MCParticleFormat} or a
\texttt{RecParticleFormat} object, according to the ordering observable
\texttt{OrderingObs}. The implemented choices for the latter 
are \texttt{ETAordering}
(pseudorapidity ordering), \texttt{ETordering} (transverse-energy ordering),
\texttt{Eordering} (energy ordering), \texttt{Pordering} (momentum ordering),
\texttt{PTordering} (transverse-momentum ordering), \texttt{PXordering}
(ordering according to the $x$-component of the momentum), \texttt{PYordering}
(ordering according to the $y$-component of the momentum) and
\texttt{PZordering} (ordering according to the $z$-component of the momentum).

The last method included in the physics services that can be useful when
implementing an analysis is related to the reference frame in which the
four-momenta of the particles included in the event are computed. When reading
the event files, all the four-momenta are, by convention, given in the laboratory
reference frame. However, for specific observables, it is necessary to
recalculate one or
several of the four-momenta in the rest frame of a given particle, which is
equivalent to applying a Lorentz boost to these four-vectors. This task can be done
automatically by means of the method
\begin{verbatim}
  PHYSICS->ToRestFrame(prt,prt1)
\end{verbatim}
where the objects \texttt{prt} and \texttt{prt1} are two instances of the
\texttt{MCParticleFormat} or \texttt{RecParticleFormat} classes. 
The four-momentum of the particle \texttt{prt} is boosted to
the rest frame of the particle \texttt{prt1} and overwritten. Let us note that
the Lorentz transformation is chosen in such a way that the tri-dimensional 
$(x,y,z)$ axes are kept unchanged.

\subsection{A detailed example}
In this Section, we give a detailed example about how to implement an
analysis within the expert mode of \madanalysis\ 5. We choose to investigate 
the polarization of the $W$-boson issued from a top quark decaying leptonically,
\bed\label{eq:topdecay}
  t \to W^+ b \to \ell^+\ \nu_l\ b \ .
\eed
This property of the $W$-boson is usually
studied through the shape of a particular angular distribution,
${\rm d}\sigma/{\rm d}\cos\theta^\ast$. The $\theta^\ast$ angle is defined as
the angle between the momentum of the $W$-boson evaluated in the rest frame of
the originating top quark and the one of the lepton $\ell^+$ 
evaluated in the rest frame of the $W$-boson.

To investigate this observable, we focus on the production of a 
top-antitop pair where one
of the final state top quarks undergoes a leptonic decay and the other one decays
hadronically,
\be
  p p \to t \bar t \to (b\ j\ j) (\bar b\ \ell^-\ \bar \nu_l)
  \qquad\text{or}\qquad
  p p \to t \bar t \to (b\ \ell^+\ \nu_l) (\bar b\ j\ j) \ , 
\ee
where $\ell$ stands for an electron or a muon and $j$ for a light jet. 
From the parton-level samples stored in the directory
\texttt{samples} of \madanalysis\ 5\footnote{If the \texttt{samples} directory
is absent, we recall that it can be created by issuing, in the command line
interface of \madanalysis\ 5, the command \texttt{install samples}.}, 
\texttt{ttbar\textunderscore sl\textunderscore 1.lhe.gz} and
\texttt{ttbar\textunderscore sl\textunderscore 2.lhe.gz}, 
we illustrate the implementation, within the \sampleanalyzer\ framework, of an
analysis code leading to the creation of a histogram representing the 
${\rm d}\sigma/{\rm d}\cos\theta^\ast$ distribution introduced above.
We refer to Section \ref{sec:firststeps} concerning details on the generation of
these two Monte Carlo samples.

When launching \madanalysis\ 5 in expert mode by issuing in a shell
\begin{verbatim}
  bin/ma5 -e
\end{verbatim}
the user is asked to enter the name of the directory to be created and the label
of the analysis. For the sake of the example, the name of the directory is
chosen to be \texttt{Wpol} whilst the string \texttt{W polarization from a top
decay} is entered as the label, or title, of the analysis.
The three files \texttt{analysisList.cpp}, \texttt{user.cpp} and \texttt{user.h}
are automatically created and stored in the directory
\texttt{Wpol/SampleAnalyzer/Analysis}, as mentioned in Section
\ref{sec:em_sampleana}. Whilst the first of these files does not
have to be modified by the user, the two others must be updated to include the
analysis which has to be implemented.

The file \texttt{user.h} strictly 
follows the structure presented in Section \ref{sec:em_template},
\begin{verbatim}
  #ifndef analysis_user_h
  #define analysis_user_h
  
  #include "Core/AnalysisBase.h"
  
  class user: public AnalysisBase
  {
    INIT_ANALYSIS(user,"W polarization from a top decay")
  
   public:
    virtual void Initialize();
    virtual void Finalize(const SampleFormat& summary, 
                const std::vector<SampleFormat>& files);
    virtual void Execute(const SampleFormat& sample, 
                         const EventFormat& event);
  
   private:
    TH1F* myHisto;
  };
  #endif
\end{verbatim}
The label of the analysis has been automatically included into
the arguments of the \texttt{INIT\textunderscore ANALYSIS} function at the
execution of \madanalysis\ 5, together with the
declaration of the three main functions \texttt{Initialize}, \texttt{Execute}
and \texttt{Finalize}. Since we are interested in the creation of 
a single histogram, we therefore need to declare, as a private member of the
\texttt{user} class, an
instance of the \texttt{TH1F} class\footnote{We refer to the \rooot\ manual
\cite{Brun:1997pa} for
the definition of the \texttt{TH1F} class and its attributes.} 
which we denote by \texttt{myHisto}.

The \cpp\ source file \texttt{user.cpp} contains the implementation of the
three core functions \texttt{Initialize}, \texttt{Execute}
and \texttt{Finalize}. Its structure reads thus 
\begin{verbatim}
#include "Analysis/user.h"

void user::Initialize()
  { ... }

void user::Execute(const SampleFormat& sample, 
                   const EventFormat& event)
  { ... }

void user::Finalize(const SampleFormat& summary, 
     const std::vector<SampleFormat>& files)
  { ... }
\end{verbatim}
where the dots are specified in the remainder of this Section. 
The function \texttt{Initialize} contains on the one hand the initialization of
the physics services as well as, on the other hand, the one of the histogram
to be drawn, following the \rooot\ syntax,
\begin{verbatim}
  void user::Initialize()
  {
    // Initializing Physics services
    PHYSICS->mcConfig().Reset();
  
    // Initializing the histogram
    myHisto = new TH1F("myHisto","cos#theta^{*}",15,-1.,1.);
  }
\end{verbatim}
We here ask for the creation of a histogram containing 15 bins
ranging from -1 to +1, the minimal and maximal values for the cosine of the
$\theta^*$-angle.

The implementation of the function \texttt{Execute} is a bit more complicated.
First of all, the observable of interest can only be computed once the three
relevant particles have been identified, \ie, the final-state lepton $\ell$, the 
intermediate $W$-boson decaying into this lepton $\ell$ and the originating top
quark. Then, the four-momentum of the $W$-boson has to be boosted into the rest
frame of the top quark and the one of the lepton into the rest frame of the
$W$-boson. Once this is done, the cosine of the angle $\theta^*$ can be computed
and the value filled into the histogram. The skeleton of the function
\texttt{Execute} reads thus
\begin{verbatim}
  void user::Execute(const SampleFormat& sample, 
    const EventFormat& event)
  {
    // Initialization of three pointers to the lepton, 
    // W and top quark 
    { ... }

    // Identification of the three particles of interest
    { ... }

    // Computing the observable; filling the histogram
    { ... }
  }
\end{verbatim}

The first series of dots concerns the declaration of the three objects which
contain the three particles of interest, once identified, and which are 
necessary to compute the observable $\cos\theta^*$,
\begin{verbatim}
  const MCParticleFormat* top    = 0;
  const MCParticleFormat* w      = 0;
  const MCParticleFormat* lepton = 0;
\end{verbatim}
In the lines above, three pointers to a \texttt{MCParticleFormat} structure are
created and initialized to the zero value. 

The second series of dots are related
to the identification of the top quark, the $W$-boson and the lepton.
In particular, the analysis code needs to
check that the mother-to-daughter relations given by the decay chain of Eq.\
\eqref{eq:topdecay} are fulfilled. Otherwise, the event must be rejected. In
practice, this task is performed through a loop over the
particle content of the event. The lepton is firstly selected as a final-state
particle whose PDG-id reads $\pm 11$ (electron) or $\pm 13$ (muon). The
$W$-boson is then further identified by requiring that the mother particle of
this lepton has a PDG-id equal to $\pm 24$. The top quark is finally identified
by requiring that the `grandmother' of the lepton has a PDG-id equal to $\pm
6$. The implemented code is given by
\begin{verbatim}
  for (unsigned int i=0;i<event.mc()->particles().size();i++)
  {
    const MCParticleFormat* prt=&(event.mc()->particles()[i]);

    // The lepton is a final state particle
    if (!PHYSICS->IsFinalState(prt)) continue;

    // Lepton selection based on the PDG-id
    if ( std::abs(prt->pdgid())!=11 && 
         std::abs(prt->pdgid())!=13 ) continue;

    // Getting the mother of the lepton and
    // checking if it is a W-boson
    const MCParticleFormat* mother = prt->mother1();
    if (mother==0) continue;
    if (std::abs(mother->pdgid())!=24) continue;

    // Getting the grand-mother of the lepton and
    // checking if it is a top quark
    const MCParticleFormat* grandmother = mother->mother1();
    if (grandmother==0) continue;
    if (std::abs(grandmother->pdgid())!=6) continue;

    // Saving the selected particles
    lepton = prt;
    w      = mother;
    top    = grandmother;

    // Particles are found: breaking the loop
    break; 
  }
\end{verbatim}
In the case the three particles are not identified, the event must be rejected,
which is done by implementing
\begin{verbatim}
  // Rejection of the event if the decay chain is not found
  if (lepton==0)
  {
    WARNING << "(t > b W > b l nu) decay chain not found!" 
            << std::endl;
    return;
  }
\end{verbatim}
where we have used the message services included in the \sampleanalyzer\
framework to print a warning message to the screen.

In the third and last series of dots included in the skeleton of the
\texttt{user.cpp} file, the observable $\cos\theta^*$ is computed and the
histogram filled, this last task being performed by means of standard
\rooot\ commands,
\begin{verbatim}
  // Boosting the lepton four-momentum to the W rest frame
  MCParticleFormat lepton_new = *lepton;
  PHYSICS->ToRestFrame(lepton_new,w);

  // Boosting the W four-momentum to the top rest frame
  MCParticleFormat w_new = *w;
  PHYSICS->ToRestFrame(w_new,top);

  // Computing the observable; filling the histogram 
  myHisto -> Fill( cos(lepton_new.angle(w_new)) );
\end{verbatim}

The histogram must now be created and exported to a human-readable format. This
is done at the level of the function \texttt{Finalize} by means of a series
of \rooot\ commands. For the sake of the example, we have decided to normalize
the histogram to the number of events ${\cal N}$ expected for 
an integrated luminosity of 10 fb$^{-1}$,
\be
  {\cal N} = \sigma {\cal L}_{\rm int} / N \ ,
\ee
where $\sigma$ is the cross section, in pb, corresponding to the process under
consideration, ${\cal L}_{\rm int}$ the integrated luminosity 
(thus set to 10000 pb$^{-1}$) and $N$ the
number of events included in the samples. As stated above, the value of the cross
section read from the \lhe\ files is 
stored in the \texttt{SampleFormat} object \texttt{summary} (see Section
\ref{sec:em_sampleformat}) where the average over both input samples has been
performed. The corresponding
\cpp\ implementation of the function \texttt{Finalize} reads 
\begin{verbatim}
  void user::Finalize(const SampleFormat& summary,
    const std::vector<SampleFormat>& files)
  {
    // Color of the canvas background
    gStyle->SetCanvasColor(0);

    // Turning off the border lines of the canvas
    gStyle->SetCanvasBorderMode(0); 

    // Configuring statistics printing  
    gStyle->SetOptStat(111110);
  
    // Creating the output root file
    TCanvas* myCanvas = new TCanvas("myCanvas","");
  
    // Setting background color
    myHisto->SetFillColor(kBlue);
  
    // Normalization of the histogram: L = 10 fb-1
    double nrm = summary.mc()->xsection() * 10000. / 
       static_cast<float>(summary.nevents());
    myHisto->Scale(nrm);
  
    // Setting axis title
    myHisto->GetXaxis()->SetTitle("cos#theta^{*}");
  
    // Drawing histogram
    myHisto->Draw();
  
    // Saving plot
    myCanvas->SaveAs((outputName_+".eps").c_str());
  }
\end{verbatim}

\begin{figure}[t]
  \center \includegraphics[width=0.90\textwidth]{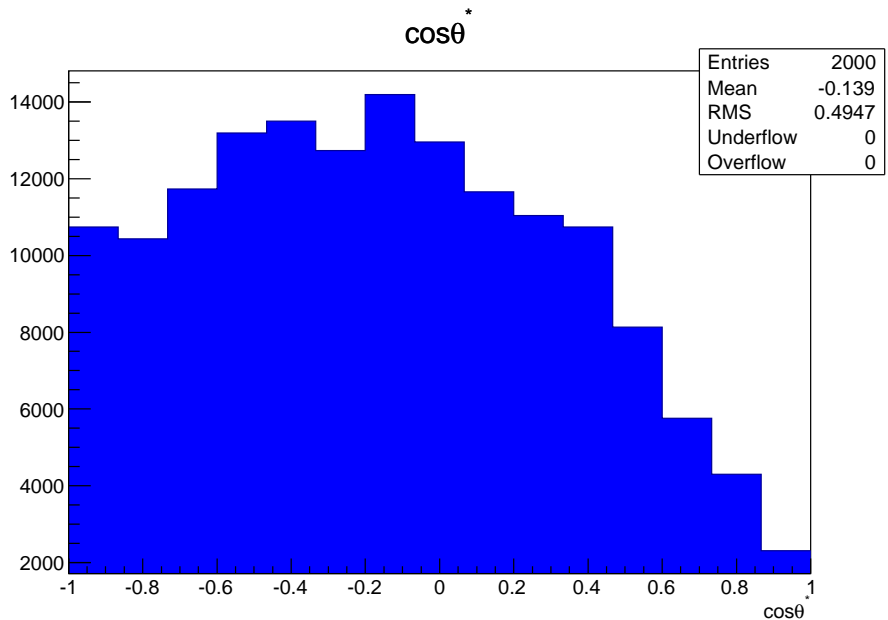}
  \caption{\label{fig:exp}$\cos\theta^*$ distribution for the
production of a top-antitop pair semi-leptonically decaying. The $\theta^*$
angle is defined as in the text.}
\end{figure}

After compiling the analysis and linking it to the external libraries with the
help of the provided \texttt{Makefile} located in the \texttt{SampleAnalyzer}
directory, the analysis can be executed,
\begin{verbatim}
  SampleAnalyzer --analysis="W polarization from a top decay" \
    list.txt
\end{verbatim}
where \texttt{list.txt} is a text file containing the absolute paths to the
two event samples under consideration. A figure, named \texttt{list.eps}, is
generated in the
directory where the executable is stored. The results, shown in Figure
\ref{fig:exp}, agree, for instance, with Ref.\ \cite{Giammanco:2005ta}.

\section{Conclusions}\label{sec:conclusions}

The task of performing a phenomenological analysis based on event files such
as those generated by Monte Carlo
generators can be divided into three stages. Firstly, the event samples must be
read and loaded into the memory of the computer. The format of these files
depends in general on the level of sophistication of the analysis
(at the parton-level, hadron-level or reconstructed-level). Secondly, the
analysis itself must be performed, \ie, selection cuts are applied on the
signal and background event samples with the aim of being able to extract
information on the signal from the often overwhelming background. 
Finally, the results are outputted as
histograms and/or cut-flow charts to improve their readability and
interpretation.

In this work, we have presented \madanalysis\ 5, a new user-friendly and
efficient framework aiming to facilitate the implementation of phenomenological
analyses such as the one described above. We have explained how to implement and
run an analysis within this framework in a straightforward way and have given
several detailed examples addressing the different facets of the program.

\madanalysis\ 5 is based on a multi-purpose
\cpp\ kernel, \sampleanalyzer, which uses the \rooot\ platform. Compatible with
most of the event file formats commonly used for analyzing parton-level,
hadron-level and reconstructed events, \madanalysis\ 5 offers two modes of
running according to the needs and the expertise of the user. 

A highly-user-friendly mode, the normal running mode of the program, uses the
strengths of a \python\ interface to reduce the implementation of an analysis to
a set of intuitive commands whose syntax has been inspired by the \python\
programming language. Therefore, rather complex analyses can be achieved without
too much effort.

For users with more advanced \cpp\ and \rooot\ programming skills, \madanalysis\
5 can also be run in its expert mode. This overcomes the limitations of the
normal mode of running in which the scope is restricted by the set of
functionalities that have been implemented. The user is in this case required to
directly implement his analysis in \cpp\ within the \sampleanalyzer\ framework,
rendering the possibilities, at the analysis level, only limited by the
imagination and the skills of the user. However, even if this mode of running is
in principle more complicated to handle, the existence of many built-in
functions and methods renders the task of implementing the analysis easier and
more straightforward.

\section*{Acknowledgments}
The authors are extremely grateful to J.\ Andrea for being the first user and
beta-tester of this program. We also thank the \madgraph\ 5 development team
(J.\ Alwall, F.\ Maltoni, O.\ Mattelaer and T.\ Stelzer) and R.\ Frederix for
commenting and supporting the development of \madanalysis\ 5 as well as our
colleagues from Strasbourg for their help in testing and debugging the code, 
J.L.\ Agram, A.\ Alloul, A.\ Aubin, E.\ Chabert, C.\ Collard, A.\ Gallo, P.\
Lansonneur and S.\ Marrazzo. Finally, we acknowledge V.\ Boucher, J.\ de
Favereau and P.\ Demin for their help in administrating our web and {\sc svn}
server. This work has been supported by the Theory-LHC France-initiative of the
CNRS/IN2P3 and a Ph.D.\ fellowship of the French ministry for education and
research.
\appendix 
\section{Installation of the program}\label{sec:install}
\subsection{Requirements}
For a proper running, \madanalysis\ 5 requires several mandatory external
libraries. In addition, the full set of functionalities of the program can be
made available by installing optional external dependencies on the computer of
the user. If these optional libraries are absent, several of the functionalities
of \madanalysis\ 5 are deactivated but analyses of
event files can still be performed. In contrast, the absence of one of the
mandatory external libraries simply does not allow use of the program.

In order to run \madanalysis\ 5 locally, \python\ 2.6 or a more recent
version (however not from the 3.x series) must be installed on the computer of
the user \cite{python}. We recall that in order to check the version of  
\python\ present on a system, it is enough to type in a shell 
\begin{verbatim}
  python --version
\end{verbatim}
The installation of the \python\ package is one of the three mandatory
requirements without which the program cannot run. 
The two other external packages that have to be installed are related to \cpp\
and \rooot.

The \sampleanalyzer\ kernel requires the installation of a \cpp\ compiler
together with the associated Standard Template Libraries (\textsc{STL}). Since
the validation procedure of \madanalysis\ 5 has only been achieved
within the context of the {\sc GNU gcc} compiler and since this compiler is
available on most operating systems \cite{gcc}, we have adopted the choice of
requiring the installation of {\sc gcc}. The program has been validated with the
versions 4.3.x and 4.4.x. 
We recall that the version of the {\sc gcc}
installed on a system can be obtained by issuing in a shell
\begin{verbatim}
  g++ --version
\end{verbatim}
Let us however stress that compatibility with any other \cpp\
compiler is in principle ensured, but requires the modification of several core
files of \madanalysis\ 5. Therefore, it is currently not supported.

Concerning the
\rooot\ package, a version more recent than version 5.27 has to be installed
\cite{rootweb}, and the user has to check that the \python\ functionalities of
the \rooot\ library are available. We remind that in order to install a version
of \rooot\ including its \python\ library, the {\sc Linux} package {\sc
Python-devel} has to be present on the system of the user and the \rooot\
configuration script must be run as 
\begin{verbatim}
  ./configure --with-python
\end{verbatim} 
Very importantly, the version of \python\ employed to start \madanalysis\ 5 must
match the one used to generate the \python\ library of \rooot. Finally, the
version of \rooot\ present on a computer, together with the compatibility with
the \python\ language, can be checked by issuing in a shell the commands
\begin{verbatim}
  root-config --version
  root-config --has-python
\end{verbatim}

The three above-mentioned programs, \ie, \python, the {\sc gcc}
compiler and \rooot, must be available from any location of the computer. If
this is not the case, the user has to modify the system variable
\texttt{\$PATH}. If these programs have been installed at standard
location on the system, the necessary environment variables are set
automatically by \madanalysis\ 5. Contrary, 
it is left to the user to check that the paths to the associated
header and library files are included in the environment variables of the {\sc
gcc} compiler, \texttt{\$CPLUS\textunderscore INCLUDE\textunderscore PATH} and
\texttt{\$LIBRARY\textunderscore PATH}. Finally, for a proper
linking of the external libraries, the variable \texttt{\$LD\textunderscore 
LIBRARY\textunderscore PATH} (or, on {\sc MacOS} operating systems,
\texttt{\$DYLD\textunderscore LIBRARY\textunderscore PATH})
must also contain the paths to the libraries to be linked.

We now turn to the optional libraries that can be linked to \madanalysis\ 5.
In order to handle zipped Monte Carlo event samples, the \textsc{zlib} library
has to be installed on the system \cite{zlib}. However, if \textsc{zlib} is not
locally installed, the possibility of analyzing zipped event samples is simply
deactivated, \ie, the user will have to unzip the event files manually before
running the code.

\subsection{Downloading the program}
It is recommended to always use the latest stable version of the \madanalysis\ 5
package, which contains, when downloaded from the web, both the \python\ command
line interface and the \sampleanalyzer\ framework. It can always be found
together with an up-to-date manual on the webpage

\maweb
\medskip

The package does not require any compilation or configuration. After having
downloaded the tar-ball from the website, it can be unpacked either in the
directory where \madgraph\ 5 is installed,
\begin{verbatim}
  cd <path-to-madgraph>
  tar -xvf ma5_xxxx.tgz
\end{verbatim}
 or in any location on the computer of the user,
\begin{verbatim}
  mkdir <ma5-dirname>
  cd <ma5-dirname>
  tar -xvf ma5_xxxx.tgz
\end{verbatim}
where \texttt{xxxx} stands for the version number of the \madanalysis\ 5 release
which has been downloaded. When \madanalysis\ 5 is installed as a dependency of
\madgraph\ 5, the list of predefined particle and multiparticle labels is
directly updated from the \madgraph\ 5 standards. Moreover, this allows to
perform analyses at the time of the event generation by \madgraph\ 5 which,
in this case, pilots \madanalysis\ 5. We emphasize that several predefined
analyses exist, but the user has the freedom to
tune the desired analysis to be performed according to his own aims.

Once installed and unpacked, \madanalysis\ 5 can be immediately laun\-ched by
issuing in a shell
\begin{verbatim}
  bin/ma5
\end{verbatim}
as presented in Section \ref{sec:uf_starting}.

\subsection{Running \madanalysis\ 5}

When started, \madanalysis\ 5 first checks that all the dependencies ({\sc gcc},
\python, \rooot, {\sc zlib}) are present on the system and that compatibility is
ensured with the installed versions. In the case of any problem, a message is
printed to the screen and the code exists if it is found that it cannot properly
run. On the first session of \madanalysis\ 5, the \sampleanalyzer\ core is
compiled behind the scene as a static library stored in the directory
\texttt{lib}. For the next sessions, the kernel is only recompiled if the
configuration of the system has changed (new version of the dependencies or of
the main program).


\bibliographystyle{elsarticle-num}
\bibliography{biblio}

\end{document}